\documentclass[
 reprint,
 superscriptaddress,
 amsmath,amssymb,
 aps,
 prx,
]{revtex4-2}

\usepackage{graphicx}
\usepackage{dcolumn}
\usepackage{bm}
\usepackage{hyperref}
\usepackage{longtable}

\usepackage{hyperref}
\hypersetup{
    colorlinks=true,
    linkcolor=blue,
    urlcolor=blue,  
}

\usepackage{xcolor}
\usepackage{amsfonts, amsmath, amssymb, amsthm}
\usepackage{braket}

\graphicspath{ {./figures/} }

\usepackage[normalem]{ulem}

\begin{document}
\preprint{APS/123-QED}

\title{Unraveling the switching dynamics in a quantum double-well potential}

\author{Qile Su}
\email[Contact author: ]{q.su@yale.edu}

\author{Rodrigo G. Cortiñas}
\altaffiliation[Present affiliation: ]{Google Quantum AI}

\author{Jayameenakshi Venkatraman}
\altaffiliation[Present affiliation: ]{Department of Physics, University of California Santa Barbara, Santa Barbara, CA 93106, USA}

\author{Shruti Puri}

\affiliation{Department of Physics and Applied Physics, Yale University, New Haven, CT 06511, USA}
\affiliation{Yale Quantum Institute, Yale University, New Haven, CT 06511, USA}

\date{\today}

\begin{abstract}
The spontaneous switching of a quantum particle between the wells of a double-well potential is a phenomenon of general interest to physics and chemistry. It was broadly believed that the switching rate decreases steadily as a function of the size of the energy barrier. This view was challenged by a recent experiment on a driven superconducting Kerr nonlinear oscillator (often called the \textit{Kerr-cat qubit} or the \textit{Kerr parametric oscillator}), whose energy barrier can be continuously increased by ramping up the drive. Remarkably, as the drive amplitude increases, the switching rate exhibits a step-like decrease termed the ``staircase". The view challenged by the experimental staircase demands a deep review of our understanding of the role of quantum effects in double wells. In this work, we use a Lindbladian model of dissipation to derive a semi-analytical formula for the switching rate, resolving a continuous transition between tunneling-dominated dynamics and dissipation-dominated dynamics. These two dynamics are observed respectively in the flat part and the steep part of each step in the staircase. Our formula exposes two distinct dissipative processes that limit tunneling: the dephasing of inter-well superpositions freezes tunneling via the quantum Zeno effect, and the decay from one excited state to another limits the time during which tunneling can be attempted. This physical understanding allows us to pinpoint the critical drive amplitude that separates the flat and steep part of each step using an equation involving the rates of the two dissipative processes and the tunnel splitting. In addition, analyzing the transition matrix elements in the formula shows that in the regime of a few ($\lesssim 10$) states in the well and under moderate to low temperatures, highly excited states are populated predominantly via cascaded and direct thermal heating rather than quantum heating. At very low temperatures, we find that the perturbation induced by the nonhermitian Hamiltonian part of the Lindbladian model becomes increasingly important and facilitates a form of quantum heating that has not been identified before. We numerically map the activation mechanism as a function of drive amplitude, damping rate, and temperature. Our theory deepens the understanding of switching dynamics between metastable quantum states, highlights the importance of a general interplay between tunneling and dissipation, and identifies a novel quantum regime in activated transitions.
\end{abstract}

\maketitle

\section{Introduction}
The quantum mechanical double-well potential is ubiquitous in physics and has a long history as the prototypical model for quantum tunneling and activated transitions between metastable states. Many examples of the double-well potential can be found in nuclear \cite{coleman_uses_1979}, atomic \cite{albiez_direct_2005}, chemical \cite{hanggi_reaction-rate_1990, minissale_quantum_2013}, optomechanical \cite{buchmann_macroscopic_2012}, and solid-state systems \cite{lisenfeld_decoherence_2016}. One particular example is the parametrically-driven Kerr/Duffing nonlinear oscillator, where the interplay between the Kerr nonlinearity and the parametric drive (two-photon drive from here on) creates an effective energy double well that is tunable via the drive \cite{milburn_quantum_1991, wielinga_quantum_1993, marthaler_quantum_2007, dykman_fluctuating_2012, guo_phase_2013, zhang_preparing_2017}. Enabled by the large nonlinearity and low loss achievable in the superconducting circuit architecture \cite{schreier_suppressing_2008, kirchmair_observation_2013, blais_circuit_2021, bland_2d_2025}, this type of driven oscillator has entered the quantum regime and found applications in gate-based quantum computing \cite{goto_universal_2016, puri_engineering_2017, royer_qubit_2018, puri_stabilized_2019, puri_bias-preserving_2020, darmawan_practical_2021, masuda_controls_2021, xu_engineering_2022, xue_fast_2022, kanao_quantum_2022, masuda_fast_2022, chono_two-qubit_2022, kang_nonadiabatic_2022} and quantum annealing \cite{goto_bifurcation-based_2016, nigg_robust_2017, puri_quantum_2017, zhao_two-photon_2018, onodera_quantum_2020, goto_quantum_2020, kewming_quantum_2020, kanao_high-accuracy_2021, miyazaki_effective_2022}, where it is sometimes known under the name of the Kerr-cat qubit \cite{grimm_stabilization_2020} or the Kerr parametric oscillator \cite{goto_bifurcation-based_2016}. In these applications, information is encoded in the oscillator as a coherent superposition between the two metastable states at the bottom of the effective energy double well \cite{cochrane_macroscopically_1999}. Steady experimental progress has been made toward these applications of the oscillator. In addition to parametric oscillations in the quantum regime \cite{yamaji_spectroscopic_2022, yamaji_correlated_2023}, experiments have not only observed by also demonstrated remarkable control over manifestly quantum effects such as a discrete spectrum, tunneling, interference, and entanglement 
\cite{wang_quantum_2019, grimm_stabilization_2020, yamaguchi_spectroscopy_2024, frattini_observation_2024, iyama_observation_2024, venkatraman_driven_2024, ding_quantum_2025, hajr_high-coherence_2024-2, albornoz_oscillatory_2024}.

A fundamental property relevant to the quantum computing and annealing applications of the oscillator is the spontaneous inter-well switching rate between the wells of the double well because it limits the lifetime of the information encoded in the oscillator. A semiclassical analysis is insufficient to capture the switching rate between metastable states in the quantum regime \cite{andersen_quantum_2020}. Interestingly, it was discovered in one of the experiments \cite{frattini_observation_2024} that the spontaneous switching rate displays a step-like dependence as a function of the depth of the double well, rather than a smooth exponential dependence predicted by Arrhenius law \cite{hanggi_reaction-rate_1990}. This effect was also reproduced in subsequent experiments \cite{hajr_high-coherence_2024-2, albornoz_oscillatory_2024}. The effect, dubbed the ``staircase", was overlooked by previous theoretical works \cite{dykman_theory_1978, dykman_quantum_1988, marthaler_switching_2006, dykman_fluctuating_2012, puri_stabilized_2019, puri_bias-preserving_2020, gautier_combined_2022, putterman_stabilizing_2022}. A full understanding of the physical origin of the staircase would benefit the quantum information and annealing applications of this system.

In this work we bridge the gap between theory and experiments by finding a semi-analytical formula for the spontaneous switching rate of the two-photon-driven Kerr nonlinear oscillator. The formula resolves a continuous transition between tunneling-dominated dynamics and dissipation-dominated dynamics. Our formula shows that the physics behind the staircase is explained by a two-stage process that involves activation from the ground state manifold of the double well into the excited state manifolds followed by tunneling. Specifically, each step of the staircase can be attributed to tunneling within one excited state manifold. The shape of the step is caused by quantum Zeno-like competition between tunneling and dissipation in this manifold. In the flat part of the step, the dynamics in the manifold is dominated by tunneling, whereas in the steep part it is dominated by dissipation. Within the latter regime, two distinct dissipative processes limit tunneling: the dephasing of inter-well superpositions freezes tunneling via the quantum Zeno effect \cite{misra_zenos_1977, itano_quantum_1990, facchi_quantum_2008, greenfield_unified_2025}, and the decay from one excited state to another limits the time during which tunneling can be attempted.

For each step of the staircase, we quantify the critical two-photon drive amplitude that separates the tunneling-dominated and the dissipation-dominated regimes. The smallest of such critical two-photon drive amplitudes has important practical implications because it must be exceeded for the switching rate to not be limited by tunneling in the first excited state manifold. This condition marks the onset of the overall exponential suppression of the switching rate as a function of drive amplitude.

We also use our theory of the switching rate to address an important question: what is the dominant activation mechanism in the two-photon-driven Kerr nonlinear oscillator? Previous work has shown that in similar systems, the dominant activation mechanism is quantum heating \cite{marthaler_switching_2006, dykman_critical_2007, vijay_invited_2009, dykman_quantum_2011, dykman_fluctuating_2012, ong_quantum_2013}, which can excite the population in the double well even when the temperature of the environment is zero. In contrast, we show that in our system the dominant physical mechanisms for the activation step of the aforementioned two-stage switching process are direct and cascaded thermal activation unless the temperature is very low (quantified numerically). The difference in activation mechanism from similar systems is ultimately because under the drive condition that we consider, the double well potential is shaped such that the ground and excited states in the wells are approximately displaced Fock states. For such states the matrix elements associated with quantum heating are negligible. For very low temperatures, however, our theory reveals that the perturbation induced by the nonhermitian effective Hamiltonian part \cite{dum_monte_1992, dalibard_wave-function_1992, carmichael_open_1993, molmer_monte_1993, plenio_quantum-jump_1998, carmichael_statistical_2008} of the Lindblad master equation \cite{carmichael_statistical_1999, breuer_theory_2002, gardiner_quantum_2010} becomes important and facilitates a new form of quantum heating that has not been identified nor observed before. The temperature at which this effect is observable can be raised by increasing the damping rate $\kappa/K$. Finally, we numerically map the activation mechanism as a function of drive amplitude, damping rate, and temperature.

In section~\ref{sec:background} we provide background on the two-photon-driven Kerr nonlinear oscillator by introducing its effective Hamiltonian, energy levels, and energy eigenstates. We introduce the Lindbladian that models the effect of the environment, and then define what we mean by the spontaneous switching rate. In section~\ref{sec:solving-for-the-spontaneous-switching-rate} we make approximations, derive the formula for the switching rate, interpret the formula, and then compare it to numerical simulations. In section~\ref{sec:location-of-steps} we calculate the critical drive amplitudes at which the step-like decreases of the spontaneous switching rate occur as the result of quantum Zeno-like inhibition of coherent tunneling. In doing so, we identify two distinct dissipative processes that limit tunneling. In section~\ref{sec:activation-mechanism} we show that the activation mechanism for switching is dominated by direct and cascaded thermal activation for moderate to low temperatures. Then we discuss the new form of quantum heating in the case of very low temperatures and show a map of activation mechanisms as a function of drive amplitude, damping rate, and temperature. In section~\ref{sec:comparison-to-previous-work} we compare our work to previous works. Finally in section~\ref{sec:conclusions} we provide concluding remarks.

\begin{longtable*}{p{1.0in}p{6.0in}}
\caption{Table of symbols.} \\
\hline \hline \\[-2ex]
	\multicolumn{1}{l}{Symbol} &
	\multicolumn{1}{l}{Definition / Result} \\ \hline
\\[-1.8ex]
\endhead
\multicolumn{2}{c}{\textit{Section \ref{sec:background}: Background}} \\

$K$ & The size of the Kerr nonlinearity in the Kerr nonlinear oscillator, expressed in units of angular frequency. \\

$\epsilon_2$ & The two-photon drive amplitude, expressed in units of angular frequency. \\

$\hat{H}$ & The Hamiltonian of the two-photon-driven Kerr nonlinear oscillator under the rotating-wave approximation and in a frame rotating at the oscillator frequency. It is written in units of angular frequency: $\hat{H} = -K\hat{a}^{\dagger2}\hat{a}^2 + \epsilon_2(\hat{a}^{\dagger2} + \hat{a}^2)$ [Eq.~\eqref{eq:hamiltonian}], where $[\hat{a}, \hat{a}^\dagger] = 1$. \\

$\ket{\pm\alpha}$ & The two coherent states that span the highest energy manifold (also the ground state manifold) of $\hat{H}$. The amplitude is $\alpha = \sqrt{\epsilon_2/K}$ [Eq.~\eqref{eq:ground-state}]. \\

$E_0$ & The energy eigenvalue of the highest energy manifold of $\hat{H}$. Its value is $E_0 = \epsilon_2^2/K$ [Eq.~\eqref{eq:ground-state}]. \\

$\ket{\mathcal{C}_{\alpha}^\pm}$ & The even and odd parity Schrödinger cat states that span the highest energy manifold of $\hat{H}$. They are defined as $\ket{\mathcal{C}_{\alpha}^\pm} = \mathcal{N}_{\alpha}^{\pm}(\ket{+\alpha} \pm \ket{-\alpha}$) [Eq.~\eqref{eq:cat-states}], where $\mathcal{N}_{\alpha}^{\pm} = (2 \pm 2e^{-2|\alpha|^2})^{-1/2}$. \\

$\ket{\psi_n^\pm}$ & Even and odd parity energy eigenstates of $\hat{H}$ [Eq.~\eqref{eq:energies}]. The $n = 0$ case reduces to $\ket{\psi_0^\pm} = \ket{\mathcal{C}_{\alpha}^\pm}$. \\

$E_n^\pm$ & Energy eigenvalues of $\hat{H}$ corresponding to the eigenstates $\ket{\psi_n^\pm}$ [Eq.~\eqref{eq:energies}]. The $n = 0$ case reduces to $E_0^\pm = E_0$. \\

$U(\beta)$ & The semiclassical double-well potential in the phase space of the nonlinear oscillator obtained from the negative expectation value of $\hat{H}$ on coherent states $\ket{\beta}$ [Eq.~\eqref{eq:double-well}]. \\ 

$\delta_n$ & The angular frequency of the tunnel splitting between the even and odd parity states in manifold $n$, defined as $\delta_n = E_n^+ - E_n^-$ [Eq.~\eqref{eq:tunnel-splitting-definition}]. \\

$E_{\text{gap}, n}$ & The energy gap between neighboring energy manifolds, written in units of angular frequency [Eq.~\eqref{eq:energy-gap-definition}]. \\

$\kappa$ & The energy damping rate (single photon loss rate) of the nonlinear oscillator due to the environment. \\

$n_{\text{th}}$ & The thermal occupation number of the environment. \\

$C$ & A set of collapse operators [Eq.~\eqref{eq:collapse-operators}] to describe the environment effects. It includes single photon loss and single photon gain: $C = \{ \sqrt{\kappa(n_{\text{th}} + 1)}\hat{a}, \sqrt{\kappa n_{\text{th}}}\hat{a}^\dagger \}$. \\

$\mathcal{D}[\hat{O}]$ & The dissipator. It is defined as $\mathcal{D}[\hat{O}]\hat{\rho} = \hat{O}\hat{\rho}\hat{O}^\dagger - \frac{1}{2}\hat{O}^\dagger\hat{O}\hat{\rho} - \frac{1}{2}\hat{\rho}\hat{O}^\dagger\hat{O}$, where $\hat{O}$ is a collapse operator. \\

$\hat{\rho}$ & The oscillator's density operator. \\

$\mathcal{L}$ & The Lindbladian in the master equation [Eq.~\eqref{eq:lindblad}]. It is defined as $\mathcal{L}(\hat{\rho}) = -i[\hat{H}, \hat{\rho}] + \sum_{\hat{O} \in C} \mathcal{D}[\hat{O}]\hat{\rho} $. \\

$\Gamma$ & The spontaneous switching rate between the wells, defined as the rate of exponential decay of the inter-well population difference after the intra-well quasi-equilibrium state has been achieved [Eq.~\eqref{eq:definition-of-Gamma-0}]. \\
 & \\

\multicolumn{2}{c}{\textit{Section \ref{sec:solving-for-the-spontaneous-switching-rate}: Solving for the spontaneous switching rate}} \\

$\ket{\psi_n^p}\bra{\psi_m^q}$ & Terms in the density operator $\hat{\rho}$ when it is written in the basis $\ket{\psi_n^\pm}$. When $m = n$ and $p = q$, it is referred to as population. When $m = n$ and $p \neq q$, it is referred to as in-manifold coherence. When $m \neq n$, it is referred to as cross-manifold coherence. \\

$\hat{I}_n$ & The orthogonal projector onto manifold $n$, defined as $\hat{I}_n = \ket{\psi_n^+}\bra{\psi_n^+} + \ket{\psi_n^-}\bra{\psi_n^-}$ [Eq.~\eqref{eq:projector}]. \\

$\mathcal{P}$ & The projector that removes cross-manifold coherences from the density operator, defined as $\mathcal{P}(\hat{\rho}) = \sum_n\hat{I}_n\hat{\rho}\hat{I}_n$ [Eq.~\eqref{eq:projection}]. \\

$\hat{\rho}_{\text{proj}}$ & The density operator after the cross-manifold coherences have been neglected. It is defined as $\hat{\rho}_{\text{proj}} = \mathcal{P}(\hat{\rho})$ [Eq.~\eqref{eq:projection}]. \\

$\hat{H}_{\text{eff}}$ & The nonhermitian effective Hamiltonian part of the Lindbladian $\mathcal{L}$ at zero temperature, defined as $\hat{H}_{\text{eff}} = \hat{H} - i\kappa\hat{a}^\dagger\hat{a}/2$ [Eq.~\eqref{eq:effective-hamiltonian}]. \\

$\ket{\psi_n^{R/L}}$ & Right- and left-well states defined as $\ket{\psi_n^{R/L}} = (\ket{\psi_n^+} \pm \ket{\psi_n^-}) / \sqrt{2}$ [Eqs.~\eqref{eq:right-wellstate} and \eqref{eq:left-wellstate}]. \\

$\hat{X}_n$ & The Pauli $x$ operator in the manifold $n$. It is defined as $\hat{X}_n = \ket{\psi_n^R}\bra{\psi_n^R} - \ket{\psi_n^L}\bra{\psi_n^L}$ [Eq.~\eqref{eq:pauli-x}]. \\

$\hat{Z}_n$ & The Pauli $z$ operator in the manifold $n$. It is defined as $\hat{Z}_n = \ket{\psi_n^+}\bra{\psi_n^+} - \ket{\psi_n^-}\bra{\psi_n^-}$ [Eq.~\eqref{eq:pauli-z}]. \\

$\hat{Y}_n$ & The Pauli $y$ operator in the manifold $n$. It is defined as $\hat{Y}_n = i\hat{X}_n\hat{Z}_n$ [Eq.~\eqref{eq:pauli-y}]. \\

$\mathcal{L}_{\text{eff}}$ & The effective Lindbladian after ignoring the cross-manifold coherences. It is defined as $\mathcal{L}_{\text{eff}} = \mathcal{P}\mathcal{L}\mathcal{P}$ [Eq.~\eqref{eq:effective-lindbladian}]. \\

$\mathcal{L}_{\text{H}}$ & The Hamiltonian part of the effective Lindbladian, defined as $\mathcal{L}_{\text{H}}(\hat{\rho}) = -i\mathcal{P}[\hat{H}, \mathcal{P}(\hat{\rho})]$ [Eq.~\eqref{eq:projected-hamiltonian}]. \\

$\mathcal{L}_{\text{D}}$ & The dissipative part of the effective Lindbladian, defined as $\mathcal{L}_{\text{D}}(\hat{\rho}) = \sum_{\hat{O} \in C}\mathcal{P}\mathcal{D}[\hat{O}]\mathcal{P}(\hat{\rho}$) [Eq.~\eqref{eq:projected-dissipator}]. \\

$\hat{O}_{pq}$ & The projected collapse operator, defined as $\hat{O}_{pq} = \hat{I}_p\hat{O}\hat{I}_q$ [Eq.~\eqref{eq:O_pq-definition}]. \\

$\hat{X}_{pq}$ & The transition operator from manifold $q$ to manifold $p$ that simultaneously flips the Bloch vector around the $x$-axis. It is defined as $\hat{X}_{pq} = \ket{\psi_p^R}\bra{\psi_q^R} - \ket{\psi_p^L}\bra{\psi_q^L}$ [Eq.~\eqref{eq:definition-x-pq}]. \\

$\hat{Y}_{pq}$ & The transition operator from manifold $q$ to manifold $p$ that simultaneously flips the Bloch vector around the $y$-axis. It is defined as $\hat{Y}_{pq} = i(\ket{\psi_p^R}\bra{\psi_q^L} - \ket{\psi_p^L}\bra{\psi_q^R})$ [Eq.~\eqref{eq:definition-y-pq}]. \\

$V_{pq}$ & The inter-well transition rate, defined as $V_{pq} = \kappa(1 + n_{\text{th}})\ |\!\bra{\psi_p^L}\hat{a}\ket{\psi_q^R}\!|^2 + \kappa n_{\text{th}}\ |\!\bra{\psi_p^L}\hat{a}^\dagger\ket{\psi_q^R}\!|^2$ [Eq.~\eqref{eq:V-definition}]. \\

$W_{pq}$ & The intra-well transition rate, defined as $W_{pq} = \kappa(1 + n_{\text{th}})\ |\!\bra{\psi_p^R}\hat{a}\ket{\psi_q^R}\!|^2 + \kappa n_{\text{th}}\ |\!\bra{\psi_p^R}\hat{a}^\dagger\ket{\psi_q^R}\!|^2$ [Eq.~\eqref{eq:W-definition}]. \\

$\hat{X}$ & The total population difference operator between the wells, defined as $\hat{X} = \sum_n\hat{X}_n$ [Eq.~\eqref{eq:definition-of-X}]. \\

$\Gamma_n$ & The contribution to the spontaneous switching rate $\Gamma$ from the manifold $n$. It is defined by Eq.~\eqref{eq:switching-rate-n}. \\

$\mu_n$ & The dephasing rate due to the environment effectively measuring ``which-well" information. It is defined as $\mu_n = 2W_{nn} + \sum_{f \neq n}(W_{fn} + V_{fn})$ [Eq.~\eqref{eq:dephasing-rate}]. \\

$\lambda_n$ & The decay rate of population in the manifold $n$ due to transitions to other manifolds in the same well or to the other well (but excluding the effect of tunneling). $\lambda_n = 2\sum_f V_{fn} + \sum_{f \neq n}(W_{fn} - V_{fn})$ [Eq.~\eqref{eq:decay-rate}]. \\

$\delta_n^2/\mu_n$ & The effective tunneling rate in manifold $n$ after taking into account the quantum Zeno effect. It appears in the expression of the switching rate $\Gamma$ [Eq.~\eqref{eq:switching-rate-x-only}] and the effective equations of motion [Eq.~\eqref{eq:matrix-eq-x-effective}]. \\

$R_{mn}$ & A shorthand notation used to express the quasi-equilibrium distribution $\braket{\hat{X}_n}$, $R_{mn} = (W_{mn} - V_{mn})/(\lambda_m + \delta_m^2/\mu_m)$ [Eq.~\eqref{eq:definition-of-Rmn}]. \\

$K_n$ & The ratio between $\braket{\hat{X}_n}$ and $\braket{\hat{X}_0}$, as defined in Eq.~\eqref{eq:x-down-correction}. It can be expressed analytically as in Eq.~\eqref{eq:K-definition}, $K_{n>0} = \sum_{\text{paths}} R_{ni_{s-1}}\cdots R_{i_2i_1}R_{i_10}.$ \\

$f_n$ & The branching ratio in manifold $n$, defined in Eq.~\eqref{eq:branching-ratio}, $f_n = (\delta_n^2/\mu_n + 2\sum_fV_{fn})\ /\ (\delta_n^2/\mu_n + \lambda_n)$. \\

$J_n$ & The inter-well difference of the total probability current entering manifold $n$, defined in Eq.~\eqref{eq:probability-current}, $J_n = \sum_{i \neq n}(W_{ni} - V_{ni})(K_i/\sum_{m\geq0}K_m)$. \\

$\Gamma_n(\delta_k, W_{pq}, V_{pq})$ & The semi-analytical formula for the switching rate contributed by manifold $n$, written in terms of the tunnel splittings and the intra-/inter-well transitions rates. $\Gamma_0(\delta_k, W_{pq}, V_{pq}) = \left(\delta_0^2/\mu_0 + 2\sum_fV_{f0}\right)(K_0/\sum_{m\geq0}K_m)$, and $\Gamma_{n>0}(\delta_k, W_{pq}, V_{pq}) = f_nJ_n$ [Eq.~\eqref{eq:Gamma-n-solution-summarized}]. \\

$\Gamma(\delta_k, W_{pq}, V_{pq})$ & The semi-analytical formula for the total switching rate, written in terms of the tunnel splittings and the intra-/inter-well transitions rates. $\Gamma(\delta_k, W_{pq}, V_{pq}) = \sum_n \Gamma_n(\delta_k, W_{pq}, V_{pq})$ [Eq.~\eqref{eq:Gamma-final-solution}]. \\

 & \\

\multicolumn{2}{c}{\textit{Section \ref{sec:activation-mechanism}: Application of the semi-analytical formula: activation mechanism within the well}} \\

$V_{pq}^{(\text{casc})}, W_{pq}^{(\text{casc})}$ & The inter-/intra-well transition rates [Eqs.~\eqref{eq:V_pq^casc}~and~\eqref{eq:W_pq^casc}] that contain cascaded thermal heating but excludes quantum heating or any thermal heating that raises the manifold index by more than some integer $d$. \\

$V_{pq}^{(\text{dir})}, W_{pq}^{(\text{dir})}$ & The inter-/intra-well transition rates [Eqs.~\eqref{eq:V_pq^dir}~and~\eqref{eq:W_pq^dir}] that contains direct thermal heating from the ground state to an excited state but excludes quantum heating and any thermal heating from the excited states. \\

$V_{pq}^{\mathrm{(nh)}}, W_{pq}^{\mathrm{(nh)}}$ & The inter-/intra-well transition rates at zero temperature between consecutive eigenstates of the nonhermitian effective Hamiltonian [Eqs.~\eqref{eq:V_pq^p}~and~\eqref{eq:W_pq^p}]. \\

$\Gamma^{(\text{casc})}$ & The switching rate predicted by the semi-analytical formula when only cascaded thermal heating is allowed. $\Gamma^{(\text{casc})} = \Gamma(\delta_n, V_{pq}^{(\text{casc})}, W_{pq}^{(\text{casc})})$ [Eq.~\eqref{eq:Gamma-casc}]. It implicitly depends on an integer cutoff $d$, which can be chosen to limit the maximum number of consecutive manifolds that each heating transition can skip over. \\

$\Gamma^{(\text{dir})}$ & The switching rate predicted by the semi-analytical formula when only direct thermal heating is allowed. $\Gamma^{(\text{dir})} = \Gamma(\delta_n, V_{pq}^{(\text{dir})}, W_{pq}^{(\text{dir})})$ [Eq.~\eqref{eq:Gamma-dir}]. \\

$\Gamma^{(\text{g})}$ & The semianalytical switching rate predicted by transitions limited to within the ground state manifold. $\Gamma^{(\text{g})} = 2V_{00}$ [Eq.~\eqref{eq:Gamma^g}]. \\

$\Gamma^{(\text{all})}$ & The semianalytical switching rate in the presence of both thermal and quantum heating between eigenstates of $\hat{H}$. Same as $\Gamma(\delta_k, W_{pq}, V_{pq})$. The superscript distinguishes it from the general definition of $\Gamma$ as defined by Eq.~\eqref{eq:definition-of-Gamma-0}. \\

$\Gamma^{\mathrm{(nh)}}$ & The semianalytical switching rate predicted by the quantum heating between eigenstates of the nonhermitian Hamiltonian part of the Lindblad master equation. $\Gamma^{\mathrm{(nh)}} = \Gamma(\delta_k, W_{pq}^{\mathrm{(nh)}}, V_{pq}^{\mathrm{(nh)}})$ [Eq.~\eqref{eq:Gamma^p}]. \\

\end{longtable*}

\section{Background}
\label{sec:background}
\begin{figure*}[t]
    \centering
    \includegraphics[width=6.75in]{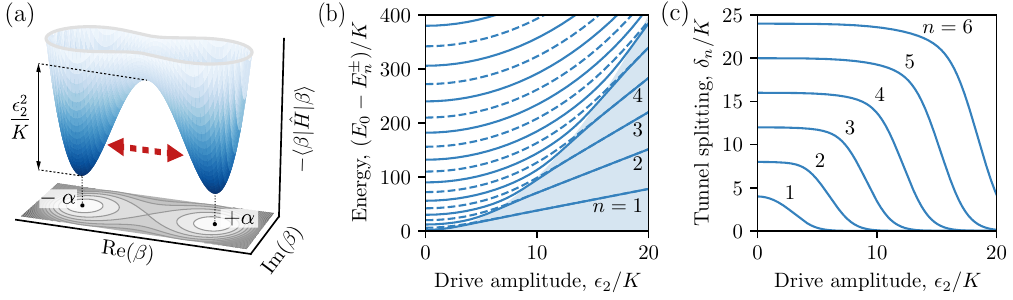}
    \caption{\textbf{The double-well shapes the spectrum of the Hamiltonian $\hat{H}$ defined in Eq.~\eqref{eq:hamiltonian}.} (a) The double-well $U(\beta) = -\bra{\beta}\hat{H}\ket{\beta}$, defined in Eq.~\eqref{eq:double-well}, in the phase space of the oscillator in the rotating frame. The depth of the well (black arrows) is $\epsilon_2^2/K$. Spontaneous switching (red arrows) reduces the inter-well population difference. The 2D projection below the double well shows the energy contours. Black dots indicate the minimum points of the double-well, $\beta = \pm \alpha$. (b) The energies $E_n^+$ of even parity excited states $\ket{\psi_n^+}$ (solid lines) and $E_n^-$ of odd parity excited states $\ket{\psi_n^-}$ (dashed lines) as functions of the two-photon drive amplitude $\epsilon_2$, where $E_n^{\pm}$ are defined by Eq.~\eqref{eq:energies}. The shaded region indicates states for which $U(\alpha) \leq -\bra{\psi_n^\pm}\hat{H}\ket{\psi_n^\pm} \leq U(0)$. (c) The excited state manifold tunnel splitting $\delta_n$, defined in Eq.~\eqref{eq:tunnel-splitting-definition}, as a function of the two-photon drive amplitude $\epsilon_2$. $\delta_n$ becomes exponentially small when the manifold $n$ drops into the double well. The states inside the well and the values of the tunnel splittings are important to determining the spontaneous switching rate.}
    \label{fig:background-combined}
\end{figure*}
In this section we introduce the system Hamiltonian and the Lindblad master equation that describes environmental decoherence. We start from the Hamiltonian of a two-photon-driven Kerr nonlinear oscillator under the rotating-wave approximation and in a frame rotating at the oscillator frequency (using units where $\hbar = 1$) \cite{dykman_fluctuating_2012, guo_phase_2013, wang_quantum_2019, grimm_stabilization_2020, frattini_observation_2024},
\begin{align}
    \hat{H} = -K\hat{a}^{\dagger2}\hat{a}^2 + \epsilon_2(\hat{a}^{\dagger2} + \hat{a}^2), \label{eq:hamiltonian}
\end{align}
where $K$ is the Kerr nonlinearity, $\epsilon_2$ is the amplitude of the two-photon drive, and $\hat{a}$, $\hat{a}^\dagger$ are harmonic oscillator ladder operators satisfying $[\hat{a}, \hat{a}^\dagger] = 1$. The two-photon drive term creates and destroys pairs of excitations in the Kerr nonlinear oscillator. The negative sign in front of the Kerr term follows the convention of recent works where the nonlinearity originates from the Josephson junction. For $K > 0$, which we assume throughout this paper without loss of generality, the free oscillation frequency of the nonlinear oscillator decreases with amplitude. As a consequence of the sign choice, the energy eigenvalues of the Hamiltonian in the rotating frame are bounded from above rather than from below. However, note that interaction with the environment tends to relax the system towards the highest energy manifold of the Hamiltonian of Eq.~\eqref{eq:hamiltonian}, and we will refer to this manifold as the ground state manifold for this reason. The ground state manifold is spanned by two coherent states $\ket{+\alpha}$ and $\ket{-\alpha}$ that have the degenerate energy eigenvalue $E_0$ \cite{cochrane_macroscopically_1999, marthaler_quantum_2007, puri_engineering_2017}, where $\alpha$ and $E_0$ are defined as
\begin{align}
    \alpha = \sqrt{\epsilon_2/K},\quad E_0 = \frac{\epsilon_2^2}{K} = K\alpha^4. \label{eq:ground-state}
\end{align}
The two coherent states are not orthogonal but have an overlap of $\braket{-\alpha|+\alpha} = e^{-2|\alpha|^2}$ which approaches zero in the limit $|\alpha| \to \infty$. From the coherent states we can always construct an orthonormal basis for the ground state manifold consisting of the even and odd parity Schrödinger cat states \cite{cochrane_macroscopically_1999},
\begin{align}
    \ket{\mathcal{C}_{\alpha}^\pm} = \mathcal{N}_{\alpha}^{\pm} (\ket{+\alpha} \pm \ket{-\alpha}), \label{eq:cat-states}
\end{align}
where the normalization constants are defined as $\mathcal{N}_{\alpha}^{\pm} = (2 \pm 2e^{-2|\alpha|^2})^{-1/2}$. We can use the Schrödinger cat states to define another orthonormal basis, $\ket{\psi_0^R}$ and $\ket{\psi_0^L}$ as follows,
\begin{align}
    \Ket{\psi_0^{R/L}} &= \frac{\ket{\mathcal{C}_{\alpha}^+} \pm \ket{\mathcal{C}_{\alpha}^-}}{\sqrt{2}} \\
    &\approx \left(1 + \frac{3}{8}e^{-4|\alpha|^2}\right)\ket{\pm\alpha} - \frac{1}{2}e^{-2|\alpha|^2}\ket{\mp\alpha} \label{eq:wellstate-approx} \\
    &\quad\quad\text{ for }|\alpha|^2 \gg 1, \nonumber
\end{align}
which are asymptotic to the coherent states $\ket{\pm\alpha}$ in the limit of large $|\alpha|$. We remark that the ground state manifold of $\hat{H}$ functions as the computational subspace when the system is used as a qubit \cite{grimm_stabilization_2020}.

Because the Hamiltonian conserves parity, we can find a basis of simultaneous eigenstates of energy and parity, but unlike the highest energy eigenvalue, the rest of the energy eigenvalues are not degenerate. Letting $+$ and $-$ indicate even and odd parity and letting $n=0, 1, 2,...$ indicate \textit{decreasing} energy in each parity sector, we denote the eigenstates and their energies by $\ket{\psi_n^\pm}$ and $E_n^\pm$. As usual, they satisfy the eigenvalue equation
\begin{align}
    \hat{H}\ket{\psi_n^\pm} = E_n^\pm\ket{\psi_n^\pm}. \label{eq:energies}
\end{align}
The $n=0$ case reduces to $\ket{\psi_0^\pm} = \ket{\mathcal{C}_{\alpha}^\pm}$ and $E_0^\pm = E_0$. Each $n$ represents a two-level manifold consisting of exactly one even parity and one odd parity eigenstate.

The energy spectrum of $\hat{H}$ and the wavefunctions of the eigenstates are molded by a semiclassical symmetric double-well potential in phase space \cite{marthaler_switching_2006, puri_stabilized_2019, venkatraman_driven_2024}. The simplest way to reveal the double-well is by taking the \textit{negative} expectation value of $\hat{H}$ on the coherent states $\ket{\beta}$,
\begin{align}
     U(\beta) = -\bra{\beta}\hat{H}\ket{\beta} &= K\beta^{*2}\beta^2 - \epsilon_2(\beta^{*2} + \beta^2), \label{eq:double-well}
\end{align}
and varying $\beta$ in the complex plane (Fig.~\ref{fig:background-combined}a). The reason we choose to define $U(\beta)$ with the added negative sign is to recover a conventional potential energy function whose graph opens upwards. Consistent with this choice, we refer to the $n = 0$ manifold as the ground state manifold, and all other manifolds as excited state manifolds. From this point forward, when we say that manifold $n$ lies ``above" another manifold $m$, we mean that the manifold index $n$ is larger than $m$, as depicted in Fig.~\ref{fig:background-combined}b.

The double well $U(\beta)$ has two global minima located symmetrically at $U(\pm\alpha) = -\epsilon_2^2/K$ and a saddle point located at $U(0) = 0$ \footnote{The number of saddle points depends on the phase-space representation we choose for $\hat{H}$. The representation used in Eq.~\eqref{eq:double-well} is proportional to the $Q$-function. Other phase space representations can also be used. The Wigner function, for example, has two saddle points for some values of $\epsilon_2$ \cite{venkatraman_driven_2024}.}. Therefore, $\epsilon_2^2/K$ is the depth of the double-well and, equivalently, the height of the energy barrier, which we can control by changing the two-photon drive amplitude $\epsilon_2$. An estimate of the number $n_{\text{bound}}$ of quantized energy levels in each well is $n_{\text{bound}} \approx \epsilon_2/(\pi K) = \alpha^2/\pi$ given by the Bohr quantization condition \cite{marthaler_switching_2006, frattini_observation_2024}.

As is commonly the case for double-well potentials, two types of energy gaps of distinct scales (Fig.~\ref{fig:background-combined}b) characterize the energy spectrum for states inside the well \cite{marthaler_quantum_2007, puri_stabilized_2019, frattini_observation_2024, venkatraman_driven_2024}. The first type consists of tunnel splittings $\delta_n$ with the following definition and scaling,
\begin{align}
    \delta_n &= E_n^+ - E_n^- \label{eq:tunnel-splitting-definition} \\
    &\sim f_n(\alpha^2)e^{-2\alpha^2}, \alpha^2 \gg n\pi,
\end{align}
where $f_n$ is asymptotically a polynomial function of $\alpha^2$ \cite{puri_stabilized_2019}. Note that while the exponent of the tunnel splitting in the WKB approximation depends on $n$, here this dependence is absorbed into $f_n$ and only the leading order term $-2\alpha^2$ is kept in $e^{-2\alpha^2}$.  In appendix \ref{sec:wkb}, we obtain the following expression for $\delta_n$ in the asymptotic limit of $\alpha^2 \to \infty$ using the WKB approximation technique presented in \cite{marthaler_quantum_2007},
\begin{align}
    \delta_n = \frac{(4e)^{2n}}{\left(n - \frac{1}{2}\right)^{n - \frac{1}{2}}\!\left(n + \frac{1}{2}\right)^{n + \frac{1}{2}}}\frac{4K}{\pi}\alpha^{4n+2}e^{-2\alpha^2}c_n(\alpha^2),
\end{align}
where $c_n(\alpha^2)$ approaches 1 as $\alpha^2 \to \infty$ [see Eq.~\eqref{eq:tunnel-splitting-analytic-lowly-excited-states} for its asymptotic expansion]. The expression for $\delta_n$ when a state is near the top of the energy barrier is complicated and given in Eq.~\eqref{eq:tunnel-splitting-analytic-highly-excited-states} of appendix \ref{sec:wkb}. The exact numerical tunnel splittings are plotted as functions of the drive amplitude in Fig.~\ref{fig:background-combined}c. The second type of energy gap consists of gaps between consecutive manifolds with the following definition and scaling,
\begin{align}
    E_{\mathrm{gap}, n} &= |\bar{E}_{n+1} - \bar{E}_n| \label{eq:energy-gap-definition} \\
    &\sim 4K\alpha^2 - K(4n + 2), \alpha^2 \gg n\pi,
\end{align}
where $\bar{E}_n$ denotes the average energy in the $n$th manifold [see also Eq.~\eqref{eq:energy-estimate} in appendix \ref{sec:matrix-element-estimates}].

\begin{figure}[t!]
    \centering
    \includegraphics[width=3.375in]{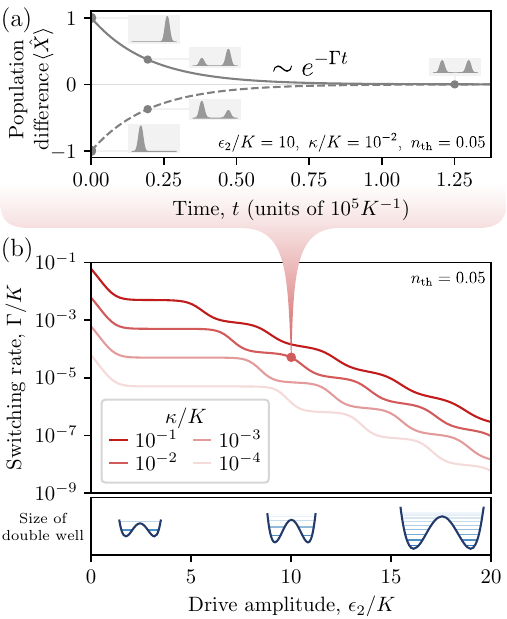}
    \caption{\textbf{The switching rate and the staircase.} (a) The population difference $\langle\hat{X}\rangle$ [see Eq.~\eqref{eq:definition-of-Gamma-0}] as a function of time, obtained via numerical simulation of the master equation Eq.~\eqref{eq:lindblad}. Insets show the population distribution over position $\hat{x} \propto \hat{a}^\dagger + \hat{a}$ at select times. No matter whether the initial state is $\ket{+\alpha}$ (solid line) or $\ket{-\alpha}$ (dashed line), the inter-well population difference decays exponentially over time. (b) We plot the numerically obtained spontaneous switching rate $\Gamma$ as a function of the two-photon drive amplitude $\epsilon_2$. The plot shows a staircase pattern. As $\epsilon_2$ increases, the double-well captures more quantized states (bottom schematics), and the spontaneous switching rate $\Gamma$ (red curves) decreases abruptly at several critical $\epsilon_2/K$ values that shift with the damping rate $\kappa$ (as well as with $n_{\text{th}}$, not shown in the plot). The red dot indicates the parameters used in (a).}
    \label{fig:numerical-staircase}
\end{figure}

The decoherence of the oscillator can be modeled in the standard way by the Lindblad master equation \cite{blais_circuit_2021} \{see also Eq.\ (20) of \cite{marthaler_switching_2006}\},
\begin{align}
    \frac{\mathrm{d}\hat{\rho}}{\mathrm{d}t} &= \mathcal{L}(\hat{\rho}) \\
    &= -i[\hat{H}, \hat{\rho}] + \kappa(1 + n_{\mathrm{th}})\mathcal{D}[\hat{a}]\hat{\rho} + \kappa n_{\mathrm{th}}\mathcal{D}[\hat{a}^\dagger]\hat{\rho} \\
    &= -i[\hat{H}, \hat{\rho}] + \sum_{\hat{O} \in C}\mathcal{D}[\hat{O}]\hat{\rho} \label{eq:lindblad}.
\end{align}
Here, $\hat{\rho}$ is the system density operator, $\mathcal{D}[\hat{O}]$ is the dissipator defined as $\mathcal{D}[\hat{O}]\hat{\rho} = \hat{O}\hat{\rho}\hat{O}^\dagger - \frac{1}{2}\hat{O}^\dagger\hat{O}\hat{\rho} - \frac{1}{2}\hat{\rho}\hat{O}^\dagger\hat{O}$, the set of collapse operators is
\begin{align}
    C = \left\{\sqrt{\kappa(n_{\mathrm{th}} + 1)}\hat{a}, \sqrt{\kappa n_{\mathrm{th}}}\hat{a}^\dagger\right\}, \label{eq:collapse-operators}
\end{align}
$\kappa$ is the energy damping rate (henceforth single photon loss rate) of the environment, and $n_{\text{th}}$ is the thermal occupation number of the environment. We work in the regime of $\kappa/K \ll 1$ and $n_{\text{th}} \ll 1$ due to its experimental relevance.

Equation~\eqref{eq:lindblad} relies on several assumptions. First, it inherits the rotating-wave approximation from Eq.~\eqref{eq:hamiltonian}, which neglects fast oscillating terms in the Hamiltonian. For a sense of scale, the frequencies of these oscillating terms are typically three orders of magnitude higher than $E_{\mathrm{gap}, 0}$ in experiments of \cite{frattini_observation_2024, hajr_high-coherence_2024-2, albornoz_oscillatory_2024}. Second, Eq.~\eqref{eq:lindblad} assumes an oscillator-environment coupling that is linear in $\hat{a}$ and $\hat{a}^\dagger$. This is commonly used to model losses in the oscillator \cite{blais_circuit_2021}. Finally, Eq.~\eqref{eq:lindblad} relies on the Born-Markov approximations \cite{gardiner_quantum_2010, breuer_theory_2002, carmichael_statistical_1999}, which assume that the oscillator-environment coupling is weak, and that the environment observable appearing in the coupling has a sufficiently flat noise power spectrum. The weakness of the coupling is supported by the fact that the transition frequencies measured in \cite{frattini_observation_2024} matches the prediction of the closed-system Hamiltonian $\hat{H}$. The flatness of the noise power spectrum is a simplifying assumption made in the absence of direct experimental observations, and is necessary for the removal of all environmental degrees of freedom from the equation of motion.

The interaction with the environment causes two types of relaxation in the double well: intra-well relaxation and inter-well relaxation. The intra-well relaxation occurs at a rate on the order of $\kappa$, while the inter-well relaxation is impeded by the presence of an energy barrier between the wells, so typically, the timescales of the two types of relaxation are well separated. We define the spontaneous switching rate $\Gamma$ between the wells to be
\begin{align}
    \Gamma = -\frac{1}{\braket{\hat{X}}}\frac{\mathrm{d}}{\mathrm{d}t}\braket{\hat{X}}, \label{eq:definition-of-Gamma-0}
\end{align}
where $\hat{X}$ is an observable that sums the inter-well population difference of all two-level manifolds. We provide the mathematical expression for $\hat{X}$ later in Eq.~\eqref{eq:definition-of-X} of section \ref{sec:perturbative-solution-of-the-effective-equations-of-motion}. The expectation value in Eq.~\eqref{eq:definition-of-Gamma-0} is evaluated at a time after the intra-well quasi-equilibrium state has been attained. We note that there is a factor of two difference between our definition of the switching rate and some previous works, where the switching rate is defined as the inverse of the average wait time between consecutive switches.

As has been observed in \cite{frattini_observation_2024}, the numerical solution of Eq.~\eqref{eq:lindblad} via the diagonalization of $\mathcal{L}$ (see appendix~\ref{sec:numerical-solution-technique} as well as \cite{johansson_qutip_2012, johansson_qutip_2013} for details) qualitatively reproduces the defining feature of the experimental staircase: as a function of the amplitude of the two-photon drive $\epsilon_2$, which controls the depth of the double well, the spontaneous switching rate $\Gamma$ between the wells shows step-like decreases at specific values of $\epsilon_2$ (Fig.~\ref{fig:numerical-staircase}). However, the simple model of Eq.~\eqref{eq:lindblad} does not capture all the qualitative features in the measured switching rate \cite{frattini_observation_2024, hajr_high-coherence_2024-2, albornoz_oscillatory_2024}. Various works have examined additional processes that affect the switching rate, such as white dephasing noise \cite{marthaler_switching_2006, frattini_observation_2024}, non-Markovian phase noise \cite{frattini_observation_2024}, effects beyond the rotating-wave approximation \cite{peano_sharp_2012, sank_measurement-induced_2016, venkatraman_static_2022-1, xiao_diagrammatic_2023-1, venkatraman_nonlinear_2024, hajr_high-coherence_2024-2, garcia-mata_effective_2024, dumas_measurement-induced_2024, chavez-carlos_driving_2024-1, benhayoune-khadraoui_how_2025, bhandari_symmetrically_2025-3}, higher-order nonlinearities \cite{benhayoune-khadraoui_how_2025, baskov_exact_2025}, and resonances with environmental modes \cite{benhayoune-khadraoui_how_2025}. While these additional processes have been shown to improve qualitative agreement with experimental measurement, quantitative agreement still has not been achieved. In this work, we choose to limit ourselves to the model of Eq.~\eqref{eq:lindblad}, which as we will see, allows us to investigate a range of physical effects despite its simplicity. It is the goal of the subsequent sections to explain the steps in the staircase within the model of Eq.~\eqref{eq:lindblad} by finding a semi-analytical formula for the spontaneous switching rate $\Gamma$.

\section{Solving for the spontaneous switching rate}
\label{sec:solving-for-the-spontaneous-switching-rate}
We solve the Lindblad master equation for the spontaneous switching rate $\Gamma$ by making a series of approximations. These approximations allow us to obtain a semi-analytical expression of $\Gamma$ in terms of the tunnel splittings $\delta_n$, the single photon loss rate $\kappa$, the thermal occupation $n_{\text{th}}$, and the matrix elements of $\hat{a}$ and $\hat{a}^\dagger$.

\subsection{Projection of the density operator onto two-level manifolds}
\label{sec:projection-of-the-density-operator}

\begin{figure*}
    \includegraphics[width=6.75in]{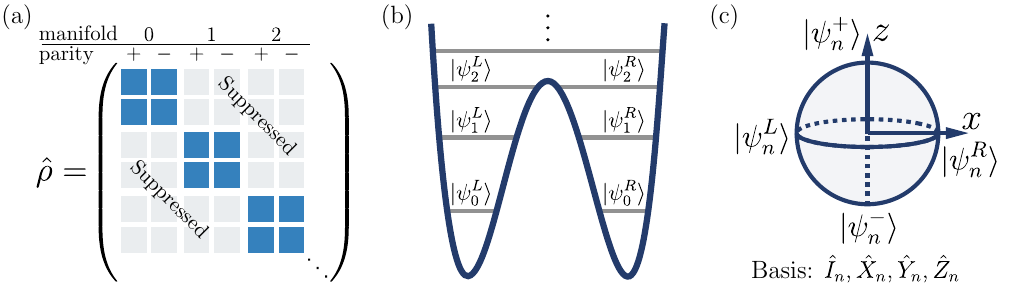}
    \caption{\textbf{Approximating the system state as a mixture between two-level manifolds.} (a) Illustration of the projection of the system density operator $\hat{\rho}$ onto a series of two-level manifolds, as described by Eq.~\eqref{eq:projection}. The matrix elements of $\hat{\rho}$ under the eigenbasis $\ket{\psi_n^\pm}$, with $\ket{\psi_n^\pm}$ defined by Eq.~\eqref{eq:energies}, are ordered by the manifold number $n$ and the parity $\pm$ as indicated above the matrix. The projection preserves the blue matrix elements (all populations and in-manifold coherences), but removes the light grey ones (cross-manifold coherences) which are perturbatively small. As a consequence of the projection of $\hat{\rho}$, the Lindbladian $\mathcal{L}$ defined in Eq.~\eqref{eq:lindblad} can be approximated by the effective Lindbladian defined in Eq.~\eqref{eq:effective-lindbladian}, which operates on the subspace formed by the blue matrix elements only. (b) Right- and left-well states $\ket{\psi_n^{R/L}}$ defined in Eqs.~\eqref{eq:right-wellstate} and \eqref{eq:left-wellstate}. (c) The definition of the Bloch sphere for manifold $n$. The axes are consistent with the Pauli operators $\hat{X}_n, \hat{Y}_n, \hat{Z}_n$ defined in Eqs.~\eqref{eq:pauli-x}, \eqref{eq:pauli-z}, and \eqref{eq:pauli-y}. Together with the projector $\hat{I}_n$ defined in Eq.~\eqref{eq:projector}, the Pauli operators form a basis for operators supported on manifold $n$.}
    \label{fig:projection-combined}
\end{figure*}

We start with a perturbation theory argument showing that given the Lindblad master equation of Eq.~\eqref{eq:lindblad} and the condition $\kappa, \kappa n_{\text{th}} \ll K$, coherent dynamics is largely confined within two-level manifolds. First we divide the terms in the expansion of $\hat{\rho}$ in the energy-parity eigenbasis $\ket{\psi_n^\pm}$ into three categories: populations are terms of the form $\ket{\psi_n^\pm}\bra{\psi_n^\pm}$; \textit{in-manifold} coherences are terms of the form $\ket{\psi_n^\pm}\bra{\psi_n^\mp}$, and \textit{cross-manifold} coherences are those of the form $\ket{\psi_n^p}\bra{\psi_m^q}$ for all $n \neq m$ and $p, q\in\{+, -\}$. In Fig.~\ref{fig:projection-combined}a, the three categories correspond to blue diagonal elements, blue off-diagonal elements, and grey off-diagonal elements.

Now, we discuss the behavior of terms in the above three categories in the absence of dissipation ($\kappa = 0$). Hamiltonian evolution causes the cross-manifold coherences to advance in phase at the angular frequency of $|E_n^p - E_m^q|$, the energy gap between manifolds $n$ and $m$. In contrast, Hamiltonian evolution leaves populations invariant, and it only causes the in-manifold coherences under the energy barrier to advance slowly in phase at the angular frequency of the tunnel splittings $\delta_n$. Therefore there is a frequency gap of roughly $\min_n(E_{\text{gap}, n})$ between the slow and fast dynamics. In appendix \ref{sec:wkb} Eq.~\eqref{eq:minimum-gap-approx}, we find the following approximate expression for the minimum gap using the WKB approximation \cite{marthaler_switching_2006},
\begin{align}
    \min_n(E_{\text{gap}, n}) &\approx \frac{4\pi K\alpha^2}{\ln(8\alpha^2) + 1.96351 + \frac{\pi}{2}}.
\end{align}
which is greater than $K$ for $\alpha^2$ not too small. Since $\kappa \ll K$, we must also have that $\kappa \ll \min_n(E_{\text{gap}, n})$.

We continue our description of the perturbation theory argument by considering the situation in the presence of dissipation $\kappa > 0$. We regard the dissipators $\mathcal{D}[\hat{O}]$ as a perturbation and perform time-dependent perturbation theory. The perturbation only creates a static coupling of order $\kappa$ and $\kappa n_{\text{th}}$ from the slow-evolving components to the fast-evolving components, so it can only off-resonantly drive the cross-manifold coherences given the frequency gap discussed in the previous paragraph. Therefore, if the initial $\hat{\rho}$ contains only populations and in-manifold coherences, then the build-up of cross-manifold coherences are off-resonantly suppressed to $O[\kappa / \min_n(E_{\text{gap}, n})]$. We remark that the confinement of coherent dynamics to two-level manifolds is a direct extension of the fact that energy eigenstates become pointer states of decoherence under the combination of a strong non-degenerate system self-Hamiltonian and a weak interaction with the environment \cite{paz_quantum_1999}.

Given our preceding perturbation theory argument, in the limit $\kappa, \kappa n_{\mathrm{th}} \ll K$, the system density operator $\hat{\rho}$ is well approximated by an incoherent mixture across the manifolds within which coherent dynamics may occur. Equivalently speaking, the manifold index $n$ can be thought of as a classical random variable. To zeroth-order in the perturbation, the density operator is projected onto two-level manifolds (Fig.~\ref{fig:projection-combined}a),
\begin{align}
    \hat{\rho}_{\text{proj}} &= \mathcal{P}(\hat{\rho}) = \sum_n\hat{I}_n\hat{\rho}\hat{I}_n, \label{eq:projection}
\end{align}
where
\begin{align}
    \hat{I}_n = \ket{\psi_n^+}\bra{\psi_n^+} + \ket{\psi_n^-}\bra{\psi_n^-} \label{eq:projector}
\end{align}
is the projector onto the $n$th manifold. The projection preserves the populations and the in-manifold coherences while projecting out the cross-manifold coherence.

We remark that this projection step, as was done in Eq.~(21) of \cite{marthaler_switching_2006}, neglects the matrix elements of $\hat{\rho}$ between different intra-well eigenstates and is correct to lowest order in $\kappa/E_{\mathrm{gap}, n}$. However, numerics show that the projection reduces the switching rate below temperatures $n_{\mathrm{th}} \lesssim 10^{-8}$ (see appendix \ref{sec:validity-of-approximations}). We will see in section \ref{sec:activation-mechanism} that this is because the neglected matrix elements contribute to an important activation mechanism that dominates the switching rate in the limit of very small $n_{\text{th}}$. We show in appendix \ref{sec:validity-of-approximations} that we can recover the correct switching rate in this very low temperature regime using the nonhermitian effective Hamiltonian from quantum trajectory theory \cite{dum_monte_1992, dalibard_wave-function_1992, carmichael_open_1993, molmer_monte_1993, plenio_quantum-jump_1998, carmichael_statistical_2008}. This nonhermitian effective Hamiltonian generates the evolution of the state conditioned on no quantum jumps. For the Lindbladian $\mathcal{L}$ at temperature $n_{\mathrm{th}} \to 0$, it is defined as
\begin{align}
    \hat{H}_{\text{eff}} = \hat{H} - i\kappa\hat{a}^\dagger\hat{a}/2 \label{eq:effective-hamiltonian}.
\end{align}
In this section, we restrict ourselves to temperatures where this treatment is not necessary, and delay the discussion of the activation mechanism at very low temperatures until section \ref{sec:activation-mechanism}.

The projected density operator $\hat{\rho}_{\text{proj}}$ can be visualized as a collection of Bloch spheres (Fig.~\ref{fig:projection-combined}c). The north and south poles of each Bloch sphere represent the even and odd parity eigenstates in a particular manifold, and the equator represents a family of equal-weight superpositions between the parity eigenstates parameterized by their phase difference. Although each of $\ket{\psi_n^+}$ and $\ket{\psi_n^-}$ has an equal probability of occupying either well due to their being parity eigenstates, they can interfere with each other constructively or destructively. We define the right-well state $\ket{\psi_n^R}$ (left-well state $\ket{\psi_n^L}$) to be the particular equal-weight superposition of $\ket{\psi_n^+}$ and $\ket{\psi_n^-}$ that maximizes (minimizes)
$\mathrm{Re}\braket{\hat{a}}$ (Fig.~\ref{fig:projection-combined}b). Given the symmetries in the Hamiltonian, this definition ensures $\bra{\psi_n^R}\hat{a}\ket{\psi_n^R} = -\bra{\psi_n^L}\hat{a}\ket{\psi_n^L} > 0$. Once the phase difference used to define the right- and left-well states are absorbed into $\ket{\psi_n^+}$ and $\ket{\psi_n^-}$, $\ket{\psi_n^R}$ and $\ket{\psi_n^L}$ can simply be expressed as
\begin{align}
    \ket{\psi_n^R} &= \frac{1}{\sqrt{2}}\left(\ket{\psi_n^+} + \ket{\psi_n^-}\right), \label{eq:right-wellstate} \\
    \ket{\psi_n^L} &= \frac{1}{\sqrt{2}}\left(\ket{\psi_n^+} - \ket{\psi_n^-}\right) \label{eq:left-wellstate},
\end{align}
hence fixing them to the $x$-axis of the Bloch sphere.
The components of the Bloch vector along the cardinal axes are given in the usual way by the expectation values of the following Pauli operators,
\begin{align}
    \hat{X}_n &= \ket{\psi_n^R}\bra{\psi_n^R} - \ket{\psi_n^L}\bra{\psi_n^L}, \label{eq:pauli-x} \\
    \hat{Z}_n &= \ket{\psi_n^+}\bra{\psi_n^+} - \ket{\psi_n^-}\bra{\psi_n^-}, \label{eq:pauli-z} \\
    \hat{Y}_n &= i\hat{X}_n\hat{Z}_n \label{eq:pauli-y} \\
    &= i(\ket{\psi_n^R}\bra{\psi_n^L} - \ket{\psi_n^L}\bra{\psi_n^R}). \nonumber
\end{align}

Our approximation of ignoring the cross-manifold coherence in Eq.~\eqref{eq:projection} simplifies the Lindbladian dynamics because it reduces the degrees of freedom in the system. The effective Lindbladian restricted to the remaining degrees of freedom is 
\begin{align}
    \mathcal{L}_{\text{eff}} &= \mathcal{P}\mathcal{L}\mathcal{P} = \mathcal{L}_{\text{H}} + \mathcal{L}_{\text{D}}, \label{eq:effective-lindbladian}
\end{align}
where at the second equality sign we separated the effective Lindbladian into the Hamiltonian part $\mathcal{L}_{\text{H}}$ and the dissipative part $\mathcal{L}_{\text{D}}$. After some algebra, we can show that the explicit expression for $\mathcal{L}_{\text{H}}$ is
\begin{align}
    \mathcal{L}_{\text{H}} &= -i\mathcal{P}[\hat{H}, \mathcal{P}\ \cdot\ ] = -i\sum_n \frac{\delta_n}{2}[\hat{Z}_n, \ \cdot\ ],
    \label{eq:projected-hamiltonian}
\end{align}
and the explicit expression for $\mathcal{L}_{\text{D}}$ is
\begin{align}
    \mathcal{L}_{\text{D}} &= \sum_{\hat{O} \in C}\mathcal{P}\mathcal{D}[\hat{O}]\mathcal{P} = \sum_{\hat{O} \in C}\sum_{pq}\mathcal{D}[\hat{O}_{pq}], \label{eq:projected-dissipator}
\end{align}
where we define $\hat{O}_{pq} = \hat{I}_p\hat{O}\hat{I}_q$, the projection of the collapse operator onto given initial and final manifolds $q$ and $p$. As a reminder, the true collapse operators $\hat{O}$ are drawn from the set $C$ defined in Eq.~\eqref{eq:collapse-operators}, which includes single photon loss and gain, and the projected collapse operators $\hat{O}_{pq}$ are a construction that facilitates analysis regardless of the experimental setup. Consistent with the approximation of Eq.~\eqref{eq:projection}, the dissipators $\mathcal{D}[\hat{O}_{pq}]$ do not create superpositions (coherence) between different manifolds.

Physically, the new dissipators $\mathcal{D}[\hat{O}_{pq}]$ cause random transitions across the Bloch spheres as well as random flips of the Bloch vector within each Bloch sphere. To illustrate this, we write the projected collapse operator $\hat{O}_{pq}$ in terms of its matrix elements
\begin{align}
    \hat{O}_{pq} &= \sum_{i, j=L/R}\ket{\psi_p^i}\bra{\psi_p^i}\hat{O}\ket{\psi_q^j}\bra{\psi_q^j}. \label{eq:O_pq-definition}
\end{align}
Since the operator $\hat{O} \in C$ is proportional to either $\hat{a}$ or $\hat{a}^\dagger$, it has odd parity symmetry $e^{i\pi\hat{a}^\dagger\hat{a}}\hat{O}e^{-i\pi\hat{a}^\dagger\hat{a}} = -\hat{O}$. The states $\ket{\psi_n^R}$ and $\ket{\psi_n^L}$ transform into each other under the parity operation, so the expression for $\hat{O}_{pq}$ can be simplified into, 
\begin{align}
    \hat{O}_{pq} &= \bra{\psi_p^R}\hat{O}\ket{\psi_q^R}\left(\ket{\psi_p^R}\bra{\psi_q^R} - \ket{\psi_p^L}\bra{\psi_q^L}\right) \nonumber \\
    &\quad + \bra{\psi_p^L}\hat{O}\ket{\psi_q^R}\left(\ket{\psi_p^L}\bra{\psi_q^R} - \ket{\psi_p^R}\bra{\psi_q^L}\right) \\
    &= \bra{\psi_p^R}\hat{O}\ket{\psi_q^R} \hat{X}_{pq} + i \bra{\psi_p^L}\hat{O}\ket{\psi_q^R} \hat{Y}_{pq} \label{eq:opq-matrix-element},
\end{align}
where we define new operators
\begin{align}
    \hat{X}_{pq} &= \ket{\psi_p^R}\bra{\psi_q^R} - \ket{\psi_p^L}\bra{\psi_q^L} \label{eq:definition-x-pq}, \\
    \hat{Y}_{pq} &= i\left(\ket{\psi_p^R}\bra{\psi_q^L} - \ket{\psi_p^L}\bra{\psi_q^R}\right) \label{eq:definition-y-pq}.
\end{align}
The operators $\hat{X}_{pq}$ and $\hat{Y}_{pq}$ generate transitions across Bloch spheres when $p\neq q$ and flips the Bloch vector within the $p$th Bloch sphere when $p = q$. The notation for the two operators indicates the axis around which the Bloch vector is flipped when a transition happens.

\subsection{Effective equations of motion of the density operator components}
\label{sec:effective-equations-of-motion}

In this section we derive the effective equations of motion of the density operator components and develop an intuitive understanding of the terms involved. Under the approximation of Eq.~\eqref{eq:projection}, the density operator can be written as
\begin{align}
    \hat{\rho}_{\text{proj}} = \frac{1}{2}\sum_{\hat{S}\in \mathcal{S}}\braket{\hat{S}}\hat{S}, \label{eq:expansion-of-rho-proj}
\end{align}
where $\mathcal{S} = \bigcup_n\{\hat{I}_n, \hat{X}_n, \hat{Y}_n, \hat{Z}_n\}$. The equation of motion of the expectation value of any $\hat{S} \in \mathcal{S}$ can be written as
\begin{align}
    \frac{\mathrm{d}}{\mathrm{d}t}\braket{\hat{S}} &= \mathrm{Tr}[\hat{S}\mathcal{L}_{\text{eff}}(\hat{\rho}_{\text{proj}})] \\
    &= \frac{1}{2}\sum_{\hat{S}' \in \mathcal{S}}\mathrm{Tr}[\hat{S}\mathcal{L}_{\text{eff}}(\hat{S}')]\braket{\hat{S}'}. \label{eq:effective-dynamics}
\end{align}
It can be shown via direct evaluation that the following coefficients vanish,
\begin{align}
    &\mathrm{Tr}[\hat{X}_n\mathcal{L}_{\text{eff}}(\hat{I}_m)] = \mathrm{Tr}[\hat{X}_n\mathcal{L}_{\text{eff}}(\hat{Z}_m)] = 0, \label{eq:decouple-x}\\
    &\mathrm{Tr}[\hat{Y}_n\mathcal{L}_{\text{eff}}(\hat{I}_m)] = \mathrm{Tr}[\hat{Y}_n\mathcal{L}_{\text{eff}}(\hat{Z}_m)] = 0, \label{eq:decouple-y}
\end{align}
indicating that the equations of motion for $\braket{\hat{X}_n}$ and $\braket{\hat{Y}_n}$ do not depend on $\braket{\hat{I}_n}$ or $\braket{\hat{Z}_n}$. We remark that the vanishing coefficients are a consequence of the weak parity symmetry of the Lindbladian \cite{gravina_critical_2023}. This is convenient because for the purpose of finding the spontaneous switching rate between the wells, we are only interested in processes that change the $x$ component of the Bloch vector. Hence we only need the equations of motion for $\braket{\hat{X}_n}$ and $\braket{\hat{Y}_n}$,
\begin{align}
    \frac{\mathrm{d}}{\mathrm{d}t}\braket{\hat{X}_n} &= \sum_m\left\{\frac{1}{2}\mathrm{Tr}[\hat{X}_n\mathcal{L}_{\text{eff}}(\hat{X}_m)]\braket{\hat{X}_m}\right. \nonumber \\
    &\quad + \left.\frac{1}{2}\mathrm{Tr}[\hat{X}_n\mathcal{L}_{\text{eff}}(\hat{Y}_m)]\braket{\hat{Y}_m}\right\}, \label{eq:eom-x-start} \\
    \frac{\mathrm{d}}{\mathrm{d}t}\braket{\hat{Y}_n} &= \sum_m\left\{\frac{1}{2}\mathrm{Tr}[\hat{Y}_n\mathcal{L}_{\text{eff}}(\hat{Y}_m)]\braket{\hat{Y}_m}\right. \nonumber \\
    &\quad + \left.\frac{1}{2}\mathrm{Tr}[\hat{Y}_n\mathcal{L}_{\text{eff}}(\hat{X}_m)]\braket{\hat{X}_m}\right\}. \label{eq:eom-y-start}
\end{align}

To proceed, we must evaluate and interpret the action of the effective Lindbladian $\mathcal{L}_{\text{eff}} = \mathcal{L}_{\text{H}} + \mathcal{L}_{\text{D}}$, defined in Eqs.~\eqref{eq:effective-lindbladian}~to~\eqref{eq:projected-dissipator}, on $\hat{X}_n$ and $\hat{Y}_n$. For the Hamiltonian part $\mathcal{L}_{\text{H}}$, we have
\begin{align}
\mathcal{L}_{\text{H}}(\hat{X}_n) = +\delta_n\hat{Y}_n,\quad \mathcal{L}_{\text{H}}(\hat{Y}_n) = -\delta_n\hat{X}_n.
\end{align}
This shows that the Hamiltonian part causes inter-well coherent tunneling in the $n$th manifold. Now we evaluate and interpret the action of the dissipative part $\mathcal{L}_{\text{D}}$. Because $\mathcal{L}_{\text{D}}$ can be expressed as a sum over a collection of dissipators following Eq.~\eqref{eq:projected-dissipator}, it suffices to look at the action of $\sum_{\hat{O} \in C}\mathcal{D}[\hat{O}_{pq}]$ on $\hat{X}_n$ and $\hat{Y}_n$. After some algebra, we obtain the following identities,
\begin{align}
    \sum_{\hat{O} \in C}\mathcal{D}[\hat{O}_{pq}]\hat{X}_n  &= \left\{ V_{pq}\mathcal{D}[\hat{Y}_{pq}] + W_{pq}\mathcal{D}[\hat{X}_{pq}] \right\} \hat{X}_n, \label{eq:action-dissipative-part-x} \\
    \sum_{\hat{O} \in C}\mathcal{D}[\hat{O}_{pq}]\hat{Y}_n &= \left\{ V_{pq}\mathcal{D}[\hat{Y}_{pq}] + W_{pq}\mathcal{D}[\hat{X}_{pq}] \right\} \hat{Y}_n, \label{eq:action-dissipative-part-y}
\end{align}
where the rates $V_{pq}$ and $W_{pq}$ are defined as
\begin{align}
    V_{pq} &= \sum_{\hat{O} \in C}\left|\bra{\psi_p^L}\hat{O}\ket{\psi_q^R}\right|^2 \\
    &= \kappa (1 + n_{\text{th}})\left|\bra{\psi_p^L}\hat{a}\ket{\psi_q^R}\right|^2 + \kappa n_{\text{th}}\left|\bra{\psi_p^L}\hat{a}^\dagger\ket{\psi_q^R}\right|^2, \label{eq:V-definition} \\
    W_{pq} &= \sum_{\hat{O} \in C}\left|\bra{\psi_p^R}\hat{O}\ket{\psi_q^R}\right|^2 \\
    &= \kappa (1 + n_{\text{th}})\left|\bra{\psi_p^R}\hat{a}\ket{\psi_q^R}\right|^2 + \kappa n_{\text{th}}\left|\bra{\psi_p^R}\hat{a}^\dagger\ket{\psi_q^R}\right|^2, \label{eq:W-definition}
\end{align}
and represent the \textit{inter-well} and \textit{intra-well} transition rates, respectively. We caution that $\sum_{\hat{O} \in C}\mathcal{D}[\hat{O}_{pq}]$ is not equal to $V_{pq}\mathcal{D}[\hat{Y}_{pq}] + W_{pq}\mathcal{D}[\hat{X}_{pq}]$, but they have identical effect on $\hat{X}_n$ and $\hat{Y}_n$, the operators of our interest.

\begin{figure}[t]
    \centering
    \includegraphics{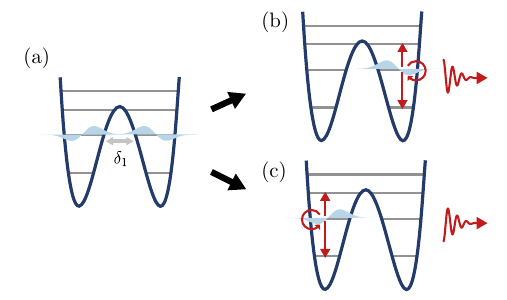}
    \caption{\textbf{Entanglement with the environment measures ``which-well" information.} (a) The wave function of some inter-well superposition state created by tunneling. The black arrows indicate that under the interaction Hamiltonian Eq.~\eqref{eq:interaction-hamiltonian}, the components of the wavefunction in the right and left wells drive the environment in opposite directions (b) and (c). The two squiggly red arrows indicate orthogonal radiation states of the environment. The straight red arrows and circular red arrows indicate possible transitions that may accompany this radiation. After tracing out the environment, the evolution becomes dissipative, and the wavefunction collapses to either the right (b) or the left well (c). }
    \label{fig:tunneling}
\end{figure}

Equations~\eqref{eq:action-dissipative-part-x}~to~\eqref{eq:W-definition} shows that the action of the dissipator $\mathcal{D}[\hat{O}_{pq}]$ has two components, $V_{pq}\mathcal{D}[\hat{Y}_{pq}]$ and $W_{pq}\mathcal{D}[\hat{X}_{pq}]$, with distinct physical meanings:
\begin{itemize}
    \item The component $V_{pq}\mathcal{D}[\hat{Y}_{pq}]$ describes direct inter-well transitions. This is because the collapse operator $\hat{Y}_{pq}$, defined in Eq.~\eqref{eq:definition-y-pq}, flips the $x$ component of the Bloch vector, which measures the inter-well population difference. Corroborating our interpretation is the fact that the rate $V_{pq}$ defined in Eq.~\eqref{eq:V-definition} only contains inter-well matrix elements.
    
    \item The component $W_{pq}\mathcal{D}[\hat{X}_{pq}]$ describes the dephasing (in the sense of loss of coherence) of superpositions between the right and left wells due to continuous measurement of ``which-well" information by the environment (Fig.~\ref{fig:tunneling}). To understand this statement, we trace the origin of $W_{pq}\mathcal{D}[\hat{X}_{pq}]$ back to the typical system-environment coupling Hamiltonian that gives rise to the Lindblad master equation [Eq.~\eqref{eq:lindblad}]. The coupling Hamiltonian is
    \begin{align}
        \hat{H}_{\text{int}} = \hat{a} \hat{B}^\dagger + \hat{a}^\dagger \hat{B}, \label{eq:interaction-hamiltonian}
    \end{align}
    where $\hat{B}$ and $\hat{B}^\dagger$ are environment operators that can be mathematically realized as bosonic raising and lowering operators \cite{haroche_exploring_2006}. The projection of $\hat{H}_{\text{int}}$ onto transitions within the right- or left-wells are
    \begin{align}
        &\bra{\psi_m^{R}}\hat{H}_{\text{int}}\ket{\psi_n^{R}} \nonumber \\
        &\quad = \bra{\psi_m^{R}}\hat{a}\ket{\psi_n^{R}} \hat{B}^\dagger + \bra{\psi_m^{R}}\hat{a}^\dagger\ket{\psi_n^{R}} \hat{B}, \\
        &\bra{\psi_m^{L}}\hat{H}_{\text{int}}\ket{\psi_n^{L}} \nonumber \\
        &\quad = \bra{\psi_m^{L}}\hat{a}\ket{\psi_n^{L}} \hat{B}^\dagger + \bra{\psi_m^{L}}\hat{a}^\dagger\ket{\psi_n^{L}} \hat{B}.
    \end{align}
    These projections can be viewed as effective drives on the environment modes. Due to parity symmetry, the matrix elements $\bra{\psi_m^{R}}\hat{a}\ket{\psi_n^{R}}$ and $\bra{\psi_m^{L}}\hat{a}\ket{\psi_n^{L}}$ differ by a sign (and similarly for $\hat{a}^\dagger$). This means that the environment is driven in opposite directions depending on which well the state is in and slowly becomes entangled with the system. Therefore we can say that the $\hat{a}$ and $\hat{a}^\dagger$ dissipators correspond to the environment continuously measuring ``which-well" information. Continuous measurement produces dephasing in the right- and left-well state basis when we trace out the environment degrees of freedom. This explains our term of interest, $W_{pq}\mathcal{D}[\hat{X}_{pq}]$, where the dissipator $\mathcal{D}[\hat{X}_{pq}]$ indeed dephases superpositions of right- and left-well states per the definition of $\hat{X}_{pq}$ in Eq.~\eqref{eq:definition-x-pq}. The dephasing rate $W_{pq}$ is related to matrix elements of $\hat{a}$ and $\hat{a}^\dagger$ between intra-well states.
\end{itemize}

Although the dephasing term $W_{pq}\mathcal{D}[\hat{X}_{pq}]$ does not directly cause any inter-well transitions, we are interested in this term because it inhibits coherent inter-well tunneling. One way to see this is through the continuous version of the quantum Zeno effect \cite{misra_zenos_1977, itano_quantum_1990, facchi_quantum_2008, greenfield_unified_2025}. Here, the environment performs a weak measurement of the ``which-well" information, and tracing out the environment reduces the inter-well coherence \{see for example Eq.(3.30) in \cite{clerk_introduction_2010}\}. Then, in the limit when the rate of measurement is much faster than the rate of coherent inter-well tunneling, the induced dephasing rate on the right- and left-well state basis is high, and the system becomes frozen in either the right- or left-well.

At this point we have enough information to qualitatively understand the origin of the staircase. The staircase is created by a competition between coherent tunneling driven by $\mathcal{L}_{\text{H}}$, which generates inter-well superpositions, and dissipation due to the coupling to the environment. One primary effect of dissipation is the continuous measurement of ``which-well" information, which dephases the superpositions created by coherent tunneling. Depending on the relative rates of these two processes, each two-level manifold can either be dominated by tunneling (regime T) or dissipation (regime D):
\begin{itemize}
    \item In regime T, the tunnel splitting $\delta_n$ of the manifold is much larger than the dephasing rate due to the continuous measurement of ``which-well" information, so a state in this manifold oscillates freely between the two wells. The contribution to the spontaneous switching rate $\Gamma$ [as defined by Eq.~\eqref{eq:definition-of-Gamma-0}] from a manifold in this regime is limited by the transition rate into the manifold and is insensitive to the value of the tunnel splitting $\delta_n$. Therefore, in this regime, the switching rate changes slowly with the drive amplitude, giving rise to the flat part of each step in the staircase.
    
    \item In regime D, the tunnel splitting $\delta_n$ is much smaller than the dephasing rate. The quantum Zeno effect freezes the population in the initial well, and the contribution to the spontaneous switching rate is suppressed. Switching in the manifold is limited by the tunneling rate, which decreases as the tunnel splitting becomes exponentially suppressed as a function of drive amplitude. Therefore, in this regime, the switching rate contributed by this manifold decreases rapidly as the drive amplitude increases, giving rise to the steep part of each step in the staircase.
\end{itemize}

\begin{figure}[t]
    \centering
    \includegraphics{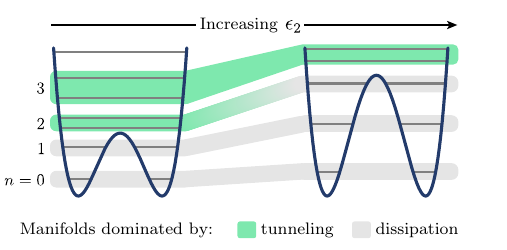}
    \caption{\textbf{Illustration of a manifold transitioning from tunneling-dominated to dissipation-dominated.} The quantum Zeno effect prevents dissipation-dominated manifolds from contributing significantly to the total spontaneous switching rate $\Gamma$. As the two-photon drive amplitude $\epsilon_2$ increases, a manifold (in the pictured example, the one with $n = 2$) can abruptly transition from being tunneling-dominated to being dissipation-dominated, leading to a step-like decrease in the spontaneous switching rate. Note that while tunneling typically refers to the penetration of a classically forbidden region, we use tunneling to refer to any coherent inter-well population transfer due to non-zero $\delta_n$ because the boundary between states below and states above the semiclassical energy barrier is fuzzy.}
    \label{fig:qualitative-understanding}
\end{figure}

The signature of the two disparate regimes T and D can be seen in the staircase. At any fixed value of the two-photon drive amplitude $\epsilon_2$ and therefore fixed depth of the double well, only the manifolds in regime T contribute significantly to the total spontaneous switching rate $\Gamma$. Such manifolds must have energies sufficiently close to or above the top of the energy barrier between the wells because the tunnel splitting is exponentially suppressed below the energy barrier. As $\epsilon_2$ increases, the double well becomes deeper and captures more excited state manifolds, and the tunnel splittings in all captured manifolds decrease exponentially. So long as a manifold stays in regime T, its contribution to the total switching rate is insensitive to the decreasing tunnel splitting. Eventually, the lowest lying manifold in regime T has such a small tunnel splitting that it abruptly transitions into regime D (Fig.~\ref{fig:qualitative-understanding}), where the quantum Zeno effect strongly inhibits the manifold's contribution to the spontaneous switching rate. As a result, we see a step-like decrease in the total spontaneous switching rate $\Gamma$. Each step-like decrease in the staircase therefore reflects one additional pair of levels that fall sufficiently below the energy barrier.

A quantitative analysis will complete the qualitative discussion in the previous paragraphs and reveal the full set of parameters that determine the boundary between the tunneling-dominated regime and the dissipation-dominated regime. To that end we write down the complete set of equations of motion necessary for finding the switching rate. Carrying out the traces in Eqs.~\eqref{eq:eom-x-start} and \eqref{eq:eom-y-start} produces the following equations of motion,
\begin{align}
    \frac{\mathrm{d}}{\mathrm{d}t}\braket{\hat{X}_n} &= -\delta_n\braket{\hat{Y}_n} - \sum_f(W_{fn} + V_{fn})\braket{\hat{X}_n} \nonumber \\
    &\quad + \sum_i(W_{ni} - V_{ni})\braket{\hat{X}_i}, \label{eq:equations-of-motion-x}\\
    \frac{\mathrm{d}}{\mathrm{d}t}\braket{\hat{Y}_n} &= + \delta_n\braket{\hat{X}_n} - \sum_f(W_{fn} + V_{fn})\braket{\hat{Y}_n} \nonumber \\
    &\quad + \sum_i(V_{ni} - W_{ni})\braket{\hat{Y}_i}. \label{eq:equations-of-motion-y}
\end{align}
We solve these equations of motion in the next subsection.

\subsection{Perturbative solution of the effective equations of motion}
\label{sec:perturbative-solution-of-the-effective-equations-of-motion}
In this subsection we extract the switching rate from the solution of Eqs.~\eqref{eq:equations-of-motion-x} and \eqref{eq:equations-of-motion-y}. As long as the switching timescale is the longest and well separated from the intra-well timescales, switching will dominate in the long time limit where it manifests as an exponential decay in the population difference between the right and left wells. In section \ref{sec:background} we have already defined the switching rate by Eq.~\eqref{eq:definition-of-Gamma-0}. Here we restate the definition for ease of reference and provide the definition for $\hat{X}$ that we will use for the rest of the text. The spontaneous switching rate $\Gamma$ is defined by
\begin{align}
    \Gamma = -\frac{1}{\braket{\hat{X}}}\frac{\mathrm{d}}{\mathrm{d}t}\braket{\hat{X}}, \label{eq:definition-of-Gamma}
\end{align}
where $\hat{X}$ is an operator that sums the inter-well population difference for all manifolds,
\begin{align}
    \hat{X} = \sum_n\hat{X}_n, \label{eq:definition-of-X}
\end{align}
and the expectation values are evaluated in the long time limit where intra-well equilibrium has been established. Substituting Eq.~\eqref{eq:equations-of-motion-x} into the definition of $\Gamma$ allows us to express it in terms of the contributions from each manifold $n$,
\begin{align}
    \Gamma = \sum_n\Gamma_n, \label{eq:Gamma-expression}
\end{align}
where the contribution $\Gamma_n$ from manifold $n$ is expressed as follows,
\begin{align}
    \Gamma_n = \frac{1}{\sum_m\braket{\hat{X}_m}}\left[\delta_n\braket{\hat{Y}_n} + \left(2\sum_fV_{fn}\right)\braket{\hat{X}_n}\right]. \label{eq:switching-rate-n}
\end{align}
The term $-\delta_n\braket{\hat{Y}_n}$ describes tunneling and the term $\left(2\sum_fV_{fn}\right)\braket{\hat{X}_n}$ describes direct inter-well transitions, as discussed in section~\ref{sec:effective-equations-of-motion}. To calculate the spontaneous switching rate, it remains to find $\braket{\hat{X}_n}$ and $\braket{\hat{Y}_n}$ via the equations of motion Eqs.~\eqref{eq:equations-of-motion-x} and \eqref{eq:equations-of-motion-y}.

\subsubsection{Exponential ansatz for the population relaxation in time}
The first step towards solving for $\braket{\hat{X}_n}$ and $\braket{\hat{Y}_n}$ is to assume an ansatz for their time dependence. In the long time limit of interest, all expectation values decay exponentially at the switching rate $\Gamma$ towards their steady-state values. Assuming a unique steady state, which we know to be the case from numerics, we must have $\braket{\hat{X}_n} \to 0$ and $\braket{\hat{Y}_n} \to 0$ in the long time limit by parity symmetry considerations. Therefore we can adopt the exponential ansatz, $\braket{\hat{X}_n}, \braket{\hat{Y}_n} \propto e^{-\Gamma t}$. This ansatz converts the equation of motion into an eigenvalue problem,
\begin{align}
    -\Gamma\braket{\hat{X}_n} &= -\delta_n\braket{\hat{Y}_n} - \sum_f(W_{fn} + V_{fn})\braket{\hat{X}_n} \nonumber \\
    &\quad + \sum_i(W_{ni} - V_{ni})\braket{\hat{X}_i}, \label{eq:matrix-equation-x} \\
    -\Gamma\braket{\hat{Y}_n} &= +\delta_n\braket{\hat{X}_n} - \sum_f(W_{fn} + V_{fn})\braket{\hat{Y}_n} \nonumber \\
    &\quad + \sum_i(V_{ni} - W_{ni})\braket{\hat{Y}_i}. \label{eq:matrix-equation-y}
\end{align}
We can solve the eigenvalue equations numerically to get $\Gamma$, but to get an analytic expression that explains the physics of the staircase we make additional approximations.

\subsubsection{Solving for the quantum Zeno dynamics via the adiabatic elimination of \texorpdfstring{$\braket{\hat{Y_n}}$}{Yn}}
\label{sec:solving-for-the-quantum-zeno-dynamics}
To obtain a solution that reflects the quantum Zeno effect that we anticipated in section~\ref{sec:effective-equations-of-motion}, we solve the equation of motion for $\braket{\hat{Y}_n}$, the coherence between the right- and left-well states. Recall that, while writing down the expression for the spontaneous switching rate $\Gamma$ in Eq.~\eqref{eq:definition-of-Gamma} and imposing the exponential ansatz for $\braket{\hat{X}_n}$ and $\braket{\hat{Y}_n}$, we have assumed that spontaneous switching is the slowest dynamics and therefore the only process occurring in the long time limit. Based on this assumption, the system should be in a quasi-equilibrium state that changes slowly in time. Therefore, in Eq.~\eqref{eq:matrix-equation-y}, we can invoke the adiabatic approximation and ignore the term proportional to $\Gamma$, which arose from $\frac{\mathrm{d}}{\mathrm{d}t}\braket{\hat{Y}_n}$ (see appendix~\ref{sec:validity-of-approximations} for numerical verification). The condition for adiabaticity can be found by comparing $\Gamma\braket{\hat{Y}_n}$ to terms on the right-hand side of Eq.~\eqref{eq:matrix-equation-y} as follows. If we denote the coefficient in front of  $\braket{\hat{Y}_n}$ on the right-hand side of Eq.~\eqref{eq:matrix-equation-y} as $\mu_n$, i.e.,
\begin{align}
    \mu_n = 2W_{nn} + \sum_{f \neq n}(W_{fn} + V_{fn}), \label{eq:dephasing-rate}
\end{align}
then Eq.~\eqref{eq:matrix-equation-y} becomes
\begin{align}
    -\Gamma\braket{\hat{Y}_n} &= +\delta_n\braket{\hat{X}_n} - \mu_n\braket{\hat{Y}_n} + \sum_{i \neq n}(V_{ni} - W_{ni})\braket{\hat{Y}_i}.
\end{align}
A sufficient condition for the adiabaticity of $\braket{\hat{Y}_n}$ is then $\Gamma \ll \mu_n$. For a sense of scale, the rate $\mu_n$ can be estimated to be the decoherence rate $\mu_n \sim 2\kappa\alpha^2$ of a cat-like superposition \cite{haroche_exploring_2006} in the limit of $\alpha \gg 1$ and $n_{\text{th}} \ll 1$ (see appendix~\ref{sec:matrix-element-estimates}). We note that because $\mu_n$ controls the rate at which the coherence between the right- and the left-well states decays, it plays the central role of the dephasing rate as discussed in section~\ref{sec:effective-equations-of-motion}. We will from now on refer to $\mu_n$ as the dephasing rate.

As a result of the adiabatic approximation, the equation for $\braket{\hat{Y}_n}$ becomes independent of $\Gamma$,
\begin{align}
    0 &= \mu_n\braket{\hat{Y}_n} + \sum_{i \neq n}(W_{ni} - V_{ni})\braket{\hat{Y}_i} - \delta_n\braket{\hat{X}_n}. \label{eq:implicit-y-in-terms-of-x}
\end{align}
We can derive an analytic approximate solution by rewriting Eq.~\eqref{eq:implicit-y-in-terms-of-x} into the following recursive form,
\begin{align}
    \braket{\hat{Y}_n} &= \frac{1}{\mu_n}\left(\delta_n\braket{\hat{X}_n} - \sum_{i\neq n}(W_{ni} - V_{ni})\braket{\hat{Y}_i}\right). \label{eq:implicit-y-in-terms-of-x-rearranged}
\end{align}
If we treat the term that describes transitions across manifolds (the second term in the parenthesis in the above expression) as a small perturbation, then the zeroth-order solution for $\braket{\hat{Y}_n}$ is
\begin{align}
      \braket{\hat{Y}_n} &\approx \frac{\delta_n}{\mu_n}\braket{\hat{X}_n}. \label{eq:explicit-y-in-terms-of-x-zeroth-order}
\end{align}
Physically, the zeroth-order solution describes a quasi-static balance between the generation of coherence due to tunneling and the dephasing of coherence due to the environment, occurring independently in each manifold.

We can evaluate the error committed by the zeroth-order approximation Eq.~\eqref{eq:explicit-y-in-terms-of-x-zeroth-order} by recursively inserting the solution back into Eq.~\eqref{eq:implicit-y-in-terms-of-x-rearranged}, yielding
\begin{align}
      &\braket{\hat{Y}_n} \nonumber \\
      &= \frac{1}{\mu_n}\delta_n\braket{\hat{X}_n} \nonumber \\
      & - \sum_{i\neq n}\frac{1}{\mu_n}(W_{ni} - V_{ni})\frac{1}{\mu_i}\delta_i\braket{\hat{X}_i} \nonumber \\
      & + \sum_{i \neq n}\sum_{j\neq i}\frac{1}{\mu_n}(W_{ni} - V_{ni})\frac{1}{\mu_i}(W_{ij} - V_{ij})\frac{1}{\mu_j}\delta_j\braket{\hat{X}_j} \nonumber \\
      & + \cdots. \label{eq:explicit-y-in-terms-of-x}
\end{align}
In this series solution, the perturbative parameter $(W_{ni}-V_{ni})/\mu_i$ compares the importance of cross-manifold transitions to that of dephasing and controls the size of the higher-order corrections. The estimates in Eq.~\eqref{eq:matrix-element-Wfn-estimate} in appendix~\ref{sec:matrix-element-estimates} lead to a bound on this perturbative parameter $(W_{ni}-V_{ni})/\mu_i = O(\alpha^{-2|i-n|}) \ll 1$ for all $i \neq n$ in the limit of $\alpha \gg 1$ and $n_{\text{th}} \ll 1$. This means that for a given manifold $n$, the only potential way for the first-order correction to become significant compared to the zeroth-order approximation is when the strength of tunneling $\delta_n\braket{\hat{X}_n}$ occurring in this manifold is much smaller than the strength of tunneling $\delta_i\braket{\hat{X}_i}$  occurring in some other manifold $i \neq n$. If this is the case, the contribution to the switching rate from manifold $n$ can anyways be neglected, and Eq.~\eqref{eq:explicit-y-in-terms-of-x-zeroth-order} remains a good approximation for the purpose of evaluating the total switching rate $\Gamma$. To make an even stronger case for the validity of this approximation, in appendix~\ref{sec:validity-of-approximations} we show numerically that each individual contribution $\Gamma_n$ to switching rate is not significantly altered by the approximation of Eq.~\eqref{eq:explicit-y-in-terms-of-x-zeroth-order}. The analytical argument and numerical evidence together justify the use of the approximation of Eq.~\eqref{eq:explicit-y-in-terms-of-x-zeroth-order} uniformly for all manifolds.

With Eq.~\eqref{eq:explicit-y-in-terms-of-x-zeroth-order} we can re-express $\Gamma_n$ as expressed by Eq.~\eqref{eq:switching-rate-n} in terms of $\braket{\hat{X}_n}$ only,
\begin{align}
    \Gamma_n = \frac{1}{\sum_m\braket{\hat{X}_m}}\left[\frac{\delta_n^2}{\mu_n}\braket{\hat{X}_n} + \left(2\sum_fV_{fn}\right)\braket{\hat{X}_n}\right]. \label{eq:switching-rate-x-only}
\end{align}
The term $(\delta_n^2/\mu_n)\braket{\hat{X}_n}$ arises from the term $-\delta_n\braket{\hat{Y}_n}$ in Eq.~\eqref{eq:switching-rate-n}, which describes tunneling. The rate $\delta_n^2/\mu_n$, a combination of the tunnel splitting and the dephasing rate \{cf. Eq.~(16) of \cite{dykman_theory_1978} and Eq.~(29) of \cite{greenfield_unified_2025}\}, can be regarded as a manifestation of the quantum Zeno effect for the following reason. Suppose we replace the environment's weak measurement, which is responsible for the dephasing rate, by periodic projective measurements of $\hat{X}_n$ at an interval of $\tau_n = 2 / \mu_n$. This creates a competition between the tunneling and measurement. In the segment of free evolution under the Hamiltonian $\hat{H}$, the state coherently tunnels from the right well to the left well at the rate $\delta_n$, creating an inter-well superposition. The superposition is periodically collapsed onto the right- or left-well states by the projective measurements. In the limit of $\delta_n \tau_n \ll 1$, the transition probability between $\ket{\psi_n^R}$ and $\ket{\psi_n^L}$ is small, and the outcome of the projective measurements will show long sequences of the same outcome, randomly interrupted by infrequent transitions. On average, the expectation value of $\braket{\hat{X}_n}$ decays by a factor of $\cos(\delta_n\tau_n) \approx \exp(-\delta_n^2\tau_n^2/2)$ per round of Hamiltonian evolution plus projective measurement. Over time, the expectation value of $\hat{X}_n$ evolves as
\begin{align}
    \braket{\hat{X}_n} &\approx e^{-\frac{1}{2}\delta_n^2\tau_n^2 \cdot \frac{t}{\tau_n}} = e^{-\frac{\delta_n^2}{\mu_n}t}.
\end{align}
We thus recover an incoherent tunneling rate of $\delta_n^2 / \mu_n$. The limit of $\mu_n \gg \delta_n$ corresponds to the well-known quantum Zeno effect where tunneling is frozen by frequent measurement.

\subsubsection{Solving for the quasi-equilibrium population distribution via the adiabatic approximation on \texorpdfstring{$\braket{\hat{X}_{n > 0}}$}{Xn>0}}
\label{sec:solving-for-xn}

Our expression for the switching rate, Eq.~\eqref{eq:switching-rate-x-only}, still depends on the unknown inter-well population difference, $\langle \hat{X}_n\rangle$, in the long time limit. To solve for $\braket{\hat{X}_n}$, we substitute Eq.~\eqref{eq:explicit-y-in-terms-of-x-zeroth-order} into Eq.~\eqref{eq:matrix-equation-x} and eliminate $\braket{\hat{Y}_n}$ to get,
\begin{align}
    -\Gamma\braket{\hat{X}_n} &= -\frac{\delta_n^2}{\mu_n}\braket{\hat{X}_n} - \lambda_n\braket{\hat{X}_n} + \sum_{i \neq n}(W_{ni} - V_{ni})\braket{\hat{X}_i}, \label{eq:matrix-eq-x-effective}
\end{align}
where we use $\lambda_n$ to denote the following rate
\begin{align}
   \lambda_n &= 2V_{nn} + \sum_{f \neq n}(W_{fn} + V_{fn}) \\
   &= 2\sum_f V_{fn} + \sum_{f \neq n}(W_{fn} - V_{fn}).\label{eq:decay-rate}
\end{align}
$\lambda_n$ has the interpretation of the decay rate of the population in the $n$th manifold due to intra-well transitions to other manifolds or inter-well transitions (but excluding the effect of tunneling).

To analytically solve Eq.~\eqref{eq:matrix-eq-x-effective}, we can invoke the adiabatic approximation again for $\braket{\hat{X}_{n > 0}}$, leaving $\braket{\hat{X}_0}$ as the only independent degree of freedom (see appendix~\ref{sec:validity-of-approximations} for numerical verification of this adiabatic approximation). A sufficient condition for the adiabaticity of $\braket{\hat{X}_{n > 0}}$ is $\Gamma \ll \lambda_n$, i.e., the switching rate that we compute must also be much slower than the decay rate of populations in the well, which returns the system into an intra-well quasi-equilibrium state. Using the estimates for $W_{mn}$ and $V_{mn}$ in appendix \ref{sec:matrix-element-estimates} we find that the decay rate scales as $\lambda_n \sim \kappa n$ in the limit of $\alpha \gg 1$ and $n_{\text{th}} \ll 1$. Using this estimate, our criterion for adiabaticity can be written as $\Gamma \ll \kappa$. After the adiabatic approximation, we can rewrite Eq.~\eqref{eq:matrix-eq-x-effective} for $\braket{\hat{X}_{n > 0}}$ into the following form that separates upward transitions from downward transitions:
\begin{align}
    \braket{\hat{X}_{n>0}} &= \frac{\sum_{i < n} (W_{ni} - V_{ni})\braket{\hat{X}_i}}{\lambda_n + \frac{\delta_n^2}{\mu_n}} \nonumber \\
    &\quad + \frac{\sum_{i > n}(W_{ni} - V_{ni})\braket{\hat{X}_i}}{\lambda_n + \frac{\delta_n^2}{\mu_n}}. \label{eq:x-population-recursive}
\end{align}
The first term consists of transitions coming from lower manifolds, and the second consists of those coming from higher manifolds. If we can neglect the second term, then
\begin{align}
    \braket{\hat{X}_{n>0}} \approx \frac{\sum_{i < n} (W_{ni} - V_{ni})\braket{\hat{X}_i}}{\lambda_n + \frac{\delta_n^2}{\mu_n}}, \label{eq:x-neglect-down-transition}
\end{align}
which constitutes a triangular set of equations that can be solved one by one analytically, starting from the $n=1$ case.

To argue that the second term of Eq.~\eqref{eq:x-population-recursive} is negligible, we estimate the magnitude of its two terms. First, we establish that $\braket{\hat{X}_n}$ decreases with $n$ using the following argument. The estimates for $W_{mn}$ and $V_{mn}$ from appendix \ref{sec:matrix-element-estimates} show that between two manifolds $m$ and $n$, the upward transition rate is always smaller than the downward transition rate as long as $n_{\text{th}} \ll 1$. They also show that the rate of upward transitions decreases exponentially with the degree of separation, $|m - n|$, between the manifolds. These two observations suggest that highly excited states are rarely populated, and the quasi-equilibrium population differences should satisfy $\braket{\hat{X}_0} \gg \braket{\hat{X}_1} \gg \braket{\hat{X}_2} \gg \cdots$. Next, we establish that the dominant downward transition into manifold $n$ comes from manifold $n + 1$ in the following way. We observe that given the estimates of appendix \ref{sec:matrix-element-estimates}, the rate of downward transitions also decreases exponentially with the degree of separation, $|m - n|$. Combined with the decrease of $\braket{\hat{X}_n}$ with $n$, we have that the largest term in the sum for downward transitions $\sum_{i > n}(W_{ni} - V_{ni})\braket{\hat{X}_i}$ is the term $i = n + 1$, i.e.,
\begin{align}
    \sum_{i > n}(W_{ni} - V_{ni})\braket{\hat{X}_i} \approx (W_{n,n+1} - V_{n,n+1})\braket{\hat{X}_{n+1}}.
\end{align}

We continue by estimating the probability current of the dominant downward transition, $(W_{ni} - V_{ni})\braket{\hat{X}_i}$ where $i = n + 1$. The pre-factor $(W_{n,n+1} - V_{n,n+1})$ is in fact the dominant contributor to the decay rate $\lambda_{n+1}$, so we may approximate
\begin{align}
    (W_{n,n+1} - V_{n,n+1})\braket{\hat{X}_{n+1}} \approx \lambda_{n+1}\braket{\hat{X}_{n + 1}}.
\end{align}
Now, given that $\lambda_n$ and $\lambda_{n + 1}$ are of the same order and $\braket{\hat{X}_n} \gg \braket{\hat{X}_{n + 1}}$, we must have that the probability current of the dominant downward transition entering manifold $n$ is much smaller than the total decay probability current leaving $n$, that is,
\begin{align}
    \sum_{i > n}(W_{ni} - V_{ni})\braket{\hat{X}_i} \approx \lambda_{n+1}\braket{\hat{X}_{n + 1}} \ll \lambda_n\braket{\hat{X}_n}.
\end{align}
Yet, in the adiabatic approximation, the total probability current entering manifold $n$ is equal to that leaving the same manifold. So we must have that the upward transition probability current entering manifold $n$ is much larger than the downward transition probability current entering the same manifold. We numerically verify the approximation of Eq.~\eqref{eq:x-neglect-down-transition} in appendix~\ref{sec:validity-of-approximations}.

Having justified the neglect of the second term of Eq.~\eqref{eq:x-population-recursive}, we proceed to solve the triangular set of equations that Eq.~\eqref{eq:x-neglect-down-transition} generates, starting from $n=1$,
\begin{align}
    \braket{\hat{X}_1} &= \frac{(W_{10} - V_{10})\braket{\hat{X}_0}}{\lambda_1 + \frac{\delta_1^2}{\mu_1}}, \\
    \braket{\hat{X}_2} &= \frac{(W_{20} - V_{20})\braket{\hat{X}_0}}{\lambda_2 + \frac{\delta_2^2}{\mu_2}} + \frac{(W_{21} - V_{21})\braket{\hat{X}_1}}{\lambda_2 + \frac{\delta_2^2}{\mu_2}}. \\
    &\cdots \nonumber
\end{align}
The first term of each equation captures excitations from the ground state manifold, and subsequent terms capture excitations from excited state manifolds. Using the following shorthand notation,
\begin{align}
    R_{mn} = \frac{W_{mn} - V_{mn}}{\lambda_m + \frac{\delta_m^2}{\mu_m}},\quad (m \neq n), \label{eq:definition-of-Rmn}
\end{align}
the explicit solution can be written down formally as the finite sum
\begin{align}
    \braket{\hat{X}_n} = \sum_{s=1}^n\left(\sum_{\substack{0<i_1<i_2<\\...<i_{s-1}<n}} R_{ni_{s-1}}\cdots R_{i_2i_1}R_{i_10}\right) \braket{\hat{X}_0} \label{eq:x-up-solution}
\end{align}
where the inner sum goes over all $s$-step paths connecting the ground-state manifold to the $n$th manifold, and the outer sum goes over all possible number of steps.

We then compute the first-order corrections due to downward transitions from higher manifolds perturbatively by substituting the solution Eq.~\eqref{eq:x-up-solution} into the second term of Eq.~\eqref{eq:x-population-recursive} and solving the triangular set of equations again. The solution that contains corrections from downward transitions to first order has the following form,
\begin{align}
    \braket{\hat{X}_n} = K_n\braket{\hat{X}_0} = \frac{K_n}{\sum_{m \geq 0} K_m}, \label{eq:x-down-correction}
\end{align}
where $K_0 = 1$ and $K_{n>0}$ is defined as
\begin{align}
K_{n>0} = \sum_{\text{paths}} R_{ni_{s-1}}\cdots R_{i_2i_1}R_{i_10}. \label{eq:K-definition}
\end{align}
The sum now goes over all possible paths $0 \mapsto i_1 \mapsto i_2 \mapsto \cdots \mapsto i_{s - 1} \mapsto n$ that begins from the ground state manifold, does not return to the ground state manifold ($i_k \neq 0$), ends at the $n$th manifold, and involves at most 1 downward step. We numerically compare the approximations of Eq.~\eqref{eq:x-up-solution} and Eq.~\eqref{eq:x-down-correction} also in appendix~\ref{sec:validity-of-approximations}.

Having solved for $\braket{\hat{X}_{n > 0}}$, we find an explicit expression for the switching rate $\Gamma_n$ of the $n$th manifold. Plugging the solutions Eq.~\eqref{eq:x-down-correction} for the quasi-equilibrium population differences into Eq.~\eqref{eq:switching-rate-x-only} yields the following explicit expression for the contribution to the spontaneous switching rate from manifold $n$,
\begin{align}
    \Gamma_n = \left(\frac{\delta_n^2}{\mu_n} + 2\sum_fV_{fn}\right)\frac{K_n}{\sum_{m\geq0}K_m}. \label{eq:Gamma-n-solution}
\end{align}
The terms in the parenthesis has the dimension of rate and represents the switching rate for the population in the $n$th manifold. The two terms in the parenthesis capture the effective tunneling rate due to the quantum Zeno effect and the transition rate due to quantum jumps, respectively. The term outside the parenthesis weighs the rates for all the manifolds by the quasi-equilibrium population in each manifold. We numerically verify this solution in appendix \ref{sec:validity-of-approximations}. Note that while the experiments \cite{frattini_observation_2024, hajr_high-coherence_2024-2, albornoz_oscillatory_2024} do not measure each $\Gamma_n$ individually, we nevertheless can see the effect of the dominant $\Gamma_n$ at each $\epsilon_2/K$ from the staircase, which is a sum of all $\Gamma_n$.

\subsubsection{Final solution and summary of the derivation}

While we have found the explicit solution for the contribution $\Gamma_n$ to the spontaneous switching rate from ma nifold $n$ [Eq.~\eqref{eq:Gamma-n-solution}], its current form obscures the physics that gives rise to the staircase. It isn't clear which term in Eq.~\eqref{eq:Gamma-n-solution} is responsible for the step-like decrease in the total spontaneous switching rate $\Gamma$ as a function of the two-photon drive amplitude $\epsilon_2/K = \alpha^2$, because the two factors in $\Gamma_n$ change simultaneously and in opposite directions. In particular, when we increase $\alpha^2$ past a critical value that allows one additional excited state manifold to fall sufficiently deep into the double well, $\delta_n$ decreases exponentially, but the normalized population difference in this manifold $\braket{\hat{X}_n} = K_n\,/\sum_{m\geq0}\!K_m$ sharply increases precisely because tunneling, which otherwise acts to reduce $\braket{\hat{X}_n}$, is now suppressed.

\begin{figure}[ht!]
    \centering
    \includegraphics{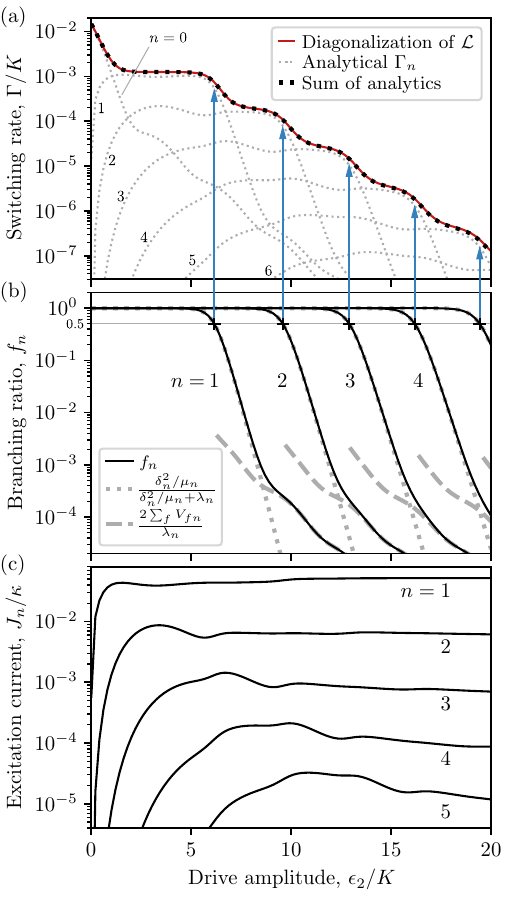}
    \caption{\textbf{The semi-analytical formula for the switching rate $\Gamma$.} (a) Comparison between the numerical total spontaneous switching rate (red solid line) and the contribution $\Gamma_n$ from each manifold (grey dotted lines) calculated using the semi-analytic formula Eq.~\eqref{eq:Gamma-n-solution-summarized}. The sum of $\Gamma_n$ (black dotted line) closely agrees with numerics. Each $\Gamma_{n>0}$ is the product of the branching ratio $f_n$, shown in (b), and the incoming probability current $J_n$, shown in (c). (b) The branching ratios $f_n$ (black solid lines) change drastically with the two-photon drive amplitude $\epsilon_2$ and can be approximated by two asymptotic expressions (grey dashed line and grey dotted line) that are valid in their respective regimes. The condition $f_n = 1/2$ (black crosses) determines the critical values of $\epsilon_2/K$ at which there is an abrupt decrease in the spontaneous switching rate (blue arrows). (c) The probability currents $J_n$ depend weakly on the two-photon drive amplitude once the manifold is captured by the double well. In (a), (b), and (c) we take $\kappa/K = 0.025$ and $n_{\text{th}} = 0.05$. Code used to generate the plot is available at \cite{su_data_2025}.}
    \label{fig:branching-ratio}
\end{figure}

To elucidate the physics of the staircase contained in the expression of $\Gamma$, we rewrite the expression for $\Gamma_n$ Eq.~\eqref{eq:Gamma-n-solution} into the following form for all excited state manifolds $n > 0$,
\begin{align}
    \Gamma_n & = f_nJ_n \quad \text{for } n > 0, \label{eq:Gamma-n-solution-manipulated} \\
    f_n &= \frac{\frac{\delta_n^2}{\mu_n} + 2\sum_fV_{fn}}{\frac{\delta_n^2}{\mu_n} + \lambda_n}, \label{eq:branching-ratio} \\
    J_n &= \sum_{i \neq n}(W_{ni} - V_{ni})\frac{K_i}{\sum_{m\geq0}K_m}, \label{eq:probability-current}
\end{align}
where $f_n$ is the branching ratio into transitions that cause switching between the wells, and $J_n$ is (the inter-well difference of) the total probability current entering manifold $n$. This rewriting is achieved by multiplying Eq.~\eqref{eq:Gamma-n-solution} by $\sum_{i \neq n} (W_{ni} - V_{ni}) \braket{\hat{X}_i}\Big/\left(\frac{\delta_n^2}{\mu_n} + \lambda_n\right)\braket{\hat{X}_n}$, which is the factor that approaches unity in the adiabatic approximation of $\Gamma \ll \lambda_n$ according to Eq.~\eqref{eq:matrix-eq-x-effective}. The difference between the two expressions for the switching rate contribution $\Gamma_n$, Eqs.~\eqref{eq:Gamma-n-solution} and \eqref{eq:Gamma-n-solution-manipulated}, is that the latter is expressed in terms of the populations in manifolds $i \neq n$ rather than the population of manifold $n$ itself. This distinction has the following benefit. For a manifold $n$ that is being captured by the double well, the probability current $J_n$ entering it predominantly depends on the quasi-equilibrium population distribution in the double well, which is insensitive to changes near the top of the energy barrier (Fig.~\ref{fig:branching-ratio}c). We therefore attribute the sharp drop in the spontaneous switching rate $\Gamma$ to the abrupt decrease in the branching ratios $f_n$ as functions of the two-photon drive amplitude $\epsilon_2/K = \alpha^2$ (Fig.~\ref{fig:branching-ratio}b). On the other hand, $f_n \approx 1$ in the flat section of the steps, indicating that the manifolds are dominated by tunneling and the limiting factor for the switching rate is the rate at which the manifold is populated. This confirms the qualitative picture that we described in section \ref{sec:effective-equations-of-motion}.

Fig.~\ref{fig:branching-ratio}a shows a plot of $\Gamma_n$ [Eq.~\eqref{eq:Gamma-n-solution-manipulated}] as functions of the drive amplitude. Although for $n_{\mathrm{th}} \ll 1$ almost all the population in the quasi-equilibrium state is in the ground state manifold, the plot shows that the dominant manifold contributing to the switching rate is near the top of the energy barrier (cf. Fig.~\ref{fig:background-combined}b). This is consistent with \cite{dykman_quantum_1988, marthaler_switching_2006}, which found that for driven oscillators, switching is dominated by over-the-barrier transitions rather than tunneling below the barrier. The plot also shows that the switching rate changes slowly within the flat part of the step despite the increasing energy barrier. This is consistent with our discussion of the tunneling-dominated regime in section \ref{sec:effective-equations-of-motion}, and agrees with the observation in \cite{dykman_theory_1978} about the case where the tunneling splitting exceeds the linewidth.

To summarize, we have expressed the spontaneous switching rate $\Gamma$ in terms of the tunnel splittings $\delta_k$, and the inter- and intra-well transition rates $W_{pq}$ and $V_{pq}$ induced by $\hat{a}$ and $\hat{a}^\dagger$,
\begin{align}
    &\Gamma = \Gamma(\delta_k, W_{pq}, V_{pq}) = \sum_n\Gamma_n(\delta_k, W_{pq}, V_{pq}) \label{eq:Gamma-final-solution} \\
    &\Gamma_n(\delta_k, W_{pq}, V_{pq}) \nonumber \\
    &\quad = \begin{cases}
        \left(\frac{\delta_0^2}{\mu_0} + 2\sum_fV_{f0}\right)\frac{K_0}{\sum_{m\geq0}K_m} & n = 0 \\
        f_nJ_n & n > 0
    \end{cases}. \label{eq:Gamma-n-solution-summarized}
\end{align}
We introduce the explicit function notation for the semi-analytic formula because later we will substitute the same functional form to compute other semi-analytical switching rates. The following is a summary of the assumptions and approximations we have made to obtain this result from the Lindblad master equation [Eq.~\eqref{eq:lindblad}]:
\begin{enumerate}
    \item We projected the density operator onto two-level manifolds;
    \item We assumed that spontaneous switching is the slowest process in the system, well separated from intra-well relaxation, and therefore is the only dynamics in the long-time limit;
    \item We assumed that the inter-well coherence is dominated by tunneling and dephasing induced by the measurement of ``which-well" information by the environment.
    \item We invoked the adiabatic approximation to eliminate all other degrees of freedom apart from the inter-well population difference.
    \item Finally, when finding the quasi-equilibrium population differences $\braket{\hat{X}_{n>0}}$, we treated downward transitions as a small perturbation and only used the first-order correction.
\end{enumerate}

\section{Location of the steps in the staircase}
\label{sec:location-of-steps}
In this section we use the semi-analytical formula for the switching rate to find the location of the steps in the staircase. To locate the critical values of the normalized two-photon drive amplitude $\alpha^2 = \epsilon_2/K$ where a sharp decrease in the spontaneous switching rate $\Gamma$ occurs, we plot the branching ratios $f_n$, defined in Eq.~\eqref{eq:branching-ratio}, as functions of $\epsilon_2/K$ for excited state manifolds $n>0$ (Fig.~\ref{fig:branching-ratio}b). At small $\alpha$, tunneling is dominant and $\delta_n^2/\mu_n \gg \lambda_n$, so $f_n \approx 1$, meaning all probability currents that enter an excited state manifold lead to spontaneous switching between wells. As $\alpha$ increases, one by one the branching ratio decreases, and the manifold enters the dissipation-dominated regime where switching is strongly inhibited by dephasing, and nearly all probability currents that enter an excited state manifold eventually stay confined within the well. Then for a given manifold $n$, we may define the critical normalized two-photon drive amplitude in terms of $\alpha_{\text{crit}, n}$, the value of $\alpha$ at which the inter- and intra-well transition rates emanating from manifold $n$ become equal. This is equivalent to the value of $\alpha$ at which $f_n = 1/2$. That is, $\alpha_{\text{crit}, n}$ is such that
\begin{align}
    f_n = \frac{\frac{\delta_n^2}{\mu_n} + 2\sum_fV_{fn}}{\frac{\delta_n^2}{\mu_n} + \lambda_n} = \frac{1}{2}.
\end{align}
Comparing Fig.~\ref{fig:branching-ratio}a and \ref{fig:branching-ratio}b shows that the location of the steps in the staircase agrees with this condition.

To obtain a simpler expression for the condition on $\alpha_{\text{crit}, n}$, we notice from Fig.~\ref{fig:branching-ratio}b that when $f_n = \frac{1}{2}$, the branching ratio can be approximated by the expression
\begin{align}
    f_n \approx \frac{\frac{\delta_n^2}{\mu_n}}{\frac{\delta_n^2}{\mu_n} + \lambda_n}.
\end{align}
Using this approximate expression, we find, after some manipulation, that $\alpha_{\text{crit}, n}$ is the solution to the following equation
\begin{align}
    \frac{\delta_n^2(\alpha)}{\mu_n(\alpha)} = \lambda_n(\alpha). \label{eq:critical-alpha}
\end{align}
Using the estimates derived in appendix~\ref{sec:matrix-element-estimates} and \ref{sec:wkb}, we can approximate this condition as
\begin{align}
    \delta_n^2(\alpha) \approx 2n\kappa^2|\alpha|^2, \label{eq:critical-alpha-approximate}
\end{align}
where $\delta_n(\alpha)$ is an explicit function of $\alpha$ given in appendix~\ref{sec:wkb}. In particular, $\alpha_{\text{crit}, 1}$ as determined by the special case $n=1$ of the condition above increases as the damping rate $\kappa$ decreases. Therefore, in this limit, the critical $\alpha$ beyond which there is an exponential suppression of the spontaneous switching rate $\Gamma$ will be delayed, as seen numerically in Fig.~\ref{fig:numerical-staircase}.

\begin{figure}[t]
    \centering
    \includegraphics[width=\linewidth]{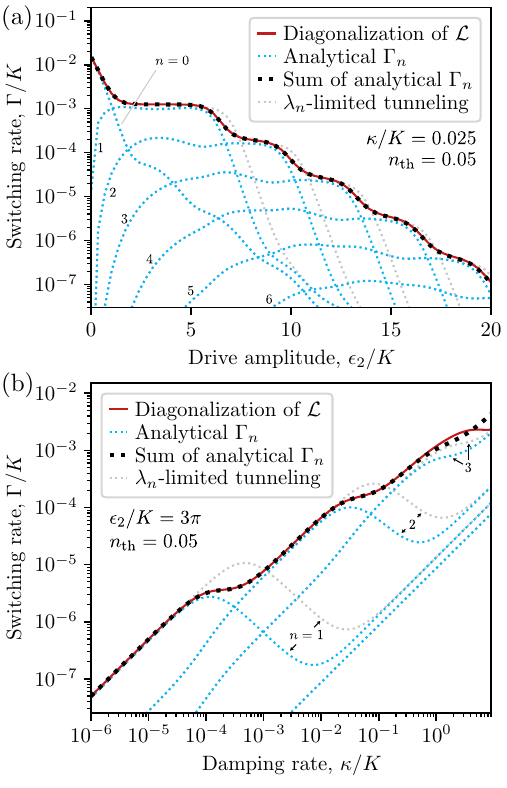}
    \caption{\textbf{The role of $\mu_n$ in the switching rate.} We compare the semi-analytical switching rates $\Gamma_n$ from Eq.~\eqref{eq:Gamma-n-solution-summarized} (dotted blue lines) and their sum (dotted black lines) to a modified version of Eq.~\eqref{eq:Gamma-n-solution-summarized}, where we replace all instance of the effective tunneling rate $\delta_n^2/\mu_n$ by $\delta_n^2/\lambda_n$ (dotted gray lines). The numerical switching rate is plotted as solid red lines for reference. In (a) the switching rate is plotted as a function of the drive amplitude $\epsilon_2/K$. The replacement shifts the drive amplitudes at which the steps occur. In (b) the switching rate is plotted as a function of the damping rate $\kappa/K$. The local maximum of the blue dotted lines can be determined from $\delta_n$, $\mu_n$ and $\lambda_n$ using Eq.~\eqref{eq:critical-alpha}, whereas the local maximum of the gray dotted lines can be determined using the condition $\delta_n = \lambda_n$. The poor agreement between the gray dotted lines and the numerical switching rate shows the importance of distinguishing $\mu_n$ and $\lambda_n$.}
    \label{fig:lamda_n-limited-tunneling}
\end{figure}

In Eq.~\eqref{eq:critical-alpha}, both $\mu_n$ and $\lambda_n$ ultimately arise from single photon loss and gain, are proportional to $\kappa$, and characterize the rates of processes that limit tunneling. So one may be led to believe that it is sufficient to keep only one of these rates. However, $\mu_n$ characterizes the dephasing rate of inter-well superpositions in manifold $n$, and $\lambda_n$ characterizes the rate of decay from manifold $n$, so $\mu_n$ and $\lambda_n$ are \textit{a priori} different. Merging $\mu_n$ and $\lambda_n$ also leads to observable and quantitative effects in the switching rate. If, for example, we merge $\mu_n$ and $\lambda_n$ and only keep $\lambda_n$, then we can simplify Eq.~\eqref{eq:critical-alpha} into $\delta_n^2(\alpha) = \lambda_n^2(\alpha)$ [or simply $\delta_n(\alpha) = \lambda_n(\alpha)$]. Owing to the different scalings $\mu_n \sim 2\alpha^2\kappa$ and $\lambda_n \sim n\kappa$ (appendix \ref{sec:matrix-element-estimates}), the modified condition approximates to $\delta_n^2(\alpha) \approx n^2\kappa^2$ instead of Eq.~\eqref{eq:critical-alpha-approximate}. The modified condition produces different solutions $\alpha_{\mathrm{crit}, n}$ and predicts shifted locations of the steps in the staircase. We numerically demonstrate this shift by changing all occurrences of $\delta_n^2/\mu_n$ in the semianalytical formula [Eq.~\eqref{eq:Gamma-n-solution-summarized}] into $\delta_n^2/\lambda_n$ (Fig.~\ref{fig:lamda_n-limited-tunneling}a). The modified formula, which we label ``$\lambda_n$-limited tunneling", yields a staircase shifted towards larger values of $\epsilon_2/K$. However, at the given damping rate $\kappa/K=0.025$, the shift of the staircase is moderate and does not grow with $\epsilon_2/K$ in the plot range. Nor does the shift modify the overall slope of the staircase. This is because $\delta_n(\alpha)$ is a rapidly changing function of $\alpha$, so the pre-factor difference between $\mu_n$ and $\lambda_n$ can be compensated by a small variation in $\alpha_{\mathrm{crit}, n}$. Consistent with this observation, several phenomena in this oscillator have been modeled when the difference between $\mu_n$ and $\lambda_n$ is omitted \cite{marthaler_switching_2006}.

Merging the rates $\mu_n$ and $\lambda_n$ into a single rate causes a more prominent effect in the dependence of the switching rate on $\kappa$. Regarded as a condition on $\kappa$ at fixed $n$ and $\alpha$, Eq.~\eqref{eq:critical-alpha} picks out the critical value of $\kappa$ that separates the tunneling- and dissipation-dominated regimes for a particular manifold. In the tunneling-dominated regime, increasing $\kappa$ increases the rate of excitation into an excited state while the branching ratios $f_n$ stays near 1, so the switching rate $\Gamma_n$ increases with $\kappa$. In the dissipation-dominated regime, $f_n$ is sharply suppressed with increasing $\kappa$, so $\Gamma_n$ decreases with $\kappa$. Therefore the critical value of $\kappa$ is a local maximum point in $\Gamma_n$ (Fig.~\ref{fig:lamda_n-limited-tunneling}b). The solution for $\kappa$ produced by the formula $\delta_n^2(\alpha) \approx n^2\kappa^2$ is a factor of $\sqrt{2|\alpha|^2 / n}$ larger than that produced by Eq.~\eqref{eq:critical-alpha-approximate}, and this factor grows in the limit of large $|\alpha|^2$ and small $n$. In Fig.~\ref{fig:lamda_n-limited-tunneling}b we numerically observe this shift in the local maximum of $\Gamma_n$ by comparing the semianalytical formula for $\Gamma_n$ [Eq.~\eqref{eq:Gamma-n-solution-summarized}] and ``$\lambda_n$-limited tunneling" defined in the previous paragraph. In particular, in the low-damping rate regime ($\kappa/K \approx 10^{-4}$), the horizontal shift in the local maximum point for $n=1$ and $|\alpha|^2 = 3\pi$ is consistent with the theoretical prediction $\sqrt{2|\alpha|^2/n} \approx 4.3$. There is also an overshoot in the local maximum value of the switching rate $\Gamma_1$ by the same factor because in the small $\kappa$ regime $\Gamma_1$ grows linearly with the $\kappa$. Therefore, distinguishing between $\mu_n$ and $\lambda_n$ is particularly important in the low-damping regime.

\section{Application of the semi-analytical formula: activation mechanism within the well}
\label{sec:activation-mechanism}
\begin{figure*}[th]
    \centering
    \includegraphics{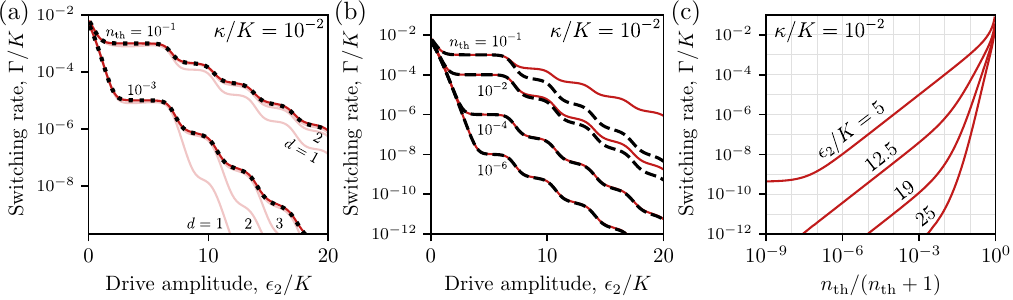}
    \caption{\textbf{The dominant activation mechanism for spontaneous switching.} We compare the spontaneous switching rate $\Gamma$ for different values of $n_{\text{th}}$ under cascaded thermal heating and direct thermal heating, as defined in section \ref{sec:activation-mechanism}. The red solid lines in all subplots are obtained numerically from the full Lindbladian $\mathcal{L}$ defined in Eq.~\eqref{eq:lindblad}. (a) The pink solid lines and the black dotted lines correspond to $\Gamma^{(\text{casc})}$, which is the semi-analytic switching rate under cascaded thermal heating as defined by Eq.~\eqref{eq:Gamma-casc}. The cutoff $d$ for allowed thermal heating transitions is set to $1, 2, 3, 4$ for the pink solid lines and $\infty$ for the black dotted lines. (b) The black dashed lines correspond to $\Gamma^{(\text{dir})}$, which is the semi-analytic switching rate under direct thermal heating as defined by Eq.~\eqref{eq:Gamma-dir}. (c) Exact numerical spontaneous switching rate $\Gamma$ as a function of $n_{\text{th}}/(n_{\text{th}} + 1)$, taken at two-photon drive amplitudes $\epsilon_2$ which correspond to the first, third, fifth, and seventh steps in the staircase.
    At $n_{\mathrm{th}} / (n_{\mathrm{th}} + 1) = 0.1$, the slopes of the curves are computed to be 1.1, 2.3, 3.6, and 4.9, respectively.}
    \label{fig:activation-mechanism}
\end{figure*}

What is the dominant activation mechanism in a Kerr parametric oscillator? This is a question we can answer with the semi-analytical formula we derived for the switching rate. Here, activation mechanism refers to any physical process that excites the population from the ground state manifold to the excited state manifold where switching occurs. One counterintuitive activation mechanism is quantum heating \cite{marthaler_switching_2006, dykman_critical_2007, dykman_quantum_2011, ong_quantum_2013}. It is a process where losing a photon creates an excitation in the oscillator \cite{dykman_fluctuating_2012}. The process is caused by the sometimes sizeable upward transition matrix elements of $\hat{a}$, and it can occur even when the temperature of the environment is zero. Previous work \cite{marthaler_switching_2006} has found quantum heating to be the dominant activation mechanism in systems similar to the one of our interest. The scaling law of the switching rate due to quantum heating has also been experimentally observed in \cite{vijay_invited_2009, ong_quantum_2013}. However, our system is a special case of the system in \cite{marthaler_switching_2006} because there is exactly zero detuning between the frequency of the two-photon drive in our system and twice the (Stark- and Lamb-shifted) resonant frequency of the nonlinear oscillator. This drive condition leads to a double well whose right- and left-well states are approximately displaced Fock states. The heating transition matrix elements of $\hat{a}$ between these states (see the estimates for $W_{mn}$ in appendix~\ref{sec:matrix-element-estimates}) are much closer to zero than when there is detuning, so it is unclear whether quantum heating remains the dominant activation mechanism. We answer this question with our theory in this section.

The activation mechanism can be deduced from the sensitivity of our semi-analytical formula to the matrix elements of $\hat{a}$ and $\hat{a}^\dagger$, which govern the rates of in- and cross-manifold transitions. First, we note that $\hat{a}$ dominates over $\hat{a}^\dagger$ for in-manifold transitions because $n_{\text{th}}\ll 1$ [see Eq.~\eqref{eq:W-definition} and take $p = q = n$]. Next, note that among all transitions towards lower manifolds (cooling), we can also neglect those induced by $\hat{a}^\dagger$ because $n_{\text{th}}\ll 1$ and the relevant matrix elements are much smaller than the corresponding ones for $\hat{a}$-induced cooling transitions (see estimates of appendix~\ref{sec:matrix-element-estimates}). The remaining cooling transitions are those induced by $\hat{a}$, among which we can also neglect the ones that lower the manifold index by more than 1. The reason for this last statement has already been discussed in section~\ref{sec:solving-for-xn} when we estimated the magnitude of downward transitions.

To verify the observations we made about the cooling transitions and to identify which heating transitions are dominant, we consider two scenarios where we set different matrix elements in the inter-well and intra-well transition rates $V_{pq}$, $W_{pq}$ to zero and see what effect the change has on the switching rate $\Gamma$ computed using the semi-analytical formula. Neither of the two scenarios which we construct will include quantum heating. In the first scenario, the modified inter-well and intra-well transition rates have the following elements,
\begin{align}
    V_{pq}^{(\text{casc})} = \begin{cases}
    \kappa (1 + n_{\text{th}})\left|\bra{\psi_p^L}\hat{a}\ket{\psi_q^R}\right|^2 & q - 1 \leq p \leq q \\
    \kappa n_{\text{th}}\left|\bra{\psi_p^L}\hat{a}^\dagger\ket{\psi_q^R}\right|^2 & q < p \leq q + d \\
    0 & \text{otherwise} \\
    \end{cases}, \label{eq:V_pq^casc}
\end{align}
\begin{align}
    W_{pq}^{(\text{casc})} = \begin{cases}
    \kappa (1 + n_{\text{th}})\left|\bra{\psi_p^R}\hat{a}\ket{\psi_q^R}\right|^2 & q - 1 \leq p \leq q \\
    \kappa n_{\text{th}}\left|\bra{\psi_p^R}\hat{a}^\dagger\ket{\psi_q^R}\right|^2 & q < p \leq q + d \\
    0 & \text{otherwise} \\
    \end{cases}, \label{eq:W_pq^casc}
\end{align}
where $d$ is some positive integer. This set of transition rates excludes all quantum heating and any thermal heating that raises the manifold index by more than $d$. This means that to reach a highly excited state, multiple thermal heating events are required if $d$ is set to a small number. We refer to this set of rates as the cascaded thermal heating scenario, hence the superscript ``(casc)". Note that setting $d = \infty$ allows all thermal heating transitions, such as those from the ground state manifold directly to any excited state manifold.

The second set of modified inter-well and intra-well transition rates we construct have the following elements,
\begin{align}
    V_{pq}^{(\text{dir})} = \begin{cases}
    \kappa (1 + n_{\text{th}})\left|\bra{\psi_p^L}\hat{a}\ket{\psi_q^R}\right|^2 & q - 1 \leq p \leq q \\
    \kappa n_{\text{th}}\left|\bra{\psi_p^L}\hat{a}^\dagger\ket{\psi_0^R}\right|^2 & q = 0 \text{ and } p > 0 \\
    0 & \text{otherwise} \\
    \end{cases}, \label{eq:V_pq^dir} \\
    W_{pq}^{(\text{dir})} = \begin{cases}
    \kappa (1 + n_{\text{th}})\left|\bra{\psi_p^R}\hat{a}\ket{\psi_q^R}\right|^2 & q - 1 \leq p \leq q \\
    \kappa n_{\text{th}}\left|\bra{\psi_p^R}\hat{a}^\dagger\ket{\psi_0^R}\right|^2 & q = 0 \text{ and } p > 0 \\
    0 & \text{otherwise} \\
    \end{cases}. \label{eq:W_pq^dir}
\end{align}
This set of transition rates excludes all quantum heating and any thermal heating that does not originate from the ground state manifold. This means that reaching a highly excited state needed for switching requires a single thermal heating event from the ground state to the target manifold. We refer to this set of rates as the direct thermal heating scenario, hence the superscript ``(dir)". The switching rates predicted by the semi-analytical formula under the scenario of cascaded thermal heating or under the scenario of direct thermal heating are defined as follows using the function notation found in Eq.~\eqref{eq:Gamma-final-solution}:
\begin{align}
    \Gamma^{(\text{casc})} &= \Gamma(\delta_n, V_{pq}^{(\text{casc})}, W_{pq}^{(\text{casc})}), \label{eq:Gamma-casc} \\
    \Gamma^{(\text{dir})} &= \Gamma(\delta_n, V_{pq}^{(\text{dir})}, W_{pq}^{(\text{dir})}). \label{eq:Gamma-dir}
\end{align}

In Fig.~\ref{fig:activation-mechanism} we compare $\Gamma^{(\text{casc})}$ and $\Gamma^{(\text{dir})}$ to the numerically obtained switching rate. Fig.~\ref{fig:activation-mechanism}a shows that to account for the switching rate at $n_{\text{th}} = 10^{-3}$ up to the $k$th step in the staircase, it is necessary and sufficient to choose the cutoff $d$ in the definition for cascaded thermal heating to be at least $k$. That is to say, thermal heating events must be allowed to raise the manifold index by at least $k$, the exact number required to directly excite population from the ground state manifold to the $k$th excited state manifold. We conclude that at low $n_{\text{th}}$, the activation mechanism is direct thermal heating rather than cascaded thermal heating or quantum heating. This is consistent with the fact that the right- and left-well states are approximately displaced Fock states, which leads to the suppression of the heating matrix elements of $\hat{a}$.

Fig.~\ref{fig:activation-mechanism}a further shows that to account for the switching rate at $n_{\text{th}} = 10^{-1}$, it is necessary and sufficient to choose a cutoff $d$ of 3. We therefore conclude that at moderate $n_{\text{th}}$, the activation mechanism includes cascaded thermal heating events in which the manifold index is raised by 3 or less at a time, rather than direct thermal heating or quantum heating. Transitions between non-nearest-neighbor levels are a consequence of the anharmonicity in the intra-well oscillations \cite{andre_emission_2012}.

The difference in activation mechanism for different $n_{\text{th}}$ is expected because the population in intermediate excited states is higher at larger $n_{\text{th}}$. Our conclusions are supported by Fig.~\ref{fig:activation-mechanism}b, where the semi-analytical switching rate $\Gamma^{(\text{dir})}$ under only direct thermal heating agrees with exact numerics for low $n_{\text{th}}$ but fails for moderate $n_{\text{th}}$.

In Fig.~\ref{fig:activation-mechanism}c, we examine the scaling of the spontaneous switching rate as a function of the ratio $n_{\text{th}} / (1 + n_{\text{th}})$. This particular ratio measures the relative strength of thermal heating and $\hat{a}$-induced cooling, which comprise the dominant cross-manifold transitions in the limit $n_{\text{th}} \ll 1$. By choosing values of $\epsilon_2/K$ that sit at the $k$th step in the staircase, we can fix the dominant manifold in which switching occurs to be manifold $k$. Then, for fixed $\epsilon_2/K$, if the switching rate scales as $\Gamma \propto [n_{\text{th}} / (1 + n_{\text{th}})]^s$, then there are roughly $s$ degrees of separation between the ground state manifold and manifold $k$, measured by the average number of thermal heating events required to complete the excitation. An alternative way to interpret the scaling is as follows. If $n_{\text{th}}$ is taken to be the average thermal photon number in the environment at the oscillator resonant frequency $\omega_0$, i.e. $n_{\text{th}} = (e^{\hbar\omega_0/k_BT} - 1)^{-1}$, then the ratio is $[n_{\text{th}} / (n_{\text{th}} + 1)]^s = \exp(-s\hbar\omega_0 / k_BT)$. This scaling implies an activation energy of $s\hbar\omega_0$.

Fig.~\ref{fig:activation-mechanism}c shows that for $\epsilon_2/K = 5$ and $n_{\text{th}}/ (1 + n_{\text{th}}) \lesssim 10^{-8}$, the switching rate is nearly independent of $n_{\text{th}}$, indicating that the switching rate is dominated by either $\hat{a}$-induced direct transitions within the ground state manifold or by quantum heating. In appendix \ref{sec:validity-of-approximations} we show numerically that for $\epsilon_2/K \lesssim 5$ the switching rate is dominated by direct transitions in the ground state manifold, whereas for $\epsilon_2/K \gtrsim 5$ it is dominated by a new type of quantum heating (see discussion in section~\ref{sec:projection-of-the-density-operator} and Eq.~\eqref{eq:Gamma^p} to follow). This type of quantum heating is induced by the nonhermitian part of the effective Hamiltonian $\hat{H}_{\text{eff}} = \hat{H} - i\kappa \hat{a}^\dagger\hat{a}/2$ (see also \footnote{A. Maiti and J. Claes et al., in preparation.}, which discusses this effect from the perspective of leakage induced by the no-jump part of the quantum trajectory). Without the nonhermitian part of the effective Hamiltonian, quantum heating would be strictly zero from the ground state. Heating into excited states is induced by the $\hat{a}$ operator present in $- i\kappa \hat{a}^\dagger\hat{a}/2$. An analytical formula that captures this effect will be presented in Eq.~\eqref{eq:Gamma^p}.

For $10^{-9} \lesssim n_{\text{th}}/ (1 + n_{\text{th}}) \lesssim 10^{-3}$, the slope of the red curves is $\approx 1$, suggesting that there is only one degree of separation between the ground state manifold and any of the first seven excited state manifolds probed by the chosen values of $\epsilon_2/K$. This confirms our conclusion that direct thermal heating is dominant for low $n_{\text{th}}$. Finally, for $10^{-3} \lesssim n_{\text{th}} / (1 + n_{\text{th}}) < 1$, the degree of separation $s$ increases above 1 (see Fig.~\ref{fig:activation-mechanism} caption for numbers) but stays below the manifold index $k$ (with the exception of $\epsilon_2/K = 5$, for which $s > k$ likely due to corrections of higher order in $n_{\text{th}}$). This means that the path of excitation from the ground state manifold to the $k$th excited state manifold is broken down into smaller steps of average size larger than one. This confirms that cascaded thermal heating is dominant for moderate $n_{\text{th}}$.

\begin{figure*}[t]
    \centering
    \includegraphics[width=\linewidth]{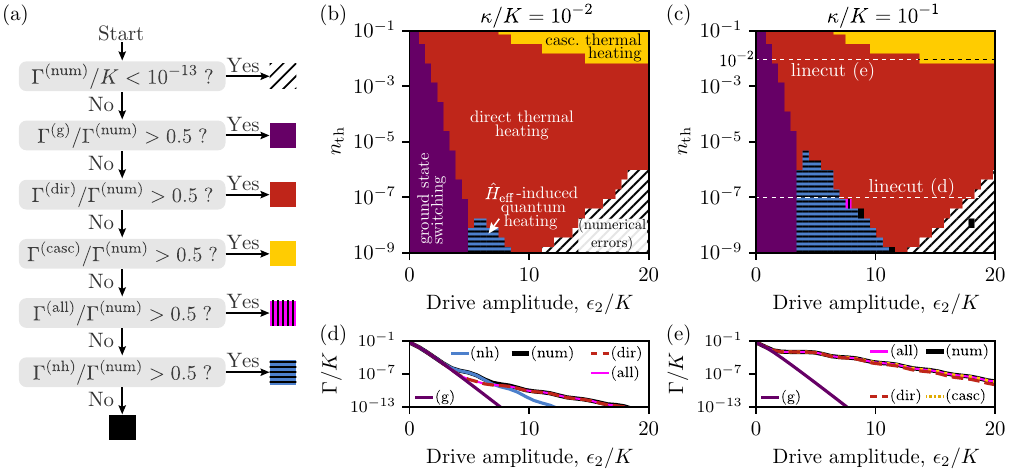}
    \caption{\textbf{Regime plot of activation mechanisms.} We numerically map out the activation mechanism as a function of $\epsilon_2/K$, $\kappa/K$, and $n_{\mathrm{th}}$ by sampling a grid of $24 \times 40$ points. See main text for details. (a) The flowchart that determines the activation mechanism and the color scheme in the regime plot. (b) The activation mechanisms at $\kappa/K=10^{-2}$. (c) The activation mechanisms at $\kappa/K=10^{-1}$. (d) A linecut through (c) at $n_{\mathrm{th}}=10^{-7}$ showing the various computed switching rates. (e) A linecut through (c) at $n_{\mathrm{th}}=10^{-2}$.}
    \label{fig:regimes}
\end{figure*}

Now we numerically map out the activation mechanism as a function of $\epsilon_2/K$, $n_{\mathrm{th}}$, and $\kappa/K$ (Fig.~\ref{fig:regimes}). At each sampled point, we determine the activation mechanism following the flowchart in Fig.~\ref{fig:regimes}a. We first compute the exact numerical switching rate, which we label $\Gamma^{\mathrm{(num)}}$, and filter out most numerical errors using a threshold for low values. Then, we compute a sequence of semi-analytical switching rates under different activation mechanisms and halt as soon as we find a mechanism that explains more than $50\%$ of $\Gamma^{\mathrm{(num)}}$. If all checks fail, then the dominant mechanism is undetermined (due to numerical errors). The activation mechanisms we consider are:
\begin{enumerate}
    \item No activation. Switching occurs entirely within the ground state manifold at a rate of
    \begin{align}
        \Gamma^{\mathrm{(g)}} = 2V_{00}. \label{eq:Gamma^g}
    \end{align}
    \item Direct thermal heating from the ground state to an excited state. Switching occurs at the rate $\Gamma^{\mathrm{(dir)}}$ defined in Eq.~\eqref{eq:Gamma-dir}.
    \item Direct and cascaded thermal activation. In this case, we use the rate $\Gamma^{\mathrm{(casc)}}$ as defined in Eq.~\eqref{eq:Gamma-casc}, but set the cutoff $d$ to $\infty$ to allow for all thermal heating.
    \item Thermal and quantum heating. Switching occurs at the rate predicted by the semi-analytical formula Eq.\eqref{eq:Gamma-final-solution}, which we label as $\Gamma^{\mathrm{(all)}}$ to distinguish from other semi-analytical rates.
    \item Quantum heating between eigenstates of the nonhermitian effective Hamiltonian $\hat{H}_{\mathrm{eff}}$ rather than $\hat{H}_0$. The nonhermitian effective Hamiltonian is the part of the Lindblad master equation that describes the evolution of the state in the absence of quantum jumps. Ignoring photon gain, the effective Hamiltonian corresponding to Eq.~\eqref{eq:lindblad} is perturbed away from $\hat{H}_0$ by the nonhermitian term $-i\kappa\hat{a}^\dagger\hat{a}/2$. To analytically capture the non-Hermitian effect, we replace $V_{pq}, W_{pq}$ in the semianalytical formula by new transition rates $V_{pq}^{\mathrm{(nh)}}, W_{pq}^{\mathrm{(nh)}}$ calculated from the matrix element of $\hat{a}$ between the perturbed eigenstates [see appendix~\ref{sec:validity-of-approximations} Eqs.~\eqref{eq:V_pq^p} and ~\eqref{eq:W_pq^p}], and label the switching rate
    \begin{align}
        \Gamma^{\mathrm{(nh)}} = \Gamma(\delta_n, V_{pq}^{\mathrm{(nh)}}, W_{pq}^{\mathrm{(nh)}}). \label{eq:Gamma^p}
    \end{align}
    This value is independent of $n_{\mathrm{th}}$. We stress that this form of quantum heating lies strictly beyond a rate model for intra-well transitions where the transition rates are specified using the matrix elements of $\hat{a}$ and $\hat{a}^\dagger$ between the unperturbed eigenstates of $\hat{H}_0$. The scaling of $\hat{H}_{\mathrm{eff}}$-induced quantum heating with $\kappa/K$ must be different for ordinary quantum heating because $\kappa/K$ enters into the perturbation of the eigenstates as well as the strength of photon loss. This effect is related to second-order perturbation theory, which has been shown to significantly modify the switching rate in a similar double-well system \cite{dubovitskii_bit-flip_2025}.
\end{enumerate}
We limit $\epsilon_2/K$ such that there are a few ($\lesssim 10$) states in the well and work in the low-damping ($\kappa/K \ll 1$) and low-temperature regime ($n_{\mathrm{th}} < 10^{-1}$).

Figs.~\ref{fig:regimes}b and \ref{fig:regimes}c show the resulting regime plots for activation mechanisms. Switching in the ground state manifold dominates at small enough $\epsilon_2/K$ (purple region). Linecuts in Figs.~\ref{fig:regimes}d and \ref{fig:regimes}e show that the value of $\epsilon_2/K$ at the border of this regime can be determined by equating $\Gamma^{\mathrm{(g)}}$ to the switching rate in the first excited state manifold, populated either by direct thermal heating or $\hat{H}_{\mathrm{eff}}$-induced quantum heating.

For very low temperatures (blue region in Figs.~\ref{fig:regimes}b and \ref{fig:regimes}c), $\hat{H}_{\mathrm{eff}}$-induced quantum heating dominates when $n_{\mathrm{th}}$ falls below a critical value that depends on $\kappa/K$ and $\epsilon_2/K$. However, at such low temperatures, switching may instead be dominated by other processes beyond the model of Eq.~\eqref{eq:lindblad}. For example, quasielastic scattering \cite{marthaler_switching_2006} can lead to oscillator dephasing, and $1/f$ frequency noise can cause oscillator detuning and enhance ordinary quantum heating. Therefore, we are interested in whether the new form of quantum heating can happen at higher temperatures.

As $\kappa/K$ increases from Figs.~\ref{fig:regimes}b to \ref{fig:regimes}c, the regime dominated by $\hat{H}_{\mathrm{eff}}$-induced quantum heating extends to higher temperatures. This is because increasing $\kappa$ perturbs the ground state of the well further from a coherent state. The linecut in Fig.~\ref{fig:regimes} shows that the maximal $n_{\mathrm{th}}$ for which $\hat{H}_{\mathrm{eff}}$-induced quantum heating dominates can be determined by increasing $n_{\mathrm{th}}$ until the purple, blue and red curves meet at a single point. We expect that to experimentally observe $\hat{H}_{\mathrm{eff}}$-induced quantum heating, it is sufficient to increase the damping rate to a high level such that the effect is not washed out by direct thermal heating, nor by other effects outside of the Lindblad master equation model. These are favorable conditions for an experiment with state-of-the-art devices.

Direct and cascaded thermal heating dominates switching in other regimes (red and yellow regions in Figs.~\ref{fig:regimes}b and \ref{fig:regimes}c). We infer from earlier discussion in this section that in the yellow region, cascaded thermal heating dominates over direct thermal heating. Note that $\hat{H}_{\mathrm{eff}}$-induced quantum heating is present but negligible in this regime, so it is not necessary to use the nonhermitian effective Hamiltonian ($\hat{H}_0$ was used in sections \ref{sec:background}-\ref{sec:location-of-steps}).

Finally, the lack of magenta-colored regions in Figs.~\ref{fig:regimes}b and \ref{fig:regimes}c and the overlap between the magenta and the red or yellow lines indicate that quantum heating between the eigenstates of $\hat{H}_0$ can be neglected in this regime. Note that in Fig.~\ref{fig:regimes}c, the isolated magenta sample and black samples on the boundary between regimes are due to ambiguities that arise when $\Gamma^{\mathrm{(nh)}}$ and $\Gamma^{\mathrm{(dir)}}$ each explains roughly 50\% of the exact numerical rate (as revealed by the linecut in Fig.~\ref{fig:regimes}d). The isolated black sample in Fig.~\ref{fig:regimes}c nearest to $\epsilon_2/K=20$ is due to a spurious numerical fluctuation above the $\Gamma^{\mathrm{(num)}}/K = 10^{-13}$ threshold.

\section{Comparison to previous work}
\label{sec:comparison-to-previous-work}
In this section, we compare our work to previous works where theoretical predictions about the switching rate were obtained. We will first focus the comparison to theoretical works that deal specifically with the two-photon-driven Kerr nonlinear oscillator, at zero two-photon drive detuning, and under the influence of single photon loss and gain but no dephasing. Note that the situation with detuning or dephasing has been studied in \cite{marthaler_switching_2006, frattini_observation_2024, ruiz_two-photon_2023, venkatraman_driven_2024}. Then, we broaden our scope and compare our results to a previous theory of resonant tunneling in the magnetization of molecular magnets \cite{garanin_thermally_1997}. Finally, we will discuss how our result relates to the famous Kramers problem (see review in \cite{hanggi_reaction-rate_1990}).

The most important distinction of this work from previous work on the two-photon-driven Kerr nonlinear oscillator is our identification of the role played by the dephasing rate $\mu_n$ in determining the switching rate $\Gamma_n$. This was previously overlooked in \cite{marthaler_switching_2006}, \cite{gautier_combined_2022} and appendix I of \cite{putterman_stabilizing_2022}. As already discussed in section \ref{sec:location-of-steps}, it is important to differentiate between $\mu_n$ from $\lambda_n$.

A related distinction of this work from previous work is that our condition for the critical manifold $n_{\text{crit}}$ that dominates switching depends on the measurement-induced dephasing rate $\mu_n$ and the decay rate $\lambda_n$, both of which depend on the parameters of dissipation. In \cite{putterman_stabilizing_2022} and \cite{marthaler_switching_2006} however, $n_{\text{crit}}$ was asserted to be the manifold that lies at the top of the energy barrier between the double wells. From our semi-analytics we see that the assertion need not be the case. Using Eq.~\eqref{eq:critical-alpha-approximate} we find that for a fixed $\alpha$, the critical manifold $n_{\text{crit}}$ decreases with $\kappa$, so with a small enough $\kappa$ it is possible for a manifold significantly below the energy barrier to dominate switching. When this manifold sits between the top and bottom of the energy well, the mechanism is typically known as thermally-assisted tunneling (see for example \cite{harris_thermally_1991}). Indeed in \cite{gautier_combined_2022}, \cite{putterman_stabilizing_2022} and in our Fig.~\ref{fig:numerical-staircase}b, when $\kappa$ takes on the relatively small value of $10^{-3}K$ or $10^{-4}K$, the switching rate $\Gamma$ remains on the first plateau for $\epsilon_2/K=\alpha^2$ as large as 8 or 9.
This happens despite an estimated the number of states in the well $\alpha^2/\pi \approx 2.5 > 1$.

A particular difference between our work and \cite{gautier_combined_2022} is that, much like \cite{marthaler_switching_2006}, we compute the switching rate after the intra-well quasi-equilibrium population distribution has already been reached through thermalization and relaxation. This distribution holds for all $t$ if the initial state is already the thermalized state, or for $t \gg \kappa^{-1}$ if the initial state is some general state. However, \cite{gautier_combined_2022} computes the average switching rate for an initial state confined to the ground state in the right-well in the transient limit, i.e. $t \ll \kappa^{-1}$, when the quasi-equilibrium has yet to be reached. The latter choice has interesting consequences. In \cite{gautier_combined_2022}, the short-time average switching rate increases as the excited state population builds up, and at a fixed time $t = t_{\text{eval}}$ the average switching rate also has a staircase dependence as a function of the two-photon drive amplitude. We emphasize that this alternative staircase is not the same as the one we consider in our work or in the experiment. In particular, the condition that determines the critical $\alpha$ at which the steps occur is
\begin{align}
    \delta_n(\alpha) = \frac{\pi}{t_{\text{eval}}},
\end{align}
and is independent of parameters of the dissipation [cf. Eq.~\eqref{eq:critical-alpha}]. We provide two justifications for our choice of the switching rate to compute. First, in the experiments \cite{frattini_observation_2024, hajr_high-coherence_2024-2, ding_quantum_2025}, the switching rate was extracted from fitting the exponential decay of the inter-well population difference over a wait time that is much longer than $\kappa^{-1}$, so the measured switching rate is dominated by the dynamics after the intra-well quasi-equilibrium has been reached. Second, because the measurement protocol (known as \textit{cat-quadrature readout} \cite{grimm_stabilization_2020}) used in these experiments projects the system onto either the right- or left-well without disturbing the intra-well population distribution, the quasi-equilibrium distribution will persist over many rounds of measurement and play the role of the initial state once it has been reached (assuming constant heating and cooling rates over time).

We find that, in our case, the mechanisms responsible for exciting the population to the manifold where switching occurs are different from previous work. References \cite{gautier_combined_2022} and \cite{putterman_stabilizing_2022} only considered direct thermal heating, whereas \cite{marthaler_switching_2006} finds quantum heating to be dominant for general values of the detuning between half the two-photon drive frequency and the oscillator resonant frequency (see also experimental observation of quantum heating in \cite{vijay_invited_2009, ong_quantum_2013} in the semi-classical regime of many levels in the well). Our work shows that in our regime, with the exception of very small $n_{\text{th}}$, quantum heating from the ground and excited states can be neglected, and both direct and cascaded thermal heating are responsible for activation. For very small $n_{\text{th}}$, we have remarked in section~\ref{sec:activation-mechanism} that a new type of quantum heating that depends on the perturbed eigenstates of the nonhermitian effective Hamiltonian emerges.

Finally, we note some minor differences of this work from previous work that allow us to identify the staircase in the switching rate. Unlike \cite{puri_bias-preserving_2020}, we do not include a strong two-photon loss term $\kappa_2\mathcal{D}[\hat{a}^2]$ in the Lindbladian, which otherwise would strongly suppress the population in all excited state manifolds and hide the staircase. Another difference is that just like \cite{gautier_combined_2022}, we do not make a continuous approximation on the critical manifold $n_{\text{crit}}$ as a function of $\epsilon_2/K = \alpha^2$. This continuous approximation typically smooths out the staircase and only preserves the overall trend in the switching rate $\Gamma$ as a function of $\epsilon_2/K$, as was the case in \cite{marthaler_switching_2006, putterman_stabilizing_2022}.

We now compare our result to a previous theory on resonant magnetization tunneling in molecular magnets \cite{garanin_thermally_1997}. The mathematical technique used in this work is similar to that in \cite{garanin_thermally_1997}, where the analytically computed switching rate between two magnetization states $\ket{m = -S}$ and $\ket{m = +S}$ of a molecular spin $S$ displayed staircase-like dependence on the transverse magnetic field strength (see also experimental observation in \cite{adams_geometric-phase_2013}). When the longitudinal magnetic field is zero, the Hamiltonian of the spin is $\hat{H} = -D\hat{S}_z^2 - H_x\hat{S}_x$, whose energy levels in the zero transverse field limit are similar to the energy levels of a symmetric double well. The transverse field $H_x$ induces resonant tunneling between pairs of opposite magnetization states $m' = -m$ at rates $\Omega_{m,m'}$ that are exponentially suppressed with decreasing $m$. As $H_x$ increases, fast resonant tunneling between a pair of states $m_{\text{top}}, m_{\text{top}}'$ below the top of the energy barrier effectively reduces the depth of the double well \cite{friedman_quantum_1998}, so $H_x$ in the spin system plays the role of $\epsilon_2$ in our work, albeit in opposite directions. The difference between the switching dynamics of the spin and that found in our work is as follows. The coupling between the spin and the environment causes neither transitions that change the manifold index by more than 1 nor self-transitions within the same two-level manifold. As a consequence, the activation mechanism consists only of cascaded thermal activation, leading to a rather simple expression for the intra-well quasi-equilibrium distribution. Furthermore, the condition that determines the critical manifold $m_b$ is simply
\begin{align}
    \Omega_{m_b,m_b'} = \Gamma_{m_b,m_b'},
\end{align}
[cf. Eq.~\eqref{eq:critical-alpha}] where $\Omega_{m_b,m_b'}$ is the resonant tunnel splitting between magnetization states $m_b$ and $m_b' = -m_b$, and $\Gamma_{m_b,m_b'}$ is the decay rate from the two-level manifold consisting of $m_b$ and $m_b'$, and there is no dephasing induced from measurement of ``which-well" information.

\begin{figure}[t]
    \centering
    \includegraphics[width=\linewidth]{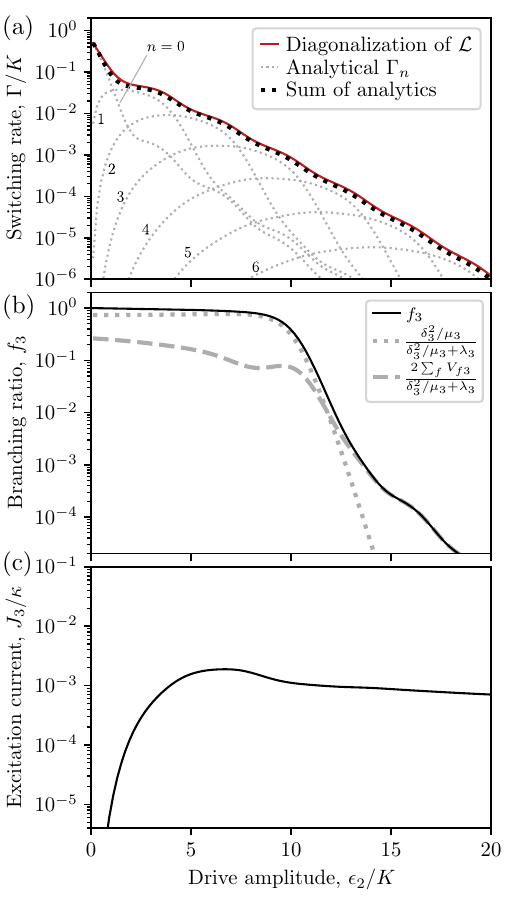}
    \caption{\textbf{The staircase when the damping rate is large.} (a) The switching rate $\Gamma$ is plotted as a function of drive amplitude $\epsilon_2$ for $\kappa/K = 1$ and $n_{\text{th}} = 0.05$. The switching rate predicted by the semi-analytical formula agrees with the numerics up to an $O(1)$ factor. The steps in the staircase are smoother than in Fig.~\ref{fig:branching-ratio}a. (b) The branching ratio $f_3$ for manifold $n = 3$ and its approximations are plotted as a function of $\epsilon_2$. The decrease of $f_3$ in the dissipation-dominated regime is slower compared to Fig.~\ref{fig:branching-ratio}b. (c) The excitation current $J_3$ into manifold $n = 3$ is plotted as a function of $\epsilon_2$. Compared to Fig.~\ref{fig:branching-ratio}c, $J_3$ decreases faster in the range of $\epsilon_2$ where the $n=3$ manifold dominates the switching rate.}
    \label{fig:large-kappa-staircase}
\end{figure}

Lastly, due to its historical relevance, let us close this section by discussing how our result relates to the famous Kramers problem \cite{kramers_brownian_1940, hanggi_reaction-rate_1990}. One phenomenon of fundamental importance is the crossover from the classical regime to the quantum regime where the effect of discrete energy levels and quantum tunneling is visible (see for example \cite{andersen_quantum_2020, yamaji_spectroscopic_2022}). Fig.~\ref{fig:large-kappa-staircase}a shows that increasing $\kappa/K$ to $1$ smooths out the staircase compared to $\kappa/K = 0.025$ (see also \cite{frattini_observation_2024}), so we can consider the $\kappa \approx K$ to be the crossover value of the damping rate. Using our semi-analytical formula for the switching rate, we identify that the smoothing is due to the following two effects: compared to Fig.~\ref{fig:branching-ratio}, the branching ratio $f_n$ decreases slower in the dissipation-dominated regime at $\kappa/K = 1$, and the excitation current $J_n$ decreases faster in the tunneling-dominated regime. These two effects lead to a less pronounced step shape for each manifold.

\begin{figure}[th]
    \centering
    \includegraphics[width=\linewidth]{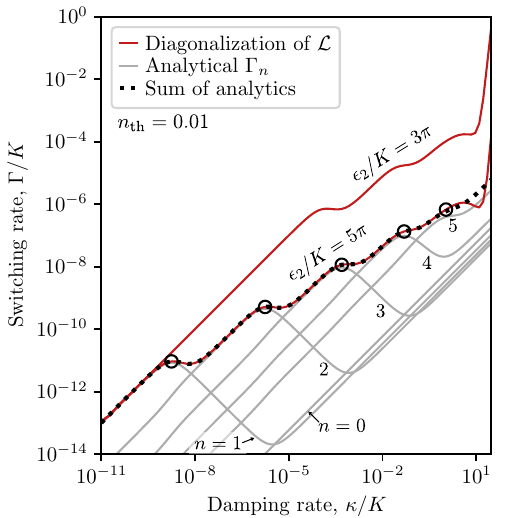}
    \caption{\textbf{Turnover behavior of the switching rate.} The numerically obtained switching rate $\Gamma$ (red solid lines) is plotted as a function of the damping rate $\kappa$ at $\epsilon_2/K = 3\pi$ or $5\pi$ and $n_{\text{th}} = 0.01$. For $\epsilon_2/K = 5\pi$, we plot for comparison the manifold contributions $\Gamma_n$ that are computed semi-analytically. The sum of $\Gamma_n$ agrees with the numerics for $\kappa/K \ll 1$. The black circles indicate the values of $\kappa$ that satisfy Eq.~\eqref{eq:critical-alpha} for each $n$. The non-monotonic behavior in the neighborhood of the black circles is what we call the ``turnover". In the small $\kappa$ limit, the switching rate is simply $\Gamma = \kappa n_{\text{th}}$.}
    \label{fig:kappa-dependence}
\end{figure}

The dependence of the switching dynamics on the damping rate $\kappa$ (see Fig.~\ref{fig:kappa-dependence}) reveals another effect that is similar to the classical effect known as Kramers turnover \cite{hanggi_reaction-rate_1990}. At fixed $\epsilon_2$, which holds fixed the size of the energy barrier, our semi-analytical switching rate as a function of the damping rate $\kappa$ exhibits many local maxima. Consider one such maximum in switching rate, dominated by the contribution from a certain manifold $n$. In the limit of small $\kappa$, the switching rate contributed by this manifold is limited by the small thermal excitation current entering the manifold, so increasing $\kappa$ increases the switching rate, much like in the underdamped regime of the Kramers problem. In the limit of large $\kappa$, the correspondingly larger measurement-induced dephasing rate $\mu_n$ and the decay rate $\lambda_n$ hinders tunneling in manifold $n$, so increasing $\kappa$ lower the switching rate, much like in the overdamped regime of Kramers problem. This turnover repeats in each excited state manifold.

Finally, a well-known effect from the research on the Kramers problem is the existence of a crossover temperature below which quantum tunneling from the ground state is important. Our results (Fig.~\ref{fig:activation-mechanism}) show that in our system, the temperature dependence of the switching rate has two crossovers instead of one. In the limit of low $n_{\text{th}}$, switching remains dominated by direct inter-well transitions within the ground state or quantum heating from the ground state of the nonhermitian effective Hamiltonian. As $n_{\text{th}}$ increases, the system crosses over and switching dominantly happens in an excited state manifold populated via direct thermal heating. Further increases in $n_{\text{th}}$ lead to another crossover, where the same excited state manifold is being populated by cascaded thermal heating rather than direct thermal heating.

\section{Conclusions}
\label{sec:conclusions}
The switching dynamics between metastable states underlies many important processes, from quantum jumps and quantum error correction to chemical and nuclear reactions. Our specific analysis of switching in the two-photon-driven Kerr nonlinear oscillator deepens previous understanding and highlights how environment-induced dephasing can freeze tunneling via the quantum Zeno effect, and how the nonhermitian effective Hamiltonian can affect the activated switching rate in a more general setting.

We summarize the specifics of our findings as follows. We have derived a semi-analytical formula [Eq.~\eqref{eq:Gamma-final-solution}] for the spontaneous switching rate in the double well of a two-photon-driven Kerr nonlinear oscillator. From this formula we find that each excited state manifold contributes one step to the staircase. In the tunneling-dominated regime, the switching rate is insensitive to the size of the tunnel splitting and varies slowly with the drive amplitude. In the dissipation-dominated regime, the tunnel splitting is small compared to the rate of dissipation, and the limiting factor for the switching rate is the effective tunneling rate $\delta_n^2/\mu_n$ due to the quantum Zeno effect freezing out the tunneling dynamics. The two regimes give rise to a flat section and a steep section in the switching rate as a function of drive amplitude. This pattern repeats when switching in a higher excited manifold becomes dominant, producing a characteristic staircase pattern. The location of the steps of the staircase pattern can be determined from the condition of Eq.~\eqref{eq:critical-alpha}, which relates the tunnel splittings, the dephasing rates, and the decay rates of the manifolds. Finally, we found a rich dependence of the phenomenology on temperature. For moderate to low values of $n_{\text{th}} \lesssim 1$, the activation mechanism that transports population from the ground state to the excited state that dominates switching is direct and cascaded thermal heating rather than quantum heating. For very low $n_{\text{th}} \lll 1$, we discover that the switching rate is dominated by a new form of quantum heating induced by the nonhermitian effective Hamiltonian. 

Our result shifts fundamental understandings of switching processes in multi-stable systems in the quantum regime and will impact the engineering of dissipation-related processes, for example, in the fluxonium qubit. The findings apply largely to parametric oscillators with multi-stability, including three- and four-legged cats \cite{xiao_diagrammatic_2023-1, puri_engineering_2017, kwon_autonomous_2022}. Other directions for future work are to apply this analysis to when the two-photon drive is detuned \cite{marthaler_quantum_2007, venkatraman_driven_2024} (see also \cite{ruiz_two-photon_2023, gravina_critical_2023}) and when a single-photon drive breaks the inversion symmetry of the double well \cite{bones_resonant-force-induced_2024, albornoz_oscillatory_2024, bones_zero-temperature_2025}.

\section{Acknowledgments}
We thank Mark Dykman for valuable scientific feedback on the manuscript. We also thank Max Schäfer, Andy Z. Ding, Benjamin L. Brock, Zhaoyou Wang, Sangil Kwon, Jahan Claes, and Aniket Maiti for fruitful discussions. This material is based upon work supported by the National Science Foundation (CAREER grant no. 2145223). Any opinions, findings, and conclusions or recommendations expressed in this material are those of the authors and do not necessarily reflect the views of the National Science Foundation. RGC acknowledges support from the Yale Quantum Institute.

\newpage
\newpage

\bibliography{202304-staircase-paper-prep-2}

\appendix

\section{Numerical solution of the spontaneous switching rate}
\label{sec:numerical-solution-technique}
In this section we describe the numerical procedure used to find the total spontaneous switching rate $\Gamma$, which is used throughout the text, and the contribution from each manifold $\Gamma_n$, which will be used in appendix \ref{sec:validity-of-approximations}.

For $\Gamma$, the main idea is to compute the switching rate $\Gamma$ using an appropriate eigenvalue of the Lindbladian $\mathcal{L}$ [Eq.~\eqref{eq:lindblad}], as was done in \cite{gravina_critical_2023}. The following describes the selection criteria for this eigenvalue intended to minimize confusion with other eigenvalues. First, notice that the Lindbladian $\mathcal{L}$ defined in Eq.~\eqref{eq:lindblad} obeys the symmetry $\hat{\Pi}\mathcal{L}(\cdot)\hat{\Pi}^\dagger = \mathcal{L}[\hat{\Pi}(\cdot)\hat{\Pi}^\dagger]$ where $\hat{\Pi} = \exp(i\pi\hat{a}^\dagger\hat{a})$ is the parity transformation operator \cite{gravina_critical_2023}. This means that the density operator can be written as the sum of two independently evolving components, $\hat{\rho}(t) = \hat{\rho}_+(t) + \hat{\rho}_-(t)$, where $\hat{\Pi}\hat{\rho}_+(t)\hat{\Pi} = \hat{\rho}_+(t)$ and $\hat{\Pi}\hat{\rho}_-(t)\hat{\Pi} = -\hat{\rho}_-(t)$. That is, $\hat{\rho}_+(t)$ is even and $\hat{\rho}_-(t)$ is odd under the parity transformation. We are only interested in the evolution of $\hat{\rho}_-(t)$ because the inter-well population difference is inverted by the parity transformation. The evolution of interest is generated by $\mathcal{L}_-$, the restriction of $\mathcal{L}$ onto the subspace of odd parity operators. Next, rather than following the time evolution, we extract the switching rate by looking at the eigenvalues of $\mathcal{L}_-$. We only consider the real eigenvalues because we expect spontaneous switching to cause an exponential decay of the inter-well population difference with no oscillations. Finally, when spontaneous switching is the slowest dynamics in the double well, we can identify $\Gamma$ as the negative of the real eigenvalue of $\mathcal{L}_-$ that is closest to 0. This concludes the numerical procedure used to find $\Gamma$.

Now, we numerically compute the contribution $\Gamma_n$ to the spontaneous switching rate from each manifold in the following way. First we define a quasi-equilibrium state $\hat{\rho}_{\text{eq}} = \hat{\rho}_{\text{ss}} + c \hat{\rho}_{\text{ex}}$, where $\hat{\rho}_{\text{ss}}$ is the steady state  of $\mathcal{L}$, $\hat{\rho}_{\text{ex}}$ is the eigenoperator of $\mathcal{L}_-$ whose eigenvalue is real and closest to 0, and $c$ is some arbitrary real number. Next we substitute the quasi-equilibrium state into the definition for $\Gamma$ Eq.~\eqref{eq:definition-of-Gamma} [same as Eq.~\eqref{eq:definition-of-Gamma-0}],
\begin{align}
    \Gamma &= -\frac{1}{\mathrm{Tr}(\hat{X}\hat{\rho}_{\text{eq}})}\frac{\mathrm{d}}{\mathrm{d}t}\mathrm{Tr}(\hat{X}\hat{\rho}_{\text{eq}}) = -\frac{\mathrm{Tr}[\hat{X}\mathcal{L}(\hat{\rho}_{\text{eq}})]}{\mathrm{Tr}(\hat{X}\hat{\rho}_{\text{eq}})}, \label{eq:numerical-gamma-intermediate}
\end{align}
where, as a reminder, $\hat{X}$ is defined by Eq.~\eqref{eq:definition-of-X} and Eq.~\eqref{eq:pauli-x}. We approximate the equilibrium state as an incoherent mixture between different manifolds as in Eq.~\eqref{eq:projection}, that is, $\hat{\rho}_{\text{eq}} \approx \sum_n\hat{I}_n\hat{\rho}_{\text{eq}}\hat{I}_n$, where $\hat{I}_n$ is a projector defined in the main text in Eq.\eqref{eq:projector}. After substituting the approximation into Eq.~\eqref{eq:numerical-gamma-intermediate}, the total spontaneous switching rate can be approximated as a sum of contributions $\Gamma_n$ from manifold $n$, defined as
\begin{align}
    \Gamma_n &= -\frac{\mathrm{Tr}[\hat{X}\mathcal{L}(\hat{I}_n\hat{\rho}_{\text{eq}}\hat{I}_n)]}{\mathrm{Tr}(\hat{X}\hat{\rho}_{\text{eq}})} \\
    &= -\frac{\mathrm{Tr}\{\hat{X}\mathcal{L}[\hat{I}_n(\hat{\rho}_{\text{ss}} + c \hat{\rho}_{\text{ex}})\hat{I}_n]\}}{\mathrm{Tr}[\hat{X}(\hat{\rho}_{\text{ss}} + c \hat{\rho}_{\text{ex}})]}
\end{align}
Because the steady-state $\hat{\rho}_{\text{ss}}$ is symmetric under the parity transformation, it does not contribute to the inter-well population difference, and $\Gamma_n$ can be equivalently expressed as
\begin{align}
    \Gamma_n = -\frac{\mathrm{Tr}[\hat{X}\mathcal{L}(\hat{I}_n\hat{\rho}_{\text{ex}}\hat{I}_n)]}{\mathrm{Tr}(\hat{X}\hat{\rho}_{\text{ex}})}. \label{eq:numerical-Gamma-n}
\end{align}
We use this expression to numerically compute the contribution of manifold $n$ to the switching rate.

\section{Validity of approximations used in the derivation}
\label{sec:validity-of-approximations}
Here we numerically demonstrate the validity of some of the approximations used in the main text. These approximations help us identify the physical processes that underlie spontaneous switching in the double-well potential of the two-photon-driven Kerr nonlinear oscillator.

\subsection{Projection onto two-level manifolds}
\begin{figure}[t]
    \centering
    \includegraphics{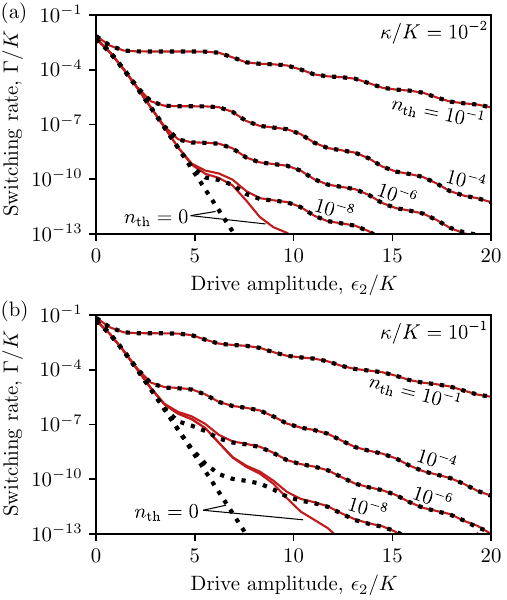}
    \caption{\textbf{Projection of the density matrix onto two-level manifolds.} The switching rate $\Gamma$ is numerically calculated using the full Lindbladian $\mathcal{L}$ (red solid lines) and the projected effective Lindbladian $\mathcal{L}_{\text{eff}}$ (black dotted lines). $\kappa/K$ is set to different values in (a) and (b).
    With the exception of the regime of large $\kappa/K$ and small $n_{\text{th}}$, the projection is a good approximation.}
    \label{fig:manifold-projection}
\end{figure}
\begin{figure}[th]
    \centering
    \includegraphics{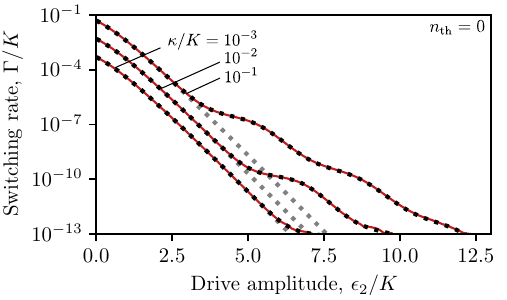}
    \caption{\textbf{The switching rate when $n_{\mathrm{th}} = 0$.} We compare the spontaneous switching rate $\Gamma$ calculated numerically using the full Lindbladian $\mathcal{L}$ (red solid lines), the semi-analytical switching rate $\Gamma$ defined in the main text (gray dotted lines), and the perturbed version $\Gamma^{\mathrm{(nh)}}$ (black dotted lines) as defined by Eq.~\eqref{eq:gamma-p} in the $n_{\text{th}} = 0$ limit. We observe good agreement between numerics and $\Gamma^{\mathrm{(nh)}}$ for various $\kappa/K$.}
    \label{fig:nonhermitian}
\end{figure}
In this subsection we discuss the effect that the projection Eq.~\eqref{eq:projection} has on the switching rate. The projection onto two-level manifolds performed in Eq.~\eqref{eq:projection} results in the effective Lindbladian $\mathcal{L}_{\text{eff}}$ defined in Eq.~\eqref{eq:effective-lindbladian}. Fig.~\ref{fig:manifold-projection} compares the switching rate $\Gamma$ obtained numerically from $\mathcal{L}_{\text{eff}}$ using the procedure described in appendix \ref{sec:numerical-solution-technique} to that from $\mathcal{L}$. The figure shows a discrepancy between the two that grows with $\kappa/K$ in the $n_{\text{th}} = 0$ limit. The reason for this discrepancy is as follows. After the projection is performed and $\mathcal{L}_{\mathrm{eff}}$ is obtained, the density operator separates into two-level manifolds, one of which is the ground state manifold spanned by coherent states $\ket{\pm\alpha}$. The coherent states are eigenstates of $\hat{a}$, so in the $n_{\text{th}} = 0$ limit where only single photon loss occurs, the dynamics is restricted to the ground state manifold, and switching can only occur in the ground state manifold at the rate $\kappa|\!\bra{\psi_0^L}\hat{a}\ket{\psi_0^R}\!|^2 \sim \kappa |\alpha|^2 e^{-4|\alpha|^2}$. This rate describes the behavior of the switching rate obtained numerically from the projected Lindbladian $\mathcal{L}_{\mathrm{eff}}$ in the limit of $n_{\text{th}} = 0$ (Fig.~\ref{fig:manifold-projection}).

However, the ground states of the nonhermitian effective Hamiltonian $\hat{H}_{\text{eff}} = \hat{H} - i\kappa\hat{a}^\dagger\hat{a} / 2$ \cite{plenio_quantum-jump_1998} are perturbed away from coherent states in the presence of non-zero $\kappa$, so the matrix elements of $\hat{a}$ from the perturbed ground state manifold to the perturbed excited state manifolds are non-zero. Such matrix elements of $\hat{a}$ between \textit{perturbed} eigenstates, which \cite{marthaler_switching_2006} did not consider, continue to activate the population to excited states despite $n_{\text{th}} = 0$, where switching can happen at a faster rate than $\kappa |\alpha|^2 e^{-4|\alpha|^2}$. This gives rise to the discrepancy seen in Fig.~\ref{fig:manifold-projection}.

Because our semi-analytical formula for the switching rate Eq.~\eqref{eq:Gamma-final-solution} is derived from $\mathcal{L}_{\text{eff}}$, it will exhibit the same discrepancy when compared to the numerics from $\mathcal{L}$ in the limit of $n_{\text{th}} = 0$. Next we show that we can capture the full numerical result by computing the transition rates between the eigenstates of the nonhermitian effective Hamiltonian.

Let the right eigenstates of the nonhermitian effective Hamiltonian $\hat{H}_{\text{eff}}$ be $\ket{\phi_n^{\pm}}$, where the sub- and superscripts have the same meaning as those of the unperturbed eigenstates $\ket{\psi_n^{\pm}}$. Let the phase and normalization of $\ket{\phi_n^{\pm}}$ be fixed by $\braket{\psi_n^{\pm}|\phi_n^{\pm}} = 1$. We additionally define a dual basis $\bra{\xi_n^{\pm}}$ that satisfies $\braket{\xi_n^\pm|\phi_{n'}^\pm} = \delta_{nn'}$ and $\braket{\xi_n^\pm|\phi_{n'}^\mp} = 0$, which will help us express the matrix elements of $\hat{a}$ and $\hat{a}^\dagger$ on the basis $\ket{\phi_n^{\pm}}$. We can define the wellstates $\ket{\phi_n^{R/L}}$ and $\bra{\xi_n^{R/L}}$ analogously as Eqs.\eqref{eq:right-wellstate} and \eqref{eq:left-wellstate}. Using the perturbed wellstates, we define the following perturbed transition rates,
\begin{align}
    V_{pq}^{\mathrm{(nh)}} &= \begin{cases}
        \kappa (1 + n_{\text{th}})\left|\bra{\xi_p^L}\hat{a}\ket{\phi_q^R}\right|^2 & q - 1 \leq p \leq q + 1 \\
        0 & \text{otherwise}
    \end{cases}, \label{eq:V_pq^p}\\
    W_{pq}^{\mathrm{(nh)}} &= \begin{cases}
        \kappa (1 + n_{\text{th}})\left|\bra{\xi_p^R}\hat{a}\ket{\phi_q^R}\right|^2 & q - 1 \leq p \leq q + 1 \\
        0 & \text{otherwise}
        \end{cases}, \label{eq:W_pq^p}
\end{align}
where the superscript $\text{(nh)}$ stands for ``nonhermitian". The restrictions on $p$ and $q$ in the definition allow cascaded transitions only between consecutive manifolds. Substituting the perturbed transition rates into our semi-analytical formula defined in Eq.~\eqref{eq:Gamma-final-solution} gives the following perturbed switching rate
\begin{align}
    \Gamma^{\mathrm{(nh)}} = \Gamma(\delta_k, W_{pq}^{\mathrm{(nh)}}, V_{pq}^{\mathrm{(nh)}}).\label{eq:gamma-p}
\end{align}
Fig.~\ref{fig:nonhermitian} shows good agreement between the semi-analytical switching rate $\Gamma^{\mathrm{(nh)}}$ and that obtained numerically from $\mathcal{L}$ when $n_{\text{th}} = 0$. Using transitions rates $V_{pq}$ and $W_{pq}$ between the unperturbed eigenstates of $\hat{H}_0$, however, leads to incorrect switching rate (grey dotted lines). This means that the pointer states of dissipation are selected by a nonhermitian Hamiltonian. This leads to an experimentally testable prediction in a regime of two-photon drive amplitudes and $n_{\text{th}}$ that has already been demonstrated. As shown in Fig.~\ref{fig:nonhermitian}, we can make the effect more pronounced by increasing the damping rate $\kappa/K$.

\subsection{Adiabatic approximation}
\label{sec:adiabatic-approximation-numerics}
\begin{figure}[t]
    \centering
    \includegraphics{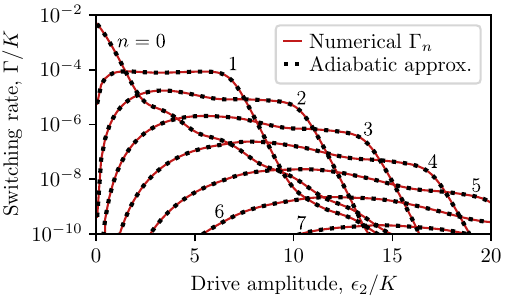}
    \caption{\textbf{Effect of the adiabatic approximation on the switching rates $\Gamma_n$.} Red solid lines show $\Gamma_n$ as functions of the drive amplitude, obtained numerically without the adiabatic approximation via Eq.~\eqref{eq:numerical-Gamma-n} following the procedure in appendix \ref{sec:numerical-solution-technique}. Black dotted lines show the numerical $\Gamma_n$ with the adiabatic approximation. The parameters used are $\kappa/K = 10^{-2}$, $n_{\text{th}} = 10^{-2}$.}
    \label{fig:adiabatic-approximation}
\end{figure}

The adiabatic approximation happens in Eqs.~\eqref{eq:implicit-y-in-terms-of-x} and \eqref{eq:x-population-recursive}. To demonstrate its validity, here we numerically examine the adiabatic approximation in isolation. Our starting point is the full set of equations of motion Eqs.~\eqref{eq:equations-of-motion-x} and ~\eqref{eq:equations-of-motion-y}, which is the result of projection onto two-level manifolds. The adiabatic approximation amounts to setting the time derivatives of all $\hat{Y}_n$ and $\hat{X}_{n>0}$ to zero, yielding
\begin{align}
    0 &= -\delta_n\braket{\hat{Y}_n} - \sum_f(W_{fn} + V_{fn})\braket{\hat{X}_n} \nonumber \\
    &\quad + \sum_i(W_{ni} - V_{ni})\braket{\hat{X}_i},\quad \text{for }n > 0 \label{eq:adiabatic-approx-appendix-x} \\
    0 &= + \delta_n\braket{\hat{X}_n} - \sum_f(W_{fn} + V_{fn})\braket{\hat{Y}_n} \nonumber \\
    &\quad + \sum_i(V_{ni} - W_{ni})\braket{\hat{Y}_i},\quad \text{for all }n \label{eq:adiabatic-approx-appendix-y}.
\end{align}
If we collect all $\braket{\hat{Y}_n}$ and $\braket{\hat{X}_{n>0}}$ into a single vector $\vec{x}$, then the above set of linear equations is of the form $0 = A\vec{x} + \braket{\hat{X}_0}\vec{b}$ with suitable identification of the matrix $A$ and $\vec{b}$. Therefore we can solve for $\braket{\hat{Y}_n}$ and $\braket{\hat{X}_{n>0}}$ in terms of $\braket{\hat{X}_0}$ by numerically  inverting the matrix $A$ and obtain $\vec{x} = -\braket{\hat{X}_0}A^{-1}\vec{b}$.
Finally, we plug the result into the expression for $\Gamma_n$ [Eq.~\eqref{eq:switching-rate-n}]. The $\braket{\hat{X}_0}$ in the numerator and the denominator will cancel, leaving us with the numerical value of $\Gamma_n$ under the adiabatic approximation. Fig.~\ref{fig:adiabatic-approximation} compares the adiabatic approximation to the numerical $\Gamma_n$. We find good agreement for all $\Gamma_n$ at the tested combination of parameters.

\subsection{Perturbative solution of \texorpdfstring{$\braket{\hat{Y}_n}$}{Yn} to zeroth order}
\begin{figure}[t]
    \centering
    \includegraphics{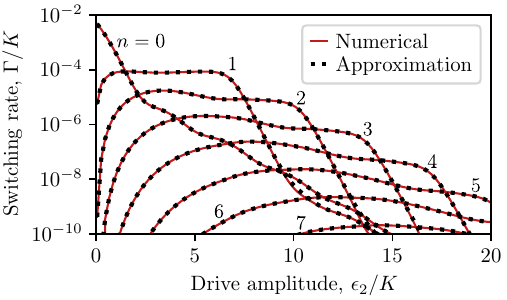}
    \caption{\textbf{Effect of the perturbative solution of $\hat{Y}_n$.} Red solid lines show $\Gamma_n$ as functions of the drive amplitude, obtained numerically via Eq.~\eqref{eq:numerical-Gamma-n} following the procedure in appendix \ref{sec:numerical-solution-technique}. Black dotted lines show the numerical $\Gamma_n$ with the adiabatic approximation and the zeroth order approximation for $\braket{\hat{Y}_n}$. The parameters used are $\kappa/K = 10^{-2}$, $n_{\text{th}} = 10^{-2}$. This plot is virtually indistinguishable from Fig.~\ref{fig:adiabatic-approximation}.}
    \label{fig:zeroth-order-y}
\end{figure}
Subsequent to the adiabatic approximation, the zeroth order solution of $\braket{\hat{Y}_n}$ in terms of $\braket{\hat{X}_n}$ were obtained in Eq.~\eqref{eq:explicit-y-in-terms-of-x-zeroth-order}. In this subsection we numerically verify the validity of this approximation using a procedure similar to that described in the previous subsection. We start from Eqs.~\eqref{eq:adiabatic-approx-appendix-x} and \eqref{eq:adiabatic-approx-appendix-y}. Next we replace Eq.~\eqref{eq:adiabatic-approx-appendix-y} by
\begin{align}
    0 &= \mu_n\braket{\hat{Y}_n} - \delta_n\braket{\hat{X}_n},
\end{align}
which encodes the zeroth order solution of Eq.~\eqref{eq:explicit-y-in-terms-of-x-zeroth-order}. After the replacement, we simply follow the procedure described in subsection \ref{sec:adiabatic-approximation-numerics} to obtain the numerical $\Gamma_n$ with both the adiabatic approximation and the zeroth order approximation for $\braket{\hat{Y}_n}$ applied. Fig.~\ref{fig:zeroth-order-y} shows good agreement between $\Gamma_n$ obtained in this way and that obtained by following section \ref{sec:numerical-solution-technique} without the adiabatic approximation or the zeroth order approximation for $\braket{\hat{Y}_n}$.

\subsection{Perturbative solution of \texorpdfstring{$\braket{\hat{X}_{n>0}}$}{Xn>0}}
\begin{figure}[t]
    \centering
    \includegraphics{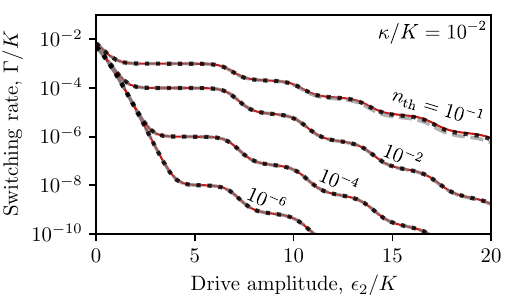}
    \caption{\textbf{Solving for $\braket{\hat{X}_{n>0}}$ perturbatively.} Red solid lines show the switching rates $\Gamma$ obtained numerically following appendix \ref{sec:numerical-solution-technique}. Grey dashed lines show the semi-analytical switching rates when downward transitions are ignored when solving for $\braket{\hat{X}_n}$. Black dotted lines show the semi-analytical switching rates when corrections due to downward transitions to first-order are added.}
    \label{fig:perturbative-x}
\end{figure}
Subsequent to the elimination of $\braket{\hat{Y}_n}$ from the equations of motion and from the expression for the switching rate $\Gamma$, the perturbative solution of $\braket{\hat{X}_{n>0}}$ was performed by ignoring downward transitions in Eq.~\eqref{eq:x-up-solution}. Then the first order correction due to downward transitions was derived in Eq.~\eqref{eq:x-down-correction}. In this subsection we numerically examine the accuracy of these two perturbative solutions. We begin from Eq.~\eqref{eq:switching-rate-x-only}, which expresses the switching rate contribution $\Gamma_n$ in terms of $\braket{\hat{X}_n}$. Next, we substitute in either the solution of Eqs.~\eqref{eq:x-up-solution} or Eqs.~\eqref{eq:x-down-correction}. Finally, we sum all $\Gamma_n$ to obtain $\Gamma$. In Fig.~\ref{fig:perturbative-x}, we compare the two switching rates obtained in this manner against the numerical switching rate obtained by following the procedure of appendix \ref{sec:numerical-solution-technique} and find good agreement. The solution that takes into account the first-order correction from downward transitions is a better approximation as expected.

\subsection{Numerical verification of the semi-analytical switching rate}
\begin{figure}[t]
    \centering
    \includegraphics{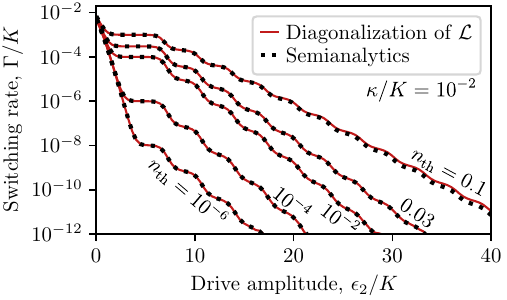}
    \caption{\textbf{Comparison between numerics and semi-analytics over a larger range of $\epsilon_2/K$}. The switching rate $\Gamma$ is plotted as a function of drive amplitude $\epsilon_2$ at various $n_{\text{th}}$. The agreement between numerics and semi-analytics is good within the plotted range.}
    \label{fig:branching-ratio-more}
\end{figure}
As a result of all preceding approximations, we obtain the semi-analytical switching rates $\Gamma_n$ in Eq.~\eqref{eq:Gamma-n-solution}, which is equivalent to its rewritten form Eq.~\eqref{eq:Gamma-n-solution-summarized}. In Fig.~\ref{fig:branching-ratio-more}, we sum the contributions from all manifolds and numerically verify the accuracy of the semi-analytics against the $\Gamma$ obtained numerically following the procedure described in appendix \ref{sec:numerical-solution-technique}. The figure shows good agreement between numerics and semi-analytics.

\section{Perturbation theory estimates of energies and matrix elements}
\label{sec:matrix-element-estimates}
This section provides perturbation theoretic estimates of the energies of $\hat{H}$ and the matrix elements of $\hat{a}$ and $\hat{a}^\dagger$, as well as quantities derived from the matrix elements in the limit of $\alpha \gg 1$ and $n_{\text{th}} \ll 1$. In the limit of $\alpha \gg 1$, the neighborhoods of the two symmetric potential minima of the double well potential are nearly harmonic. Defining a ladder operator $\hat{b}$ centered at one of the minima, i.e. $\hat{a} = \alpha + \hat{b}$, we can write the Hamiltonian in Eq.~\eqref{eq:hamiltonian} as
\begin{align}
    \hat{H} = -4\epsilon_2\left[\hat{b}^\dagger\hat{b} + \frac{1}{4\alpha^2}\hat{b}^{\dagger2}\hat{b}^2 + \frac{1}{2\alpha}(\hat{b}^{\dagger2}\hat{b} + \hat{b}^\dagger\hat{b}^2)\right] + \frac{\epsilon_2^2}{K}.
\end{align}
Treating the third term in the square bracket as a small perturbation, we find the unperturbed energies to be
\begin{align}
    E_n^{(0)} = -4K\alpha^2\left[n + \frac{1}{4\alpha^2}n(n-1)\right] + \frac{\epsilon_2^2}{K}.
\end{align}
Standard perturbation theory gives the following energy correction,
\begin{align}
    \delta E_n &= -4K\alpha^2\left(\frac{1}{2\alpha}\right)^2\left[\frac{|\!\bra{n-1}\hat{b}^\dagger\hat{b}^2\ket{n}\!|^2}{n - (n-1) + O(1/\alpha^2)}\right. \nonumber \\
    &\quad\quad - \left.\frac{|\!\bra{n+1}\hat{b}^{\dagger2}\hat{b}\ket{n}\!|^2}{(n+1) - n + O(1/\alpha^2)} + O\left(\frac{1}{\alpha^4}\right)\right] \\
    &= -K\left[|\!\bra{n-1}\hat{b}^\dagger\hat{b}^2\ket{n}\!|^2 - |\!\bra{n+1}\hat{b}^{\dagger2}\hat{b}\ket{n}\!|^2\right. \nonumber \\
    &\quad\quad + \left. O\left(\frac{1}{\alpha^2}\right) + O\left(\frac{1}{\alpha^4}\right)\right] \\
    &= Kn(3n-1) + O\left(\frac{K}{\alpha^2}\right).
\end{align}
Adding the correction to the unperturbed energy gives
\begin{align}
    E_n = \frac{\epsilon_2^2}{K} - 4K\alpha^2 n + 2Kn^2 + O\left(\frac{K}{\alpha^2}\right). \label{eq:energy-estimate}
\end{align}

The perturbation theory calculation also produces perturbed eigenstates. These eigenstates provide the following crude asymptotic bound on the intra-well norm square matrix element of $\hat{a}$,
\begin{align}
    \left|\bra{\psi_n^R}\hat{a}\ket{\psi_m^R}\right|^2 &= O\!\left(\alpha^{2-2|m-n|}\right).
\end{align}
The bound can be refined by performing Schrieffer-Wolff perturbation theory up to the eleventh order with respect to the small parameter $\alpha^{-1}$ (via the python computer algebra package SymPy \cite{meurer_sympy_2017}). The result shows that, for all $m - 12 \leq n \leq m + 4$, the norm square matrix elements of $\hat{a}$ are bounded by the following estimates,
\begin{align}
    \left|\bra{\psi_n^R}\hat{a}\ket{\psi_m^R}\right|^2 &= \begin{cases}O\left(\alpha^{2-2|m-n|}\right) & n \leq m \\ O\left(\alpha^{2-6|m-n|}\right) & n > m\end{cases}, \label{eq:matrix-element-a-estimate}
\end{align}
and in particular, $\left|\bra{\psi_n^R}\hat{a}\ket{\psi_n^R}\right|^2 \sim \alpha^2$. We will assume that this estimate continues to apply for those pairs of $m$ and $n$ that we have not verified. The inter-well norm square matrix elements $|\bra{\psi_m^L}\hat{a}\ket{\psi_n^R}|^2$ between states in different wells are exponentially suppressed in the limit of $\alpha \gg 1$ because low photon number Fock states that are displaced to the right well have exponentially small overlap with those displaced to the left well.

Using the estimates of the matrix elements of $\hat{a}$, we can assemble the following estimates for $W_{mn}$ and $V_{mn}$ in the limit of $n_{\text{th}} \ll 1$ and $\alpha \gg 1$:
\begin{align}
    W_{nn} &\sim \kappa|\bra{\psi_n^R}\hat{a}\ket{\psi_n^R}|^2 \sim \kappa\alpha^2, \label{eq:matrix-element-Wnn-estimate} \\
    W_{fn} &= \begin{cases}O(\kappa \alpha^{2-2|f-n|}) & f < n \\ O(\kappa \alpha^{2-6|f-n|})  + O(\kappa n_{\text{th}}\alpha^{2-2|f-n|})& f > n  \end{cases} \label{eq:matrix-element-Wfn-estimate},
\end{align}
and $V_{fn}$ are exponentially small as functions of $\alpha^2$. Based on the estimates of $W_{mn}$ and $V_{mn}$, the dephasing rate of inter-well coherence [Eq.~\eqref{eq:dephasing-rate}] scales as
\begin{align}
    \mu_n \sim 2W_{nn} \sim 2\kappa\alpha^2
\end{align}
in the limit of $n_{\text{th}} \ll 1$ and $\alpha \gg 1$. The scaling of the decay rate $\lambda_n$ [Eq.~\eqref{eq:decay-rate}] in the same limit can be approximated using displaced Fock states,
\begin{align}
    \lambda_n &\sim W_{n-1,n} \sim \kappa|\!\bra{\psi_{n-1}^R}\hat{a}\ket{\psi_n^R}\!|^2 \\
    &\sim \kappa|\!\bra{n-1}(\alpha + \hat{b})\ket{n}\!|^2 \\
    &\sim n\kappa.
\end{align}

\section{WKB approximation of the tunnel splitting}
\label{sec:wkb}
The tunnel splittings $\delta_n$ of the Hamiltonian in Eq.~\eqref{eq:hamiltonian} has been estimated via overlap of shifted Fock states \cite{puri_stabilized_2019}, empirical fits \cite{putterman_stabilizing_2022}, and the WKB approximation \cite{marthaler_switching_2006, marthaler_quantum_2007, venkatraman_driven_2024}. However, we are additionally interested in a regime where the tunnel splitting $\delta_n$ is still appreciable compared to the energy gap $E_{\text{gap}, n}$, which happens near the barrier top. The usual formula, though it works for excited states, fails for states near the top of the energy barrier. In this section we use the WKB approximation to find the tunnel splittings of these states, taking particular care to use a good approximation of the wavefunction in the region near the top of the central barrier. We start from the Hamiltonian of Eq.~\eqref{eq:hamiltonian} but defined with an additional detuning term $\Delta\hat{a}^\dagger\hat{a}$ that we will later set back to zero,
\begin{align}
    \hat{H} = \Delta\hat{a}^\dagger\hat{a} - K\hat{a}^{\dagger2}\hat{a}^2 - \epsilon_2(\hat{a}^{\dagger2} + \hat{a}^2),
\end{align}
where $[\hat{a}, \hat{a}^\dagger] = 1$. Using the approach developed in \cite{marthaler_switching_2006, marthaler_quantum_2007, peano_sharp_2012}, we nondimensionalize the problem by defining a rescaled Hamiltonian $\hat{g}$ and rescaled phase space coordinates $\hat{Q}$ and $\hat{P}$ as follows,
\begin{align}
    \hat{H} &= -\frac{4\epsilon_2^2}{K}\hat{g} - \frac{2\Delta + 3K}{4}, \\
    \hat{a} &= \frac{\hat{Q} + i\hat{P}}{\sqrt{2\lambda}},\quad \text{with } [\hat{Q}, \hat{P}] = i\lambda.
\end{align}
The scale conversion between eigenvalues of $\hat{H}$ and $\hat{g}$ is 
\begin{align}
    \delta\!H = -(4\epsilon_2^2/K)\,\delta\!g, \label{eq:g-H-conversion}
\end{align}
and the reference energy when $g = 0$ is $E_{\text{ref}} = -(2\Delta + 3K)/4$. Here the conversion factor $4\epsilon_2^2/K$ is four times the height of the energy barrier (or four times the depth of the double well), and the free parameter $\lambda$ defines the scale of the new phase space coordinates $\hat{Q}$ and $\hat{P}$ as well as function as an effective Planck constant. Choosing $\lambda = K/(2\epsilon_2)$ absorbs most of the $\epsilon_2$ dependence of the Hamiltonian into the effective Planck constant $\lambda$ and results in the following expression for the rescaled Hamiltonian $\hat{g}$ that was used in \cite{marthaler_switching_2006, marthaler_quantum_2007, peano_sharp_2012},
\begin{align}
    \hat{g} = \frac{1}{4}(\hat{P}^2 + \hat{Q}^2)^2 + \frac{1}{2}(1 - \mu)\hat{P}^2 - \frac{1}{2}(1 + \mu)\hat{Q}^2, \label{eq:rescaled-hamiltonian}
\end{align}
where $\mu = (\Delta + 2K)/(2\epsilon_2)$ is a rescaled detuning parameter. In the limit of zero detuning and large two-photon drive amplitude, we have $\Delta = 0$, $\epsilon_2 = K\alpha^2$, and $\alpha^2 \gg 1$. Consequently, the effective Plank constant scales as
\begin{align}
    \lambda = 1 / (2\alpha^2) \ll 1,
\end{align}
justifying the WKB approximation, and the rescaled detuning parameter scales as
\begin{align}
    \mu = 1 / \alpha^2 \ll 1,
\end{align}
only slightly perturbing the functional form of $\hat{g}$.

\subsection{Characterizing classical orbits in phase space}
We characterize the classical orbits in phase space in preparation for a WKB calculation. Removing hats from Eq.~\eqref{eq:rescaled-hamiltonian} produces the following classical version of the rescaled Hamiltonian \cite{marthaler_quantum_2007},
\begin{align}
    g = \frac{1}{4}(P^2 + Q^2)^2 + \frac{1}{2}(1 - \mu)P^2 - \frac{1}{2}(1 + \mu)Q^2. \label{eq:classical-hamiltonian}
\end{align}
The energy landscape in phase space has a double well structure for $|\mu| < 1$. In particular, for each energy $g$ within the range $-(1 + \mu)^2 / 4 < g < 0$, there are two orbits bound to the double well, one in the $Q > 0$ half-plane and one in the $Q < 0$ half-plane. For $g > 0$, the two orbits merge into a single unbound orbit.

The WKB solution expresses the wavefunction in terms of the momentum and the velocity along the orbit, and uses the action enclosed by the orbit to determine the quantization condition. We define these quantities here. The momentum along the orbit is the function $P(Q, g; \mu)$ defined as the positive real solution of Eq.~\eqref{eq:classical-hamiltonian}. The velocity along the orbit is obtained from Hamilton's equations via $\dot{Q}(Q, g; \mu) = \left.\partial g/\partial P\right|_{P = P(Q, g; \mu)}$. Finally the action enclosed by the orbit (restricted to the $Q > 0$ half-plane) is defined as $S(g, \mu) = \oint_{Q > 0} P\,\mathrm{d}Q = 2\int_{Q_{\min}}^{Q_{\max}}P(Q, g; \mu)\,\mathrm{d}Q$, where $Q_{\min} = Q_{\min}(g, \mu)$ and $Q_{\max} = Q_{\max}(g, \mu)$ denote the minimum and maximum extent of the classical orbit in the $Q > 0$ half-plane (thus setting $Q_{\min} = 0$ for $g > 0$).

To slightly simplify the integrand that defines the enclosed action, we apply the following canonical transformation $Q = \sqrt{2J}\cos(\theta), P = -\sqrt{2J}\sin(\theta)$ to Eq.~\eqref{eq:classical-hamiltonian} and arrive at the following equation for the classical orbit in terms of the new coordinates $(\theta, J)$, $g = J^2 - J(\cos2\theta + \mu)$.
Solving for $J$ in terms of $\theta$ gives the action in terms of the $\theta, J$ variables as follows,
\begin{align}
    &\quad S(g, \mu) = \oint_{|\theta| < \pi/2} \frac{\cos2\theta + \mu \pm \sqrt{(\cos2\theta + \mu)^2 + 4g}}{2}\mathrm{d}\theta \\
    &= \begin{cases} 2 \int_0^{\frac{1}{2}\arccos(\sqrt{-4g} - \mu)} \sqrt{(\cos2\theta + \mu)^2 + 4g}\ \mathrm{d}\theta & g \leq 0 \\
    2 \int_0^{\pi/2} \frac{\cos2\theta + \mu + \sqrt{(\cos2\theta + \mu)^2 + 4g}}{2}\ \mathrm{d}\theta & g > 0
    \end{cases}. \label{eq:S-action-explicit}
\end{align}

We are interested in the enclosed action for orbits near the barrier top $g \to 0$. A first approximation for the enclosed action is the value at $g = 0$, corresponding to the phase space area occupied by each well, $S(0, \mu) = \pi\mu/2 + \sqrt{1 - \mu^2} + \mu\arcsin(\mu)$. Away from $g = 0$, it is useful to find an asymptotic expression of $S(g, \mu)$ in the limit of $g \to 0$. After some calculation we find that $S(g, \mu)$ has the following asymptotics in the $g \to 0$ limit,
\begin{align}
    S(g, \mu) &= S(0, \mu) + \frac{1}{\sqrt{1 - \mu^2}}\left\{1 - \ln\left[\frac{|g|}{4(1 - \mu^2)^2}\right]\right\}g \nonumber \\
    &\quad + \frac{3 + 14\mu^2 + (2 + 4\mu^2)\ln\left[\frac{|g|}{4(1 - \mu^2)^2}\right]}{4(1 - \mu^2)^{5/2}}g^2 \nonumber \\
    &\quad + O(|g|^3\ln|g|), \label{eq:S-action-asymptotics}
\end{align}
where the use of the absolute value function allows for $g > 0$ as well as $g < 0$.

For problems involving tunneling under the barrier, we must additionally consider imaginary solutions $P(Q, g; \mu)$, which characterizes the exponential tail of the WKB wavefunction. Therefore we let $P(Q, g; \mu) = i\Pi(Q, g; \mu)$. Then Eq.~\eqref{eq:classical-hamiltonian} gives the following implicit equation for the imaginary solutions,
\begin{align}
    g = \frac{1}{4}(Q^2 - \Pi^2)^2 - \frac{1}{2}(1 - \mu)\Pi^2 - \frac{1}{2}(1 + \mu)Q^2. \label{eq:classical-hamiltonian-imaginary-p}
\end{align}
The WKB tunnel splitting is related to the tunnel action integral, which is defined for energies $-(1 - \mu)^2 / 4 < g \leq 0$ (excluding the ground state) in terms of the integral of $\Pi(Q, g; \mu)$ in the classically forbidden region, $I(g, \mu) = \int_{-Q_{\min}}^{Q_{\min}}\Pi(Q, g; \mu)\,\mathrm{d}Q$. The integrand can again be simplified by performing the canonical transformation $Q = \sqrt{2F}\cos(\theta), \Pi = -\sqrt{2F}\sin(\theta)$ on Eq.~\eqref{eq:classical-hamiltonian-imaginary-p}, which produces the equation $g = F^2\cos^22\theta - F(1 + \mu\cos2\theta)$. Solving for $F$ in terms of $\theta$ gives the tunnel action as
{\small
\begin{align}
    &I(g, \mu) = \int_0^{\pi}\frac{1 + \mu\cos2\theta - \sqrt{(1 + \mu\cos2\theta)^2 + 4g\cos^22\theta}}{2\cos^22\theta}\,\mathrm{d}\theta. \label{eq:I-action-explicit}
\end{align}
}
We remark that the above integral remains defined for $g > 0$ as well as $g \leq 0$ and is analytic at $g = 0$. With the help of Mathematica \cite{wolfram_research_inc_mathematica_2022}, the asymptotics of $I(g, \mu)$ near $g \to 0$ is found to be
\begin{align}
    I(g, \mu) &= -\frac{\pi}{\sqrt{1 - \mu^2}}g + \frac{\pi(1 + 2\mu^2)}{2(1 - \mu^2)^{5/2}}g^2 \nonumber \\
    &\quad + \frac{\pi[-3 - 8\mu^2(3 + \mu^2)]}{4(1 - \mu^2)^{9/2}}g^3 + O(g^4). \label{eq:I-action-asymptotics}
\end{align}

We also need the asymptotics of $I$ near $g \to -\frac{1}{4}(1 - \mu)^2$. This is because for the $n$th excited state in the well, we have that $E_n = \frac{\epsilon_2^2}{K} - 4n\epsilon_2 - 2Kn^2 + O(\mu K)$, so
\begin{align}
    g_n &= -\frac{K}{4\epsilon_2^2}\left(E_n + \frac{3K}{4}\right) \\
    &= -\frac{1}{4} + n\frac{1}{\alpha^2} - \left(\frac{3}{16} + \frac{n^2}{2}\right)\frac{1}{\alpha^4} + O\left(\frac{1}{\alpha^6}\right) \\
    &= -\frac{1}{4}(1 - \mu)^2 + \left(n - \frac{1}{2}\right)\mu + \left(\frac{1}{16} - \frac{n^2}{2}\right)\mu^2 \nonumber \\
    &\quad + O(\mu^3).
\end{align}
For convenience in taking the limit, we define a transformed energy $\tilde{g}$ satisfying $g = -\frac{1}{4}(1 - \mu)^2 + \tilde{g}\mu$. The the expression for $\tilde{g}$ is
\begin{align}
    \tilde{g} = n - \frac{1}{2} + \left(\frac{1}{16} - \frac{n^2}{2}\right)\mu + O(\mu^2).
\end{align}
In the limit $\mu \to 0$, the asymptotics for $I(g, \mu)$ (as a function of $\mu$) can be found to be the following after much calculation,
\begin{align}
    I = 1 - \delta I
\end{align}
where 
\begin{align}
    \delta I &= -\frac{1}{2}\left\{\mu(1 + \tilde{g})\ln[\mu(1 + \tilde{g})] + \mu\tilde{g}\ln(\mu\tilde{g})\right\} \nonumber \\
    &\quad + \left[\frac{1}{2} + \ln(2)\right](1 + 2\tilde{g})\mu \nonumber \\
    &\quad - \frac{1}{4}\mu^2\tilde{g}(1 + \tilde{g})  \ln\left[\frac{e^3}{16}\mu^2\tilde{g}(1 + \tilde{g})\right] \nonumber \\
    &\quad - \frac{1}{8}\mu^2+ O(\mu^3\ln\mu). \label{eq:I-action-asymptotics-low-energy-limit}
\end{align}
Plugging in the asymptotics for $\tilde{g}$ in terms of $n$ gives
\begin{align}
    \delta I &= \frac{1}{2}\mu(- 2n\ln\mu + \ln A_n) \nonumber \\
    &\quad + \frac{1}{2}\mu\left[\frac{1}{8}\mu\ln\mu + \mu B_n\right]  + O(\mu^3\ln\mu).
    \label{eq:I-action-asymptotics-low-energy-limit-2}
\end{align}
where
\begin{align}
    A_n &= \frac{(4e)^{2n}}{\left(n - \frac{1}{2}\right)^{n - \frac{1}{2}}\left(n + \frac{1}{2}\right)^{n + \frac{1}{2}}} \label{eq:definition-A-n} \\
    B_n &= \frac{1}{8} - \frac{3}{2} n^2 + \frac{1}{8}\ln\left[\frac{1}{4}\sqrt{n^2 - \frac{1}{4}}\right]. \label{eq:definition-B-n}
\end{align}

\subsection{Solving for approximate wavefunctions}
In the classically allowed region within $Q > 0$, standard WKB approximation gives the position basis wavefunction $\psi(Q)$ in terms of the classical momentum $P(Q, g)$ and the classical velocity $\dot{Q}(Q, g)$ as follows,
\begin{align}
    \psi(Q) = \frac{C}{\sqrt{\dot{Q}(Q, g; \mu)}}\cos\left(\frac{1}{\lambda}\int_{Q_{\max}}^{Q}P(Q', g; \mu)dQ' + \frac{\pi}{4}\right). \label{eq:WKB-wavefunction}
\end{align}
This wavefunction matches the boundary condition at $Q = Q_{\max}$. Now we obtain a wavefunction that matches the boundary condition at $Q = Q_{\min} \approx 0$. To do so, we use a quadratic approximation for the Hamiltonian $\hat{g}$,
\begin{align}
    \hat{g} \approx \frac{1}{2}(1 - \mu)\hat{P}^2 - \frac{1}{2}(1 + \mu)\hat{Q}^2. \label{eq:quadratic-bump}
\end{align}
This approximation captures the energy saddle point in phase space. It is justified because for a state whose energy is close to the barrier top, both the position and the momentum are small in the region near $Q = 0$. From Eq.~\eqref{eq:quadratic-bump}, we can identify an effective mass $(1 - \mu)^{-1}$ and an effective spring constant $(1 + \mu)$. Together they give rise to the effective frequency $\omega_b$ and an energy scale $g_b$,
\begin{align}
    \omega_b &= \sqrt{\frac{1 + \mu}{(1 - \mu)^{-1}}} = \sqrt{1 - \mu^2}, \\
    g_b &= \lambda\omega_b = \lambda\sqrt{1 - \mu^2}.
\end{align}
Since $g_b$ sets a natural energy scale for motion near the barrier top, we define a new energy variable $a$ by taking the ratio of $g$ and $g_b$,
\begin{align}
    a = -\frac{g}{g_b} = -\frac{g}{\lambda\sqrt{1 - \mu^2}}. \label{eq:definition-of-a}
\end{align}

The time-independent Schrödinger equation corresponding to Eq.~\eqref{eq:quadratic-bump} is 
\begin{align}
    -\lambda^2\frac{1 - \mu}{2}\,\frac{\partial^2\psi}{\partial Q^2} - \frac{1 + \mu}{2}\,Q^2\,\psi = g\,\psi,
\end{align}
where $\psi(Q)$ is the position basis wavefunction. Changing into the new energy variable $a$ and a new position variable as $Q = \left\{\lambda^2(1 - \mu)/[4(1 + \mu)]\right\}^{1/4} \xi$ reduces the differential equation for $\psi$ to the following standard form, $\partial^2\psi/\partial \xi^2 + \left(\xi^2/4 - a\right)\psi = 0$, for which the general solutions are $\psi = c_1W(a, \xi) + c_2W(a, -\xi)$, where $W$ is a parabolic cylinder function (see section 12.14 of the Digital Library of Mathematical Functions, DLMF \cite{olver_nist_2024}.) The wavefunction can be made to have even or odd parity,
\begin{align}
    \psi_{\pm} \propto W(a, \xi) \pm W(a, -\xi). \label{eq:parabolic-cylinder-function}
\end{align}

\subsection{Matching the two wavefunctions to produce a quantization condition}
To produce a quantization condition, we need to match the parabolic cylinder function solution in Eq.~\eqref{eq:parabolic-cylinder-function} to the WKB wavefunction in Eq.~\eqref{eq:WKB-wavefunction}. We first apply the following asymptotics of the parabolic cylinder function $W$ for large $\xi$ (see DLMF Eq. 12.14.17 and 12.14.18),
\begin{align}
    W(a, \xi) \sim \sqrt{\frac{2k}{\xi}}\cos\omega,\quad W(a, -\xi) \sim \sqrt{\frac{2}{k\xi}}\sin\omega,
\end{align}
where $\omega = \frac{1}{4}\xi^2 - a\ln\xi + \frac{1}{4}\pi + \frac{1}{2}\arg\Gamma\left(\frac{1}{2} + ia\right) + O(\xi^{-2})$, $\Gamma$ is the gamma function, and $k^{-1} = \sqrt{1 + e^{2\pi a}} + e^{\pi a}$. Substituting the asymptotics into the parabolic cylinder function solution Eq.~\eqref{eq:parabolic-cylinder-function} gives the following wavefunction in the large $\xi$ limit,
\begin{align}
    \psi_{\pm} &= \frac{C}{\sqrt{\xi}}\cos\left[\frac{1}{4}\xi^2 - a\ln\xi + \frac{1}{4}\pi\right. + \frac{1}{2}\arg\Gamma\left(\frac{1}{2} + ia\right) \nonumber \\
    &\quad\quad \left.\mp \arctan(\sqrt{1 + e^{2\pi a}} + e^{\pi a}) + O(\xi^{-2})\right]. \label{eq:parabolic-cylinder-function-asymptotics}
\end{align}

Next, we rewrite the amplitude and the phase of the wavefunction above in terms of the classical action and velocity. Using the classical Hamiltonian obtained by removing the hats from Eq.~\eqref{eq:quadratic-bump}, $g \approx \frac{1}{2}(1 - \mu)P^2 - \frac{1}{2}(1 + \mu)Q^2$, we can find the following expressions for the momentum and velocity, valid near $Q = 0$,
\begin{align}
    P(Q, g; \mu) &= \lambda\left[\frac{4(1 + \mu)}{\lambda^2(1 - \mu)}\right]^{1/4}\sqrt{\frac{1}{4}\xi^2 - a}, \\
    \dot{Q}(Q, g; \mu) &\propto \sqrt{\frac{1}{4}\xi^2 - a}, \label{eq:velocity-expression}
\end{align}
where we have change into the $(\xi, a)$ variables. Substituting the momentum into the following (normalized) action integral, integrating, and finding the asymptotics at $\xi \to \infty$ produces
\begin{align}
    &\frac{1}{\lambda}\int_{Q_{\min}}^Q P(Q', g; \mu)\,\mathrm{d}Q' = \int_{\xi_{\min}}^{\xi}\sqrt{\frac{1}{4}(\xi')^2 - a}\,\mathrm{d}\xi' \\
    &= \frac{1}{4}\xi^2 - a\ln\xi + \frac{1}{2}a\ln |a| - \frac{1}{2}a + O(\xi^{-2}),
\end{align}
where $Q_{\min}$ and $\xi_{\min}$ denote the minimum position reached by the classical orbit if we restrict to the $Q > 0$ half-plane ($Q_{\min} = \xi_{\min} = 0$ for $g > 0$). Taking the $\xi \to \infty$ asymptotics of the velocity Eq.~\eqref{eq:velocity-expression} gives $\dot{Q} \propto \sqrt{\frac{1}{4}\xi^2 - a} \sim \frac{1}{2}\xi$. Using the asymptotics of the action and the velocity above, we can rewrite the solution for the wavefunction in Eq.~\eqref{eq:parabolic-cylinder-function-asymptotics} as follows,
\begin{align}
    \psi_{\pm} &= \frac{C'}{\sqrt{\dot{Q}(Q, g; \mu)}}\cos\left[\frac{1}{\lambda}\int_{Q_{\min}}^Q P(Q', g; \mu)\ \mathrm{d}Q'\right. \nonumber \\
    &\quad\quad -\frac{1}{2}a\ln|a| + \frac{1}{2}a + \frac{1}{4}\pi + \frac{1}{2}\arg\Gamma\left(\frac{1}{2} + ia\right) \nonumber \\
    &\quad\quad \left.\mp \arctan(\sqrt{1 + e^{2\pi a}} + e^{\pi a}) + O(\xi^{-2})\right]. \label{eq:solution-left-boundary}
\end{align}
This is a solution of the wavefunction that satisfies the boundary condition near $Q = 0$.

We now have two approximate wavefunctions Eqs.~\eqref{eq:WKB-wavefunction} and \eqref{eq:solution-left-boundary}, each satisfying the boundary condition at one position extremum of the classical orbit. Now we match the two to obtain a quantization condition for the energy $g$. Since the arguments of the cosine function must differ by an integer multiple of $\pi$ for the solutions to match, we obtain the following quantization condition on the action integral (omitting higher order terms),
\begin{align}
    &\ \frac{1}{\lambda}\int_{Q_{\min}}^{Q_{\max}} P(Q', g; \mu)\ \mathrm{d}Q' \nonumber \\
    &= \pi n + \frac{1}{2}a\ln|a| - \frac{1}{2}a - \frac{1}{2}\arg\Gamma\left(\frac{1}{2} + ia\right) \nonumber \\
    &\quad\quad \pm \arctan(\sqrt{1 + e^{2\pi a}} + e^{\pi a}). \label{eq:quantization-condition-proto-1}
\end{align}
This condition can be rewritten as
\begin{align}
    \frac{1}{\lambda}S(g, \mu) &= 2\pi n + a\ln|a| - a - \mathrm{Im}\left[\ln\Gamma\left(\frac{1}{2} + ia\right)\right] \nonumber \\
    &\quad \pm \left(\pi - \arctan(e^{-\pi a})\right).
\end{align}
Adding the term $(\pi \mp \pi)$ to the right-hand side allows us to shift the index $n$ by 1 for all odd parity states without changing the physical content of the quantization condition,
\begin{align}
    \frac{1}{\lambda}S(g, \mu) &= 2\pi \left(n + \frac{1}{2}\right) + a\ln|a| - a \nonumber \\
    &\quad - \mathrm{Im}\left[\ln\Gamma\left(\frac{1}{2} + ia\right)\right] \mp \arctan(e^{-\pi a}), \label{eq:quantization-condition-proto-2}
\end{align}
where the upper sign is taken for even states, and the lower sign is taken for odd states.
\begin{figure}[t]
    \centering
    \includegraphics[width=\linewidth]{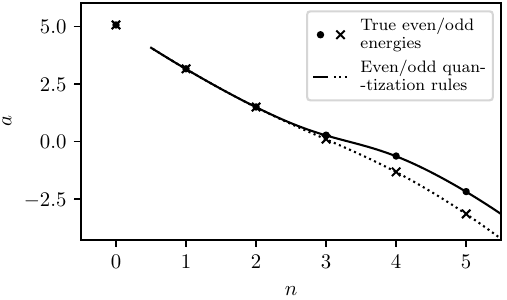}
    \caption{\textbf{Verification of quantization rule.} The (rescaled) energy $a$ of an eigenstate, defined in Eq.~\eqref{eq:definition-of-a}, is plotted as a function of the principal quantum number $n$ in the even (odd) parity sector. The dots (crosses) are the energies numerically obtained from the diagonalization of the Hamiltonian. The solid (dotted) lines plot the implicit relationship between $a$ and $n$ dictated by the quantization rule Eq.~\eqref{eq:quantization-condition-final} for the even (odd) parity sector. We restrict the lines to the energy range $g > - (1 - \mu)^2/ 4$ where the term $I(g, \mu)$ in Eq.~\eqref{eq:quantization-condition-final} is real-valued. The parameters used are $\epsilon_2/K = \alpha^2 = 10$, $\mu = 1 / \alpha^2 = 1/10$, $\lambda = 1 / (2\alpha^2) = 1/20$.}
    \label{fig:quantization-rule}
\end{figure}
In the limit of $a \gg 1$ (states deep in the well), the quantization condition reduces to the standard one,
\begin{align}
    \frac{1}{\lambda}S(g, \mu) &= 2\pi \left(n + \frac{1}{2}\right) \mp e^{-\pi a}, \label{eq:quantization-condition-limit}
\end{align}
where $n = 0$ denotes the ground state in each parity sector. The simple form of the tunnel exponent $\pi a$ is a side effect of using a quadratic approximation for the energy barrier. We can fix this by simply going back to Eq.~\eqref{eq:quantization-condition-proto-2} and replacing the tunnel exponent by the actual tunnel action integral, $\pi a \to \frac{1}{\lambda}I(g, \mu)$. Therefore, the final quantization condition for even and odd parity states are
\begin{align}
    &\ \frac{1}{\lambda}S(g, \mu) \nonumber \\
    &= 2\pi \left(n + \frac{1}{2}\right) + a\ln|a| - a - \mathrm{Im}\left[\ln\Gamma\left(\frac{1}{2} + ia\right)\right] \nonumber \\
    &\quad \mp \arctan\left\{\exp\left[-\frac{1}{\lambda}I(g, \mu)\right]\right\}, \label{eq:quantization-condition-final}
\end{align}
where $S(g, \mu)$ is defined in Eq.~\eqref{eq:S-action-explicit}, $I(g, \mu)$ is defined in Eq.~\eqref{eq:I-action-explicit}, and $a$ is defined in Eq.~\eqref{eq:definition-of-a}. This quantization condition allows a smooth interpolation between the quantization condition for states inside and outside the double well and is verified numerically in Fig.~\ref{fig:quantization-rule}. For states deep in the well and away from oscillatory behavior in the classically forbidden region, the quantization condition reduces to the following,
\begin{align}
    &\ \frac{1}{\lambda}S(g, \mu) = 2\pi \left(n + \frac{1}{2}\right) \mp \exp\left[-\frac{1}{\lambda}I(g, \mu)\right].
\end{align}
This in turn produces the following expression of the tunnel splitting $\delta$ in terms of the energy gap between manifolds,
\begin{align}
    \frac{\delta}{E_{\text{gap}}} = \frac{1}{\pi}e^{-\frac{I}{\lambda}}. \label{eq:tunnel-splitting-expression-small-delta}
\end{align}

\subsection{Solving the quantization condition for the tunnel splittings}
\begin{figure}[ht]
    \centering
    \includegraphics[width=\linewidth]{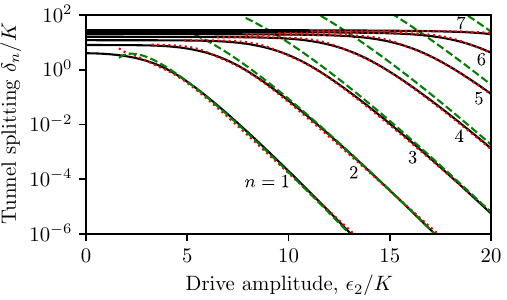}
    \caption{\textbf{WKB approximate tunnel splittings.} The tunnel splittings $\delta_n$ are plotted as functions of the two-photon drive amplitude. Black solid lines are obtained from numerical diagonalization of the Hamiltonian. Red dotted lines are analytical approximations obtained from Eq.~\eqref{eq:tunnel-splitting-analytic-highly-excited-states}. Green dashed lines are analytical approximations obtained from Eq.~\eqref{eq:tunnel-splitting-analytic-lowly-excited-states}}
    \label{fig:wkb-splitting}
\end{figure}
Suppose we are interested in the tunnel splittings of states near the top of the barrier $g = 0$ ($a = 0$). Here it is more convenient to work with the new energy variable $a$ rather than $g$. The first step is to approximately solve for $a$ in terms of the principal quantum number $n$ using the quantization condition Eq.~\eqref{eq:quantization-condition-final}. We plug the asymptotics of $S(g, \mu)$ [Eq.~\eqref{eq:S-action-asymptotics}] and $I(g, \mu)$ [Eq.~\eqref{eq:I-action-asymptotics}] into the quantization condition Eq.~\eqref{eq:quantization-condition-final}, rewrite the condition in terms of the rescaled energy $a$ [Eq.~\eqref{eq:definition-of-a}], and keep terms of order $a$. The resulting quantization condition after some manipulation is
\begin{align}
    a\ln\left[\frac{\lambda}{4(1 - \mu^2)^{3/2}}\right] = 2\pi \left(n - n_0\right) + h_{\pm}(a), \label{eq:approximate-quantization-condition}
\end{align}
where $n_0 = S(0, \mu)/(2\pi\lambda) - 1/2$ and
\begin{align}
    h_{\pm}(a) &= -\mathrm{Im}\left[\ln\Gamma\left(\frac{1}{2} + ia\right)\right] \mp \arctan\left[e^{-\pi a\,q(a)}\right], \\
    q(a) &= 1 + \frac{1 + 2\mu^2}{2(1 - \mu^2)^{3/2}}\lambda a \nonumber \\
    &\quad+ \frac{3 + 8\mu^2(3 + \mu^2)}{4(1 - \mu^2)^{6/2}}\lambda^2a^2 + O(a^3).
\end{align}
We may use this condition to iteratively solve for the rescaled energy $a_-(n, \mu)$ of odd parity states. The zeroth order approximation is obtained by approximating $h_-(a) \approx h_-(a = 0) = \frac{\pi}{4}$, giving $a_-^{(0)} = [2\pi \left(n - n_0\right) + \frac{\pi}{4}]/\ln\left\{\lambda/[4(1 - \mu^2)^{3/2}]\right\}$. We reinsert this solution into $h_-(a)$ to get the following first-order solution,
\begin{align}
    a_-^{(1)} &= \frac{2\pi \left(n - n_0\right) + h_-\left(a_-^{(0)}\right)}{\ln\left[\frac{\lambda}{4(1 - \mu^2)^{3/2}}\right]}.
\end{align}

Rather than repeating the process above for the even parity states and subtracting the even and odd parity state energies, the tunnel splitting can be approximated more accurately in the following way. Rewrite the quantization condition as
\begin{align}
    n_{\pm}(a) = \frac{a\ln\left[\frac{\lambda}{4(1 - \mu^2)^{3/2}}\right] - h_{\pm}(a)}{2\pi} + n_0.
\end{align}
For fixed $a = a_-^{(1)}$, the quantization condition for even and odd parity states gives different principal quantum numbers, whose difference is $\Delta n = (n_+ - n_-)|_{a_-^{(1)}} = \left.\frac{1}{\pi}\arctan(e^{-\pi a})\right|_{a_-^{(1)}}$. The tunnel splitting $\Delta a$ can be approximated using $\Delta n$ by dividing the derivative of $n_-(a)$,
\begin{align}
    &\Delta a = \left.\frac{\Delta n}{-n_-'(a)}\right|_{a_-^{(1)}} \\
    &= \frac{2\arctan\left(e^{-\pi a_-^{(1)}\ q(a_-^{(1)})}\right)}{\ln\left(\frac{4(1 - \mu^2)^{3/2}}{\lambda}\right) + \mathrm{Re}\left[\Psi\left(\frac{1}{2} + ia_-^{(1)}\right)\right] - \frac{\pi}{2}\mathrm{sech}\left(\pi a_-^{(1)}\right)},
\end{align}
where $\Psi(z)$ denotes the digamma function. Finally, using Eq.~\eqref{eq:definition-of-a} and Eq.~\eqref{eq:g-H-conversion}, we convert $\Delta a$ back to the real splitting $\delta_n$, and obtain
\begin{align}
    \delta_n &= (4\epsilon_2^2/K)\lambda\sqrt{1 - \mu^2}\,\Delta a \\
    &= 2K\sqrt{\alpha^4 - 1}\,\Delta a. \label{eq:tunnel-splitting-analytic-highly-excited-states}
\end{align}
Fig.~\ref{fig:wkb-splitting} compares the above analytical approximation to the numerically obtained tunnel splitting.

Finally, we consider the tunnel splittings for low excited state manifolds, in which $\delta_n$ is exponentially small and $g$ is close to $-(1 - \mu)^2/4$ (the minimum value for which the tunneling action $I(g, \mu)$ is completely real). In this regime we use Eq.~\eqref{eq:tunnel-splitting-expression-small-delta} for the tunnel splitting. We plug in the asymptotics for $I(g, \mu)$ [Eq.~\eqref{eq:I-action-asymptotics-low-energy-limit-2}]. We also use the energy eigenvalues Eq.~\eqref{eq:energy-estimate} from perturbation theory,
\begin{align}
    \left|\frac{\mathrm{d} E_n}{\mathrm{d}n}\right| = 4K\alpha^2\left[1 - \frac{n}{\alpha^2} + O\left(\frac{1}{\alpha^4}\right)\right].
\end{align}
After some manipulation, the result is
\begin{align}
    \delta_n &= \frac{1}{\pi}\left|\frac{\mathrm{d} E_n}{\mathrm{d}n}\right|e^{-2\alpha^2}e^{2\alpha^2 \delta\!I} \\
    &= \frac{1}{\pi}A_n\left|\frac{\mathrm{d} E_n}{\mathrm{d}n}\right|\alpha^{4n}e^{-2\alpha^2}e^{-\frac{\ln(\alpha^2)}{8\alpha^2}+\frac{B_n}{\alpha^2} + O\left[\frac{\ln(\alpha^2)}{\alpha^4}\right]} \\
    &=\frac{4A_n}{\pi}K\alpha^{4n+2}e^{-2\alpha^2}c_n(\alpha^2), \label{eq:tunnel-splitting-analytic-lowly-excited-states}
\end{align}
where $c_n(\alpha^2)$ denotes the following order-1 correction factor combined from that of the energy gap and of the tunnel action,
\begin{align}
    c_n(\alpha^2) &= \left[1 - \frac{n}{\alpha^2} + O\left(\frac{1}{\alpha^4}\right)\right] \nonumber \\
    &\quad \times \exp\left\{-\frac{\ln(\alpha^2)}{8\alpha^2}+\frac{B_n}{\alpha^2} + O\left[\frac{\ln(\alpha^2)}{\alpha^4}\right]\right\}.
\end{align}
The constants $A_n$ and $B_n$ are defined in Eqs.~\eqref{eq:definition-A-n} and \eqref{eq:definition-B-n}. The numerical value of the prefactor $4A_n/\pi$ is $115.9$ for $n=1$ and $980.3$ for $n=2$. Fig.~\ref{fig:wkb-splitting} compares the above analytical approximation to the numerically obtained tunnel splittings.

\subsection{Minimum energy gap}
\begin{figure}[th]
    \centering
    \includegraphics[width=\linewidth]{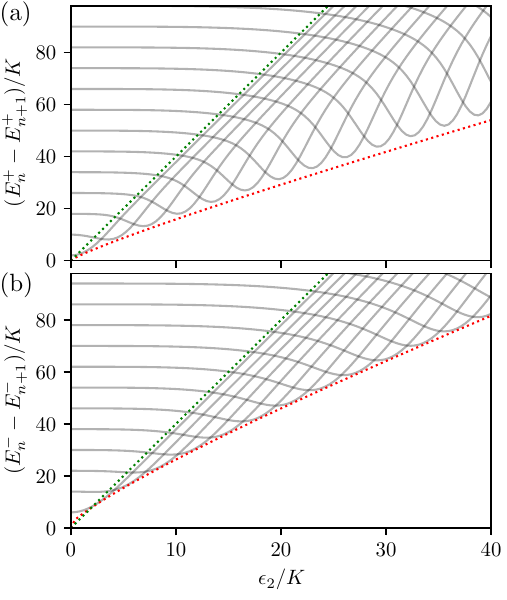}
    \caption{\textbf{Minimum energy gap.} The even and odd parity energy gaps $(E_n^{\pm} - E_{n+1}^{\pm})$ are plotted as functions of $\epsilon_2$ in (a) and (b) respectively as grey solid lines. The red dotted lines show the analytical approximation of the minimum energy gap derived in Eq.\eqref{eq:minimum-gap-approx}. The green dotted lines show the analytic approximation $E_{n}^{\pm} - E_{n+1}^{\pm} = 4\epsilon_2$.}
    \label{fig:minimum-gap}
\end{figure}
We can approximate the minimum energy gap by differentiating the approximate quantization condition Eq.~\eqref{eq:approximate-quantization-condition} with respect to $n$ at $a = 0$. This gives
\begin{align}
    \left.\frac{\mathrm{d}a_{\pm}}{\mathrm{d}n}\right|_{a=0} \ln\left[\frac{\lambda}{4(1 - \mu^2)^{3/2}}\right] = 2\pi + h_{\pm}'(0)\left.\frac{\mathrm{d}a_{\pm}}{\mathrm{d}n}\right|_{a=0}.
\end{align}
Therefore,
\begin{align}
    \left.\frac{\mathrm{d}a_{\pm}}{\mathrm{d}n}\right|_{a=0} = \frac{2\pi}{\ln\left[\frac{\lambda}{4(1 - \mu^2)^{3/2}}\right] - h_{\pm}'(0)}.
\end{align}
Applying the conversion factors defined by Eqs.\eqref{eq:definition-of-a} and \eqref{eq:g-H-conversion} between $a$ and energy $E$ gives
\begin{align}
    \left.\frac{\mathrm{d}E_n^{\pm}}{\mathrm{d}n}\right|_{g=0} &= \frac{-\frac{4\epsilon_2^2}{K}2\pi\lambda\sqrt{1 - \mu^2}}{\ln\left[\frac{\lambda}{4(1 - \mu^2)^{3/2}}\right] - h_{\pm}'(0)} \\
    &= \frac{4\pi K\alpha^2\sqrt{1 - \frac{1}{\alpha^2}}}{\ln\left[\frac{4(1 - \mu^2)^{3/2}}{\lambda}\right] -\Psi\left(\frac{1}{2}\right) \pm \frac{\pi}{2}},
\end{align}
where again $\Psi$ is the digamma function, and $\Psi\left(\frac{1}{2}\right) \approx -1.96351$. Ignoring factors of order one, we have
\begin{align}
    \min_n(E_{\text{gap}, n}) \approx \left.\frac{\mathrm{d}E_n^{\pm}}{\mathrm{d}n}\right|_{g=0} &\approx \frac{4\pi K\alpha^2}{\ln(8\alpha^2) + 1.96351 \pm \frac{\pi}{2}}. \label{eq:minimum-gap-approx}
\end{align}
We compare the predicted minimum energy gap against numerics in Fig.~\ref{fig:minimum-gap}.

\end{document}